\documentclass[
reprint,
twocolumn,nofootinbib,floatfix,
amsmath,amssymb,aps,prd,superscriptaddress]{revtex4-2}

\usepackage{xspace}
\usepackage{aas_macros}
\usepackage{graphicx}
\usepackage{graphics}
\usepackage[table,svgnames,dvipsnames]{xcolor}
\usepackage{hyperref}
\hypersetup{colorlinks=true,citecolor=blue,filecolor=blue,urlcolor=blue,}
\usepackage{orcidlink} 
\usepackage{booktabs}
\usepackage{amsmath}
\usepackage{hyperref}
\usepackage{bm}
\usepackage{tabularx}

\def\be{\begin{equation}}
\def\ee{\end{equation}}

\def\ba#1\ea{\begin{align*}#1\end{align*}}

\defcitealias{dMdB2020}{\texttt{dMdB20}}

\defcitealias{DESI2024.IV.KP6}{\texttt{DESI2024-IV}}

\defcitealias{Herbold:2026}{\texttt{H26}}
\newcommand{\validation}{\citetalias{Herbold:2026}\xspace}

\newcommand{\lya}{Ly$\alpha$\xspace}
\newcommand{\lyaf}{Ly$\alpha$ forest\xspace}

\newcommand{\lyaxlyaA}{Ly$\alpha$(A)$\times$Ly$\alpha$(A)\xspace}
\newcommand{\lyaxlyaB}{Ly$\alpha$(A)$\times$Ly$\alpha$(B)\xspace}
\newcommand{\lyaBxlyaB}{Ly$\alpha$(B)$\times$Ly$\alpha$(B)\xspace}

\newcommand{\lyaxqsoA}{Ly$\alpha$(A)$\times$QSO\xspace}
\newcommand{\lyaxqsoB}{Ly$\alpha$(B)$\times$QSO\xspace}

\newcommand{\hMpc}{$h^{-1}\;\text{Mpc}$}

\newcommand{\zeff}{z_{\rm eff}}
\newcommand{\planck}{{\it Planck}}
\newcommand{\lcdm}{{$\Lambda$CDM}}

\newcommand{\vega}{\texttt{Vega}}
\newcommand{\kmsMpc}{\,{\rm km\,s^{-1}\,Mpc^{-1}}}

\newcommand{\colore}{\texttt{CoLoRe}}

\newcommand{\lpt}{\texttt{CoLoRe 2LPT}}
\newcommand{\abacus}{\texttt{AbacusSummit}}

\newcommand{\phip}{$\phi_\mathrm{p}$}
\newcommand{\phis}{$\phi_\mathrm{s}$}

\newcommand{\alphap}{$\alpha_\mathrm{p}$}
\newcommand{\alphas}{$\alpha_\mathrm{s}$}

\newcommand{\alpac}{Alcock--Paczy\'nski}

\newcommand{\ea}[1]{{\color{orange} EA: #1}}

\newcommand{\leftparbox}[2]{\parbox{#1}{\begin{flushleft} #2 \end{flushleft}}}

\newcommand{\Planck}{\emph{Planck}}

\usepackage[nameinlink,noabbrev]{cleveref}
\crefname{equation}{Eq.}{Eqs.}
\crefname{section}{Section}{Sections}
\crefname{figure}{Fig.}{Figs.}
\crefname{table}{Tab.}{Tabs.}
\crefname{appendix}{Appendix}{Appendices}
\Crefname{figure}{Figure}{Figures}
\Crefname{equation}{Equation}{Equations}
\Crefname{section}{Section}{Sections}
\Crefname{table}{Table}{Tables}
\Crefname{appendix}{Appendix}{Appendices}

\begin{document}



\title{\textbf{DESI DR2 Results IV: \alpac\ Measurements from the Lyman Alpha Forest and Cosmological Constraints}
}%




\author{A.~G.~Adame\orcidlink{0009-0005-0594-9391}}
\affiliation{Department of Astrophysics, University of Vienna, Türkenschanzstrasse 17, 1180 Vienna, Austria}

\author{J.~Aguilar}
\affiliation{Lawrence Berkeley National Laboratory, 1 Cyclotron Road, Berkeley, CA 94720, USA}

\author{S.~Ahlen\orcidlink{0000-0001-6098-7247}}
\affiliation{Department of Physics, Boston University, 590 Commonwealth Avenue, Boston, MA 02215 USA}

\author{O.~Alves}
\affiliation{University of Michigan, 500 S. State Street, Ann Arbor, MI 48109, USA}

\author{A.~Anand\orcidlink{0000-0003-2923-1585}}
\affiliation{Inter-University Centre for Astronomy and Astrophysics, Post Bag 4, Ganeshkhind, Pune 411 007, India}

\author{U.~Andrade\orcidlink{0000-0002-4118-8236}}
\affiliation{Leinweber Center for Theoretical Physics, University of Michigan, 450 Church Street, Ann Arbor, Michigan 48109-1040, USA}
\affiliation{University of Michigan, 500 S. State Street, Ann Arbor, MI 48109, USA}

\author{E.~Armengaud\orcidlink{0000-0001-7600-5148}}
\affiliation{IRFU, CEA, Universit\'{e} Paris-Saclay, F-91191 Gif-sur-Yvette, France}

\author{S.~Avila\orcidlink{0000-0001-5043-3662}}
\affiliation{CIEMAT, Avenida Complutense 40, E-28040 Madrid, Spain}

\author{A.~Aviles\orcidlink{0000-0001-5998-3986}}
\affiliation{Instituto Avanzado de Cosmolog\'{\i}a A.~C., San Marcos 11 - Atenas 202. Magdalena Contreras. Ciudad de M\'{e}xico C.~P.~10720, M\'{e}xico}
\affiliation{Instituto de Ciencias F\'{\i}sicas, Universidad Nacional Aut\'onoma de M\'exico, Av. Universidad s/n, Cuernavaca, Morelos, C.~P.~62210, M\'exico}
\affiliation{Instituto de F\'{\i}sica, Universidad Nacional Aut\'{o}noma de M\'{e}xico,  Circuito de la Investigaci\'{o}n Cient\'{\i}fica, Ciudad Universitaria, Cd. de M\'{e}xico  C.~P.~04510,  M\'{e}xico}

\author{P.~Bansal\orcidlink{0009-0000-7309-4341}}
\affiliation{Leinweber Center for Theoretical Physics, University of Michigan, 450 Church Street, Ann Arbor, Michigan 48109-1040, USA}
\affiliation{University of Michigan, 500 S. State Street, Ann Arbor, MI 48109, USA}

% ...

\collaboration{DESI Collaboration}
\email{spokespersons@desi.lbl.gov}



\date{\today}


\begin{abstract}

We present \alpac\ (AP) measurements from the full shape of Lyman-$\alpha$ (\lya) forest correlation functions measured from the second data release (DR2) of the Dark Energy Spectroscopic Instrument (DESI). Our measurements include information from the \lyaf\ three-dimensional auto-correlation and its cross-correlation with quasar positions. The analysis was validated using two types of mocks, several data splits, and extensive variations. We constrain the AP effect with $1\%$ precision at an effective redshift $z_\mathrm{eff}=2.33$, which is twice as tight as the Baryon Acoustic Oscillation (BAO) constraint from the same data. When using the joint \lya\ AP and BAO results, we measure the ratios $D_\text{H}(z_\mathrm{eff})/r_\text{d}=8.600 \pm 0.066$ ($0.8\%$) and $D_\text{M}(z_\mathrm{eff})/r_\text{d}=39.32 \pm 0.33$ ($0.8\%$), where $D_\text{M}$ is the transverse comoving distance, $D_\text{H}$ is the Hubble distance, and $r_\text{d}$ is the size of the sound horizon at the drag epoch. Assuming \lcdm, \lyaf\ measurements combined with a big bang nucleosynthesis prior produce a constraint on the Hubble constant $H_0=66.5\pm1.3\kmsMpc$. The \lya\ AP result corresponds to a matter fraction constraint $\Omega_\text{m}=0.325\pm0.018$ in \lcdm, which is $1.4\sigma$ higher than DESI BAO. This impacts the DESI results relative to the Cosmic Microwave Background (CMB), slightly reducing their discrepancy from $2.4\sigma$ to $2.2\sigma$. We present updated constraints on several extended models using the joint DESI DR2 BAO and \lya\ forest full shape data, together with external data sets. When considering a time-evolving dark energy equation of state parametrized by $w_0$ and $w_a$, we find it is preferred over \lcdm\ at $2.7\sigma$ for the combination of DESI and CMB data, and at $3.2\sigma$ when also including supernovae. With the new \lya\ AP measurement, DESI provides its most precise anchor for the expansion history at $z > 1$ in the matter-dominated Universe.

\end{abstract}


\maketitle

\tableofcontents 


\section{Introduction}

The physics behind cosmic acceleration remains one of the most important problems in cosmology today, with progress requiring increasingly precise constraints of the cosmic expansion rate over a wide span of redshifts. Baryon acoustic oscillations (BAO) have long been central to studying the expansion history, with recent BAO measurements from the Dark Energy Spectroscopic Survey (DESI, \cite{Snowmass2013.Levi,DESI2016a.Science,DESI2022.KP1.Instr}) providing intriguing hints that dark energy might not be described by a cosmological constant \cite{DESI2024.VI.KP7A,DESIDR2:2025}. DESI BAO measurements come from galaxy clustering at $z\lesssim2$ and from the Lyman-$\alpha$ (\lya) forest at $z\gtrsim2$, with the two probes providing complementary constraints \cite[e.g.,][]{Cuceu:2019,DESI2024.VI.KP7A}. In this paper, we use the full shape of \lyaf\ correlation functions measured from the second DESI data release (DR2) to significantly improve constraints on the cosmic expansion rate at $z>2$. We also examine the cosmological implications of the new constraints when combined with DESI BAO results and with external probes.

The \lyaf\ is the pattern of absorption in spectra of high-redshift quasars (QSO) and galaxies, caused by intervening neutral hydrogen in the intergalactic medium (IGM). The forest represents a unique tracer of large-scale structure (LSS), because it allows us to continuously map the density field along each line-of-sight \cite[e.g.,][]{McQuinn2011}. Over the last few decades, it has served as the primary tracer of LSS at high redshift ($z>2$), a regime that has been difficult to access with large galaxy surveys thus far. The \lyaf\ and its cross-correlation with high-redshift quasars have been used to measure BAO starting with the Baryon Oscillation Spectroscopic Survey (BOSS) \cite{Dawson2013} more than a decade ago \cite{Slosar2013,Busca2013}, continuing with the extended BOSS (eBOSS) survey \cite{Dawson2016,dMdB2020} and recently with DESI \cite{DESI2024.IV.KP6,DR2Lya:2025}.

Over the last few years, large scales in the \lyaf\ correlation functions have also been used to extract cosmological information beyond BAO \cite{Cuceu2023a,Cuceu:2025}. These works leverage the \alpac\ (AP) effect \cite{AP1979}, which imprints a characteristic anisotropy due to differences between the true background cosmology and the fiducial cosmology used to convert observed angles and redshifts to comoving coordinates. This anisotropy signal is used in BAO analyses to constrain the ratio of the BAO size along versus across the line-of-sight. However, as the AP effect appears at all scales, using the full shape of \lyaf\ correlation functions can lead to significantly tighter constraints. Recent eBOSS and DESI full-shape \lyaf\ analyses have obtained a factor of $\sim2$ tighter constraints on AP when compared to BAO analyses of the same data sets \cite{Cuceu2023a,Cuceu:2025}. This significant improvement translates to much tighter constraints on the cosmic expansion rate.

In this work, we use the full shape of the \lyaf\ auto-correlation and the \lya-QSO cross-correlation measured from DESI DR2 data to constrain the AP effect. We use a template-based approach that allows us to isolate the AP signal while marginalizing over other effects. The data set used here comprises the first three years of DESI observations \cite{DESI2016b.Instr,DESI2023a.KP1.SV,Poppett:2020,Corrector.Miller.2023,Spectro.Pipeline.Guy.2023,SurveyOps.Schlafly.2023,Poppett:2024}. The DR2 data set represents an increase of about a factor of two in the number of \lya\ forests compared to the first data release (DR1, \cite{DESI2024.I.DR1,DESI2024.II.KP3}). Furthermore, \lya\ quasars are observed up to six times by DESI, which corresponds to significant improvements in the signal-to-noise of a large fraction of spectra in DR2. In addition to the increase in the sample size, we have also developed several improvements in our analysis methodology which are broadly relevant to \lyaf\ analyses.

This work is accompanied by a set of supporting papers that describe certain aspects of the analysis in significantly more detail. These include the validation of the method \cite{Herbold:2026} (hereafter \validation), a new set of mocks \cite{RuizHerrera:2026}, and the origin of our priors for small-scale non-linearities. \cite{ChavesMontero:2026}.

We begin by introducing the data and the \lyaf\ correlations we use in \Cref{sec:data}. Our methodology is then introduced in \Cref{sec:methods} and is followed by our validation tests using mocks and blinded data in \Cref{sec:validation}. We then present our main AP results in \Cref{sec:results}, along with the corresponding cosmic expansion rate constraints. Finally, we explore the cosmological implications of our measurements in \Cref{sec:cosmo} and conclude in \Cref{sec:conclusions}.

\section{Data}\label{sec:data}

We use the same dataset as described in detail in the DR2 \lya\ BAO
analysis \cite{DR2Lya:2025} and provide here a brief summary. The DR2 observations were taken between
May~2021 and April~2024 and include over 820,000 \lya\ forest spectra and
the positions of over 1.2 million quasars used for cross-correlations with the \lyaf, representing approximately a
factor of two increase in sample size relative to DR1. In addition, the typical 
\lya\ forest QSO has 2-3 observations in DR2, as opposed to a single 
observation in DR1. DESI collects
spectra with ten identical spectrographs covering $3600$--$9800$\,\AA\ in
three channels (blue, red, and near-infrared). The blue channel, with
spectral resolution $\mathcal{R}\simeq2000$--3000, is most relevant for the
\lya\ forest.  All data are processed through the DESI spectroscopic
pipeline \cite{Spectro.Pipeline.Guy.2023} and classified with the \textsc{Redrock} software
\cite{Anand:2024,Redrock.Bailey.2024}, supplemented by a Mg\,\textsc{ii} afterburner
\cite{QSO.TS.Chaussidon.2023} and the \textsc{QuasarNet} convolutional neural
network classifier \cite{Busca18,QN.Green.2025}.

\subsection{Catalogs}

The DR2 quasar catalog
contains 1,289,874 quasars with $z>1.77$ that can contribute to the \lya\
quasar cross-correlation and 824,989 with $z>2.09$ with a \lya\ forest
contributing to the correlations given our selection criteria, following the
same construction procedure as for DR1~\cite{DESI2024.IV.KP6}. In short, this 
catalog contains objects observed as part of the main quasar target selection \citep{QSO.TS.Chaussidon.2023} 
and the WISE variable quasar secondary targeting sample (DESI target bit 
\texttt{WISE\_VAR\_QSO}) that are spectroscopically classified as a quasar by either the
\textsc{Redrock} software, the Mg\,\textsc{ii} afterburner, or \textsc{QuasarNet}.
A quasar classification with \textsc{QuasarNet} is defined as at least one emission
line detected with a confidence greater than 0.99. The catalog is supplemented with 
quasars incidentally observed as Bright Galaxy Sample (BGS) \citep{BGS.TS.Hahn.2023} 
or Emission Line Galaxy (ELG) \citep{ELG.TS.Raichoor.2023} targets. For these 
objects, we apply stricter selection criteria than for main quasar or WISE quasar
targets. In both cases we require a quasar classification from \textsc{Redrock}:
an ELG-targeted object must additionally have at least one line detected by 
\textsc{QuasarNet} with a confidence greater than 0.6, while a BGS-targeted 
object must additionally have either at least one line detected by 
\textsc{QuasarNet} with a confidence greater than 0.6 or a quasar classification 
from the Mg\,\textsc{ii} afterburner. Including quasars from these additional targeting programs beyond the main quasar selection increases the catalog size by roughly 13\%.

For this analysis, we updated the Broad Absorption Line (BAL) quasar catalog and 
the Damped \lya\ Absorbers (DLA) catalog from the versions used in the DR2 BAO analysis. Both are used to mask out part of the
\lya\ forest spectral region when determining the quasar continuum and for the 
measurement of the correlation functions.

BAL quasars are identified in approximately 20.5\%
of quasars in the redshift range $1.77 < z < 3.8$ used for the \lya\
analysis.  We identify and mask the expected locations of BAL troughs
falling within the \lya\ forest following the same procedure as in
\cite{DESI2024.IV.KP6}.  The BAL catalog used here differs very slightly from the
one used in the DR2 BAO analysis due to minor code changes that increased
the efficiency of the BAL finding tool without altering its algorithm.

We use three methods to identify DLAs \cite{Wolfe:2005}, based respectively on a Convolutional Neural Network
(CNN, \cite{Wang2022}), Gaussian Processes (GP, \cite{Ho21}), and Template
Fitting (TMP, \cite{brodzeller2025}).  Since the BAO analysis, each of the
three methods has been improved and re-run, and the full-shape DLA catalog
is constructed from these updated results. It is worth noting that the largest
change in the behavior of the individual DLA detections is the improvement in the high SNR completeness of the TMP method (see \cite{brodzeller2025}). Using mock catalogs, we updated several combination choices relative to the BAO catalog.  First, we increased the redshift matching tolerance between catalogs from 800 to
3000\,km\,s$^{-1}$.  This slightly reduces the overall purity by 0.9\%
while increasing the completeness by 1\%, with a particular gain for DLAs
with $\log(N_\mathrm{HI}) > 22$\footnote{$N_\mathrm{HI}$ is the neutral hydrogen column density.} where completeness improves by 30\% with
minimal impact on purity. Second, we lowered the CNN detection confidence
threshold from $\mathrm{DLA\_CONF}>0.6$ to $>0.5$, which increased both
purity and completeness in the mocks.  Finally, the BAO analysis only
considered DLAs in sightlines with $\mathrm{SNR}>2$\footnote{Measured over
the rest-frame wavelength range $1420 < \lambda_\mathrm{rest} <
1480$\,\AA.}, whereas the full-shape catalog includes DLAs in all
sightlines.  Mocks show that completeness decreases below $\mathrm{SNR}=2$
while purity remains roughly constant, so the added DLAs maintain the
overall purity of the sample.

The impact of new DLA catalog changes are described in Section VB1 of \validation. In particular, 
we tested analysis variations with the DESI DR2 BAO DLA catalog \cite{brodzeller2025,DR2Lya:2025} and also with a few different SNR cuts in the new DLA catalog, and found minimal impact on our
measurements (see Figure 9 of \validation). 

\subsection{\lyaf\ measurements}
\label{subsec:deltas}

We measure the \lya\ forest in two regions of the QSO rest-frame spectrum:
region B ($920$--$1020$\,\AA) and region A ($1040$--$1205$\,\AA), following
the same procedure as in \cite{DESI2024.IV.KP6} and the DR2 BAO analysis.  After
applying masks for bad pixels, cosmic rays, BAL troughs, and DLA cores, and
discarding forests narrower than 120\,\AA, the product of mean \lya\ flux transmission
$\overline{F}\left(\lambda\right)$ times the unabsorbed quasar continuum
$C_q\left(\lambda/(1+z_q)\right)$ is estimated by a maximum-likelihood fit that
assumes this product to be a universal function of the QSO rest-frame wavelength corrected by a linear polynomial in $\log\lambda$
to account for quasar luminosity, spectral diversity, and redshift
evolution of the mean transmission~\cite{DESI2024.IV.KP6}.  The transmitted flux overdensity is then
\begin{equation}
    \delta_q(\lambda) = \frac{f_q(\lambda)}{\bar{F}(\lambda)\,C_q(\lambda/(1+z_q))} - 1.
\end{equation}
Because this procedure subtracts the mean flux and its slope from each
individual forest, it introduces distortions in the measured correlation
functions.

To model the impact of these distortions, we follow \cite{Bautista2017} and compute projection matrices $\eta_{ij}$ for each forest such that $\sum_j \eta_{ij} \delta^m_q(\lambda_j)=\sum_j \eta_{ij} \delta^t_q(\lambda_j)$, where $\delta^t_q$ and $\delta^m_q$ are the true and measured \lyaf\ flux decrements in the forest of quasar $q$. These projection matrices are given by:
\begin{equation}
    \eta_{ij} = \delta_{ij}^K - \frac{w_j}{\sum_k w_k} - \frac{w_j\kappa_i\kappa_j}{\sum_k w_k \kappa_k^2},
    \label{eq:etas}
\end{equation}
where $\delta_{ij}^K$ is the Kronecker delta, $\kappa_k=\log \lambda_k - \overline{\log \lambda_k}$, and the weights $w_i$ are the usual \lyaf\ pixel weights described in detail by \cite{Ramirez2024}. The $\eta_{ij}$ matrices multiply the measured $\delta^m_q$ before we compute the \lya\ correlation functions. On the other hand, for the theory side we model correlation functions and not $\delta^t_q$. Therefore, the $\eta_{ij}$ matrices are used to compute distortion matrices, $D_{MN}$, which multiply the model correlation functions. This is described in more detail in \Cref{subsec:ssm} below.

\subsection{\lyaf\ correlations}

Following previous DESI \lyaf\ analyses, we measure four correlation functions: \lyaxlyaA, \lyaxlyaB, \lyaxqsoA, and \lyaxqsoB, where \lya(A) and \lya(B) represent \lya\ absorption in the A and B regions defined above. We ignore the \lyaBxlyaB\ and QSO$\times$QSO correlations, as well as correlations with higher-order Lyman lines, because they carry less AP information \cite{Cuceu2021}. The correlations are computed on a rectangular grid of comoving separations $(r_\parallel, r_\perp)$ with $4\,$\hMpc\ bins extending to $200\,$\hMpc, using the fiducial flat-\lcdm\ cosmology from the DR2 BAO analysis \cite{DR2Lya:2025}. The covariance matrix is estimated from more than 1000 HEALPix sub-regions of the sky, following \cite{DESI2024.IV.KP6}. To obtain a positive definite covariance matrix, we apply the smoothing procedure described in \cite{DESI2024.IV.KP6}, which has been validated using mocks in previous DESI \lyaf\ analyses \cite[e.g.,][]{KP6s6-Cuceu,Casas2025,Cuceu:2025}. Some of these tests are replicated in the context of our DR2 full-shape measurements in \validation. The effective redshift of the combined sample is $\zeff = 2.33$, in line with previous DESI DR1 and DR2 \lyaf\ measurements \cite{DESI2024.IV.KP6,DR2Lya:2025}.

The binning results in $50\times50$ grids for the auto-correlations, and $100\times50$ grids for the cross-correlations. The larger number of bins in the cross-correlations is due to the use of negative and positive line-of-sight separations, which distinguish between \lya\ pixels in front and behind the quasar, respectively. Therefore, our measured data vector contains a total of 15,000 points. However, we impose an isotropic maximum scale cut of $r_\mathrm{max}=200$\hMpc, which reduces the number of fitted points to $11790$. Minimum-scale cuts are applied using a new small-scale marginalization method, which will be described in \Cref{subsec:ssm}. The reasoning and tests that led to our scale cuts are described in \Cref{subsec:mock_validation} below.

\section{Methodology}
\label{sec:methods}

In this section, we describe our methodology for measuring the \alpac\ effect from \lyaf\ correlation functions. Our approach closely follows that used for the same measurement in DESI DR1 \cite{Cuceu:2025} but incorporates model improvements developed for the DR2 BAO analysis \cite{DR2Lya:2025} and also two new improvements which were developed during this project.

We begin with a detailed description of the \alpac\ effect in \Cref{subsec:alpac}. After that, we give an overview of our model for the \lyaf\ correlation functions and its contaminants in \Cref{subsec:model_corr}. Finally, we present the two new methodology improvements in \Cref{subsec:ssm,subsec:uvb}.

\subsection{The \alpac\ effect}
\label{subsec:alpac}

\subsubsection{Modeling and parametrization}

The AP effect is the apparent stretching or compression of intrinsically isotropic cosmic structures that occurs when angles and redshifts are converted into distances using an incorrect cosmological model. For a 3D tracer, its correlation function in comoving coordinates, $\xi(r_\parallel, r_\perp)$, is derived from measurements of angular separations $\Delta \theta$ and redshift pairs $(z_i, z_j)$, with:
\begin{align}
    r_{||} &= [D_\mathrm{C}(z_i) - D_\mathrm{C}(z_j)] \cos{\frac{\Delta \theta}{2}}, \\
    r_\bot &= [D_\mathrm{M}(z_i) + D_\mathrm{M}(z_j)] \sin{\frac{\Delta \theta}{2}},
\end{align}
where the comoving distance, $D_C(z)$, and the comoving angular diameter distance, $D_M(z)$, are computed based on a fiducial background cosmology (see \Cref{tab:fid_cosmo}). 
When fitting a model, we can account for this effect by introducing two scale parameters, $q_{||}$ and $q_\bot$, which rescale the coordinates along and across the line-of-sight, respectively. These parameters account for the ratio between the coordinates obtained using the fiducial cosmology ($r_\parallel,r_\perp$) and the coordinates we would have obtained using the true cosmology ($r'_\parallel,r'_\perp$), such that $r'_\parallel=q_\parallel r_\parallel$ and $r'_\perp = q_\perp r_\perp$. In this work, we follow \cite{Cuceu2021} and isolate the AP anisotropy using the parameter:
\begin{align}
    \phi \equiv \frac{q_\perp}{q_\parallel} = \frac{D_M / D_H}{[D_M / D_H]_{\rm fid}},
\end{align}
where $D_H=c/H$, with the Hubble parameter $H$, and speed of light $c$. Therefore, $\phi$ measures the ratio of the true AP parameter $F_\mathrm{AP}=D_M / D_H$ with respect to its fiducial value. Note that this measurement does not depend on a specific scale, such as BAO, since the AP effect distorts all scales in our measured correlations equally. For more details on this parametrization and the motivation behind this approach, see \cite{Cuceu2021}.

\begin{table}
\centering
\begin{tabular}{cc}
\hline
\hline
Parameter                              & Planck 2018 cosmology             \\
                                       & (TT,TE,EE+lowE+lensing)                \\
\hline
$\Omega_{\rm m}h^2$ =                  & 0.14297                             \\
$+\Omega_{\rm c}h^2$                   & 0.12                                \\
$+\Omega_{\rm b}h^2$                   & 0.02237                             \\
$+\Omega_{\rm \nu}h^2$                 & 0.0006                              \\
$h$                                    & 0.6736                              \\
$n_{\rm s}$                            & 0.9649                              \\
$10^9 A_{\rm s}$                       & 2.100                               \\

\hline
$\Omega_{\rm m}$                       & 0.31509                            \\
$\Omega_{\rm r}$                       & 7.9638e-05                         \\
$\sigma_8 (z=0)$                     & 0.8119                               \\
$f(\zeff=2.33)$                     & 0.9703                               \\
$f\sigma_8 (z=2.33)$                     & 0.2979                               \\
$r_{\rm d} \; [\rm Mpc]$             & 147.09                              \\
$r_{\rm d} \; [h^{-1} \rm Mpc]$      & 99.08                               \\
$D_{\rm H}(\zeff=2.33)/r_{\rm d}$   & 8.6172                               \\
$D_{\rm M}(\zeff=2.33)/r_{\rm d}$   & 39.1879                              \\
\hline
\end{tabular}
    \caption{
    The fiducial cosmology parameters used in this work. The first set of parameters defines the cosmological model in \lcdm, while the second set contains derived quantities used in this work. The fiducial cosmology is used to convert observed angles and redshifts into comoving coordinates, and to compute the template matter power spectrum at the core of our model for \lya\ correlations.}
    \label{tab:fid_cosmo}
\end{table}

In BAO analyses~\cite[e.g.,][]{DR2Lya:2025}, the acoustic scale is constrained using a template approach based on an isotropic matter power spectrum $P_\mathrm{fid}(k)$. This template is decomposed into a peak (or wiggle) component and a smooth (or no-wiggle) component. The AP effect from the BAO scale is then measured by rescaling the peak component anisotropically, while the coordinates of the smooth components are kept fixed. Here, we extend this framework by also rescaling the coordinates of the smooth component, thus extracting the AP effect from the full shape of \lyaf\ correlations. This approach was first introduced in the context of \lyaf\ analyses by \cite{Cuceu2021}, and we follow their methodology here. Finally, we note that we use the terms smooth component and broadband interchangeably in this work.

The parametrization we use for our scale parameters is given by the $\phi$ parameter above and by an isotropic scale parameter $\alpha$, defined as $\alpha \equiv \sqrt{q_\perp q_\parallel}$. We choose this particular combination because it is close to the optimal combination that minimizes the uncertainty in the isotropic BAO signal from the \lyaf\ \cite[see e.g.,][]{Cuceu2021,DESI2024.IV.KP6}. This choice does not impact our results in any way, and is purely a matter of preference.

Because our template still uses a peak and a smooth component (as in BAO analyses), we introduce two sets of scale parameters: one for the peak component (\phip,\alphap) and one for the smooth component (\phis,\alphas). As described above, the interpretation of the AP parameters is consistent for both components. In the case of \alphap, the measurement corresponds to:
\begin{align}
\alpha_p \equiv \sqrt{q_\perp q_\parallel} = \frac{\sqrt{D_M D_H} / r_d}{[\sqrt{D_M D_H} / r_d]_{\rm fid}},
\end{align}
where $r_d$ is the size of the acoustic scale at the drag epoch. In principle, the isotropic scale from the smooth component, \alphas, also contains cosmological information, likely related to the scale of matter-radiation equality \cite{Cuceu2021}. However, this measurement is fairly weak and affected by many of the contaminants in the \lyaf. Therefore, we treat \alphas\ as a nuisance parameter and marginalize over it, in line with previous analyses \cite{Cuceu2023a,Cuceu:2025}. As the parameters for the BAO peak (\phip,\alphap) have already been measured in the DR2 BAO analysis \cite{DR2Lya:2025}, our main goal here is to also measure the AP effect from the broadband through the \phis\ parameter. Our final results will include consistent measurements of all three parameters: \phis, \phip, \alphap.

\subsubsection{\alpac\ versus RSD}

In practice, the AP effect is just one of multiple effects that introduce anisotropies in our measurements. Therefore, in order to measure the AP effect, we need to marginalize over other sources of anisotropy. The primary anisotropy present in our correlations is due to redshift-space distortions (RSD). When the anisotropy due to RSD has a functional form similar to that introduced by the AP effect, marginalizing over RSD results in a significant loss of AP information. Indeed, when the two functional forms are identical, the effects are completely degenerate, and we cannot extract information from either.\footnote{Unless an external constraint on one of the quantities is used to break the degeneracy.} On the other hand, when there is a strong scale dependence in our measured correlation functions or power spectra, the two effects are fairly distinct. This is because scale-dependent features like BAO allow us to distinguish between a coordinate change (AP) and an amplitude change (RSD), even when the two anisotropies have the same functional form.

\begin{figure}
    \centering
    \includegraphics[width=1.\columnwidth,keepaspectratio]{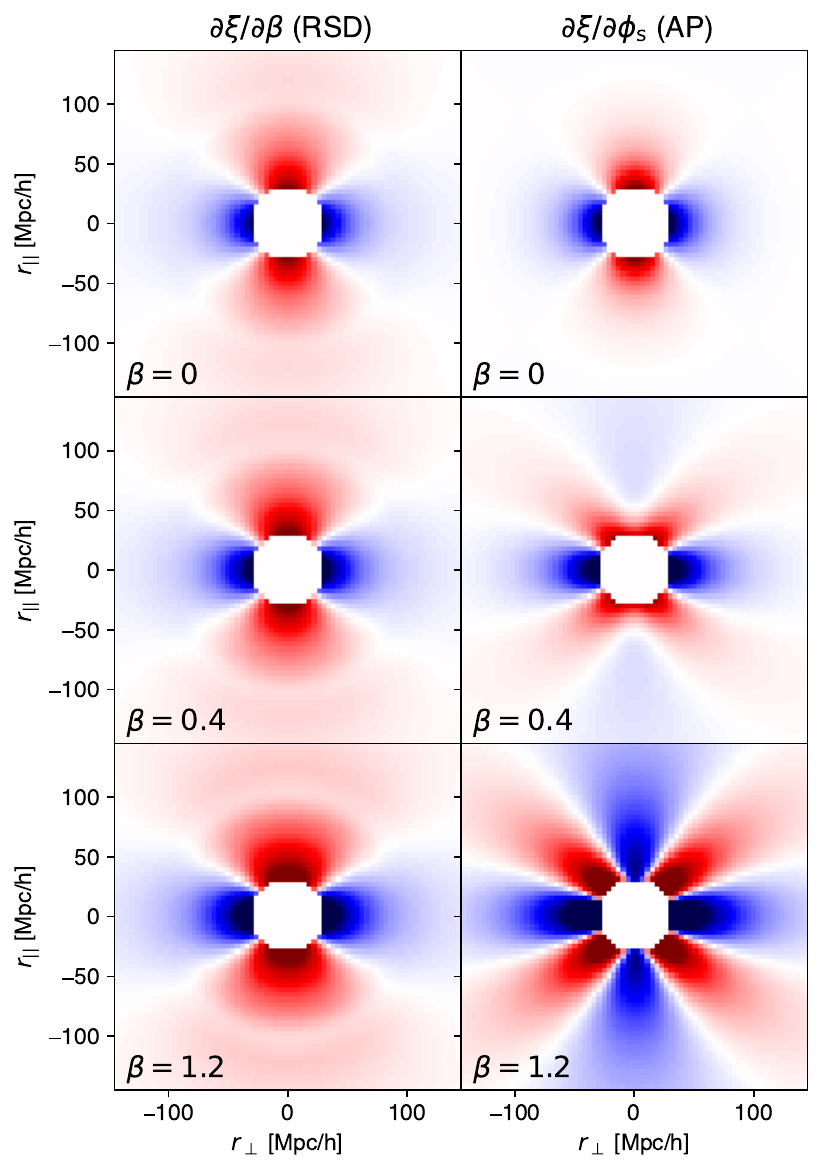}
    \caption{
        Illustration of parameter sensitivity in configuration space for a linear theory model, focusing on the smooth component only (i.e., no BAO). Each row shows a different fixed value of the RSD parameter $\beta$, while the left and right columns show the response to small changes in $\beta$ and in the broadband AP parameter $\phi_s$, plotted as functions of transverse and line-of-sight separations. The response to $\beta$ is dominated by LOS-versus-transverse anisotropy and preserves its shape with increasing RSD. On the other hand, the response to $\phi_s$ evolves from a shape similar to the one for $\beta$ at null RSD to a significantly different shape as $\beta$ increases. Thus, the RSD and AP signals become easier to disentangle for tracers with larger values of $\beta$.
    }
    \label{fig:derivatives}
\end{figure}

We illustrate the similarities and differences between RSD and the AP effect in \Cref{fig:derivatives} by plotting the response of a linear theory (Kaiser) auto-correlation model in configuration space to changes in the RSD amplitude parameter $\beta$ (left column) and to small changes in the broadband AP parameter \phis\ (right column). For this comparison, we use only the smooth component (i.e., without BAO) and show each response for models with three different values of $\beta$, including $\beta=0$ (no RSD), $\beta=0.4$ (typical value for galaxy clustering), and $\beta=1.2$ (typical value for the \lyaf). We also mask the small scales ($r < 30$\,\hMpc), as in that regime linear theory breaks down, and the comparison would need to include marginalization over non-linear effects.\footnote{The exact scale where this happens can be quite different depending on the tracer and redshift used. We only use this as an example to illustrate why AP and RSD are less degenerate for \lya, so we refrain from a more realistic investigation in the context of galaxy tracers, and use the $r > 30$\,\hMpc\ scale cut (which we have validated for the \lyaf\ at redshifts $2 < z < 4$) for all comparisons.}

In the case of null RSD, the responses to the two parameters have very similar shapes to each other, indicating that the two effects are close to degenerate. In fact, one can show that in the case of a purely scale-independent (i.e., power law) power spectrum, the two effects are exactly degenerate in the limit of null RSD.\footnote{While we use the smooth component here, this still has some scale dependency at low-$k$ due to the turn-over in the matter power, which explains why the two responses are not exactly identical here.} On the other hand, as $\beta$ increases, the response to AP evolves quite significantly, while the response to RSD preserves its shape. Furthermore, with increasing $\beta$, the amplitude of the response to AP also significantly increases, which is illustrated through darker colors in the bottom right panel versus the top right panel of \Cref{fig:derivatives} (the same color scale is used for each column). For galaxy clustering, the relevant comparison is the one in the middle row, where the two responses still have a fairly similar pattern at small separation where amplitudes of the derivatives are largest. Finally, the comparison relevant for the \lyaf\ is the one in the bottom row, where the response to AP has a very distinct pattern from the one of the response to RSD.

The comparisons in \Cref{fig:derivatives} show that AP is easier to disentangle from RSD in \lyaf\ analyses compared to galaxy clustering analyses. This is one of the key reasons why \lyaf\ analyses have so far delivered significant improvements (a factor of $2$ to $2.4$ better) in the AP constraints from full-shape analyses versus BAO analyses of the same datasets \cite{Cuceu2023a,Cuceu:2025}. On the other hand, galaxy clustering analyses have only delivered mild improvements ($10\%$ to $30\%$ better) in full-shape AP compared to BAO \cite[e.g.,][]{DESI2024.V.KP5,DESI2024.VII.KP7B}, although it is important to note that modeling non-linearities also plays a more important role in that case.\footnote{Both due to galaxies tracing highly clustered environments when compared to the \lyaf, and because large galaxy surveys have so far been done at lower redshifts.}

\subsection{Modeling Ly$\alpha$ forest correlations}
\label{subsec:model_corr}

Our model of the \lyaf\ correlation functions closely follows that used in \cite{DR2Lya:2025} and \cite{Cuceu:2025}, and is implemented in the \vega\ package\footnote{\url{https://github.com/andreicuceu/vega}} \cite{Cuceu2023b}.  The anisotropic \lya\ auto- and \lya--QSO cross-power spectra are given by:
\begin{align}
    P_F(\mathbf{k}) =& b'^2_F(k) \left(1 + \beta'_F(k) \mu_k^2\right)^2 \nonumber \\
        & \times G(\mathbf{k})\,F_{NL}(\mathbf{k})\,P_\mathrm{fid}(k),  \label{eq:pk1} \\
    P_{F\times Q}(\mathbf{k}) =& b'_F(k) \left(1 + \beta'_F(k) \mu_k^2\right)
        \left(b_Q + f \mu_k^2\right) \nonumber \\
        & \times G(\mathbf{k})\,X_{NL}(\mathbf{k})\,P_\mathrm{fid}(k),\label{eq:pk2}
\end{align}
where $\mathbf{k}=(k_\parallel,k_\perp)$, $\mu_k=k_\parallel/k$, and $P_\mathrm{fid}(k)$ is computed with \textsc{camb}~\cite{Lewis:2019} at $\zeff=2.33$ using the fiducial \planck\ cosmology in \Cref{tab:fid_cosmo}.  The function $G(\mathbf{k})$ accounts for the finite bin size ($4\,h^{-1}$\,Mpc) of the correlation function grids and is given by a sinc function \cite{dMdB2020}. The effective \lya\ bias $b'_F$ and RSD parameter $\beta'_F$ account for both \lya\ flux and undetected HCD absorption and are a function of $k_\parallel$ (see \Cref{sec:hcd}). They also have an extra scale-dependent contribution due to UV background fluctuations (see \Cref{subsec:uvb}). $b_Q$ is the linear quasar bias and $f$ is the logarithmic derivative of the linear growth factor.  $F_{NL}$ and $X_{NL}$ model small-scale non-linearities in the auto- and cross-correlations respectively, and are described below. These anisotropic power spectra are transformed into 2D model correlation functions via a multipole decomposition up to $\ell=6$ followed by a Hankel transform.\footnote{We have tested that going up to $\ell=10$ does not have a significant impact on our results.}

Note that in \Cref{eq:pk1,eq:pk2}, we made explicit the dependency of the QSO term on $f$, but used the RSD parameter $\beta_F$ for the forest terms. This is because the \lyaf\ RSD parameter is given by $\beta_F=b_\eta f / b_F$, which includes an extra unknown velocity divergence bias, $b_\eta$. The presence of $b_\eta$ precludes direct measurements of the cosmic growth rate from linear scales in the \lyaf, unless other sources of information are used to constrain it and break its degeneracy with $f$ (e.g., from hydro-simulations \cite{ForestFlow_2025}). Therefore, we follow previous \lyaf\ analyses and directly marginalize over $\beta_F$ \cite{Cuceu2023a,Cuceu:2025}. We also note that in early versions of our analysis, we attempted to measure the growth rate times the amplitude of fluctuations in $8$\,\hMpc\ regions, $f\sigma_8$, from the QSO term in the cross correlation, following \cite{Cuceu2021,Cuceu:2025}. However, as in DESI DR1 \cite{Cuceu:2025}, we decided to de-scope that measurement before unblinding because we found it was not robust in either mocks or data. Given that the constraint is fairly weak ($\sim15\%$), we chose to simply treat the growth rate as a nuisance parameter to be marginalized. See Appendix E of \validation\ for more details.

\subsubsection{Small-scale non-linearities}

For the \lya\ auto-correlation, we model small-scale deviations from linear theory with the empirical model of \cite{Arinyo2015}:
\begin{multline}
    F_{NL}(\mathbf{k}) = \exp \Biggl\{
    \left[q_1 \Delta^2(k) + q_2 \Delta^4(k) \right] \\ \times \left[1-\left(\frac{k}{k_v}\right)^{a_v} \mu^{b_v}\right] - \left(\frac{k}{k_p}\right)^2 \Biggl\},
    \label{eq:arinyo}
\end{multline}
where $\Delta^2(k)\equiv k^3 P_\mathrm{fid}(k)/(2\pi^2)$ and $\{q_1,q_2,k_v,a_v,b_v,k_p\}$ are either treated as free parameters or fixed to values measured from other datasets. This model was shown to accurately reproduce small-scale \lya\ power spectra in hydrodynamical simulations \cite{Arinyo2015,Givans2022}.

Previous \lyaf\ full-shape analyses \cite{Cuceu:2025} marginalized over various combinations of parameters in this model. However, we found that after introducing marginalization over small scales in the undistorted correlations, as described in \Cref{subsec:ssm} below, we lost sensitivity to these parameters.\footnote{By loss of sensitivity we mean that the posterior does not improve upon the prior for these parameters. See \validation\ for more details.} Therefore, we decided to fix them in our baseline model to the values measured in the companion paper \cite{ChavesMontero:2026}: $q_1=0.303$, $q_2 = 0.267$, $k_v = 0.576$, $a_v = 0.443$, $b_v = 1.66$, $k_p = 11.062$. These constraints come from measurements of the \lyaf\ 1D power spectrum \cite{Karacayli_2025,Ravoux_2025,P1DCosmo:2026} and used the ForestFlow emulator trained on hydrodynamic simulations \cite{ForestFlow_2025}.\footnote{We note that pixel pairs in the same forest are not included in our 3D correlation measurements, which means there is no overlap between the data used here and that used by \cite{ChavesMontero:2026} to obtain these constraints. Furthermore, this information comes from scales that are significantly smaller than those used in our analysis \cite{P1DCosmo:2026}.}. We tested the impact of this choice and found that it does not significantly affect either the values nor the uncertainties of our measurements. This is discussed in more detail in \Cref{subsec:data_validation}, and in \validation.

\subsubsection{Quasar redshift errors}

For the cross-correlation, the dominant small-scale effect is due to quasar
redshift errors \cite{FontRibera2013}, which we model as a Lorentzian
damping of the cross-power spectrum:
\begin{equation}
    X_{NL}^2 = [1 + (k_{||}\sigma_z)^2]^{-1},
\end{equation}
where $\sigma_z$ is a free parameter. In practice, this model is also sensitive to the impact of large peculiar velocities (finger-of-god effect). \validation\ tested using a Gaussian damping instead of the Lorentzian, and found no significant impact on our constraints. The fact that our results remain insensitive to the exact functional form of this damping is a good indication that our scale cuts in the cross-correlation, $r > 40$\,\hMpc, are conservative enough.

In addition to the damping effect, we also account for the impact of systematic redshift errors. The \lya-QSO cross-correlation has both negative and positive $r_{||}$ values to distinguish between \lya\ absorption in front versus behind the quasar it is correlated with (\Cref{sec:data}). Systematic quasar redshift errors lead to a shift of the measured cross-correlation along the $r_{||}$ direction. We account for this possible shift with an extra parameter, $\Delta r_{||}$, that shifts the coordinate grid of the cross-correlation along the line-of-sight.

\subsubsection{High column density (HCD) absorbers}\label{sec:hcd}

Strong HCDs ($\log N_\mathrm{HI}>20.3$, i.e.\ DLAs) are masked as described in \Cref{sec:data}.  Weaker HCDs, and those in low-SNR spectra which have low DLA-finder completeness \cite{brodzeller2025}, contribute to the correlations through effective \lya\ bias and RSD
parameters \cite{FontRibera2012b,dMdB2020}:
\begin{align}
    b'_F &= b_F + b_\mathrm{HCD} F_{HCD}(k_{||}), \\
    b'_F \beta'_F &= b_F \beta_F + b_\mathrm{HCD} \beta_\mathrm{HCD} F_{HCD}(k_{||}),
\end{align}
where $b_\mathrm{HCD}$ and $\beta_\mathrm{HCD}$ are the HCD linear bias and
RSD parameter.  As in the BAO measurement \cite{DR2Lya:2025}, we adopt the exponential profile
$F_\mathrm{HCD} = \exp(-L_\mathrm{HCD}\,k_\parallel)$, where
$L_\mathrm{HCD}$ is a free parameter representing the typical scale of
undetected HCDs.

\subsubsection{Metal contamination}

Metal absorption near the \lya\ rest-frame wavelength in the \lya\ forest produces spurious peaks along the line of sight in the correlation functions because we treat all forest pixels as \lya\ pixels when computing their redshift \cite{Pieri2014,Morrison2023}. We model the contribution of four silicon lines (Si\,\textsc{ii}(1207), Si\,\textsc{ii}(1190), Si\,\textsc{ii}(1193) and Si\,\textsc{ii}(1260)), modeling their cross-correlation with \lya and the other metal lines, along with their auto-correlation. We also include the CIV auto-correlation from smaller redshifts. Each metal line is treated as an independent tracer of LSS and has its own free linear bias parameter and an RSD parameter fixed to 0.5, following \cite{Bautista2017,KP6s5-Guy}.\footnote{We have tested that fixing the metal RSD parameters to 0.4 and 0.6 does not lead to significant changes in our constraints.} For each contributing pair of transitions, the contaminated model correlation is mapped from the true comoving separation to the observed grid via a {\it metal} matrix (see Eq. 10 and 11 in \cite{Cuceu:2025}).

\subsubsection{Other effects}

We also include two additional contamination terms.  First, spectra from
fibers of the same spectrograph share correlated sky noise, which
contributes primarily at purely transverse separation ($r_\parallel = 0$).
We model this with a fixed template
$\xi_\mathrm{noise}(r_\perp)$ and a free amplitude $a_\mathrm{noise}$, following \cite{KP6s5-Guy,DESI2024.IV.KP6}.

Second, for the \lya--QSO cross-correlation we include the transverse
proximity effect \cite{FontRibera2013}, modeled as an isotropic profile \\
$$\xi^\mathrm{TP} =\xi_0^\mathrm{TP}  \left((1 \, h^{-1}\mathrm{Mpc})/r\right)^{2} \exp(-r/\lambda_\mathrm{UV})$$
with a free amplitude $\xi_0^\mathrm{TP}$ and fixed scale $\lambda_\mathrm{UV}=300\,h^{-1}$\,Mpc.

\subsection{Small-scale marginalization}
\label{subsec:ssm}

As mentioned in \Cref{subsec:deltas}, the measured \lyaf\ correlations are distorted by the fact that we fit for an amplitude and slope in the continuum of each forest, which has the effect of removing LSS modes of the size of the forest and larger \cite{Blomqvist:2015}. To model the impact of this distortion on our model correlations, we compute distortion matrices, $D_{MN}$, which multiply the correlation function model, $\xi(r_{||},r_\bot)$, to give the distorted correlation:
\begin{equation}
    \hat\xi_M = \sum_N D_{MN} \xi_N,
    \label{eq:dm_xi}
\end{equation}
where $N$ indexes over undistorted model bins, and $M$ over data bins. The distortion matrices for the \lya\ auto and QSO cross-correlation are given by:
\begin{align}
    D_{MN}^\mathrm{auto} &= \frac{1}{W_M} \sum_{i,j \in M} w_i w_j \sum_{i',j' \in N} \eta_{ii'} \eta_{jj'} f_{a}(z_i,z_j,z_\mathrm{ref}), \label{eq:dmat1} \\
    D_{MN}^\mathrm{cross} &= \frac{1}{W_M} \sum_{i,j \in M} w_i w_j \sum_{i',j \in N} \eta_{ii'} f_c(z_i,z_j,z_\mathrm{ref}), \label{eq:dmat2}
\end{align}
where $W_M=\sum_{i\in M}w_i$, and the model bins $N$ are half the size (i.e., $2$\,\hMpc) of our data bins $M$. The functions $f_a$ and $f_c$ account for the evolution of the
amplitude of the correlation functions between a reference redshift $z_\mathrm{ref}$ and the redshifts of each measurement \cite{2025PhRvD.112h3514A}.

The impact of multiplying the model correlations with the distortion matrix has been studied in detail in \cite{Bautista2017,Busca:2025}. The primary effect is a mixing of information along the line-of-sight in the correlation function (i.e., as a function of $r_{||}$). During our validation process, we encountered a consequence of this mixing which, if ignored, could lead to percent-level biases in \lya\ AP constraints (see \Cref{sec:ssm_app}). In particular, we found that this mixing results in significant information from very small $r_{||}$ scales in $\xi_N$ (i.e. $r_{||,N} < 10$\,\hMpc) leaking into all scales of $\hat\xi_M$ (i.e., including $r_{||,M} > r_\text{min}$). This means that our usual scale cuts $r>r_\text{min}$ do not have the intended effect of completely removing our sensitivity to scales $r<r_\text{min}$. To resolve this issue, we introduce a new formalism for marginalizing over this small-scale contamination.

The first step in this new approach is to marginalize over small-scale bins at $r<r_\text{min}$ in the undistorted model space, $\xi_N$. We do this by adding one free amplitude parameter $\eta_T$ for each bin $T$ in the undistorted model, where the bins $T$ do not have to be the same as the native undistorted model bins $N$. Our choice of bins $T$ is such that they match the data bins $M$ that are coarser than the model bins. The amplitudes $\eta_T$ multiply a linear template in the distorted space, given by $\hat{\Delta}_{MT} = \sum_N D_{MN}\ \Delta_{NT}$, where $\Delta_{NT} = 1$ if the finer bin N is included in the coarser data bin T and 0 elsewhere. In the case of linear templates, the most efficient method to find $\eta_T$ is to analytically solve for the template amplitudes that maximize the posterior density (i.e., minimize the $\chi^2$). We denote all model parameters with $\theta$ and the chi-squared with a general Gaussian prior is:
    \begin{equation}
        \begin{split}
        \chi^2 =\;& (\xi_\mathrm{obs} - \hat\xi_\mathrm{model}(\bm{\theta}))^\mathrm{T}\
        \mathrm{C}^{-1}\
        (\xi_\mathrm{obs} - \hat\xi_\mathrm{model}(\bm{\theta})) \\
        &+ (\bm{\theta}-\bm{\mu})^\mathrm{T}\
        \mathbf{S}^{-1}\
        (\bm{\theta}-\bm{\mu}) ,
        \end{split}\label{eq:ssm}
    \end{equation}
\noindent
with a mean $\bm \mu$ and covariance matrix $\mathbf{S}$, where $S_{ii}\rightarrow \infty$ is equal to an uninformative prior on a given parameter $\theta_i$. Omitting the template indices $T$, we denote the vector of linear template amplitudes $\bm \eta$, and the associated templates with a matrix $\mathbf{\hat{\Delta}}$, such that $\mathbf{\hat{\Delta}}\bm\eta$ results in a vector of the same shape as the data vector $\xi_\mathrm{obs}$. Additionally, the model parameters $\bm \theta$ can absorb the mean $\mu$, which is set to zero for simplicity. With this definition, $\bm \eta$ parametrizes deviations from the model within each bin. A modified covariance matrix $\mathrm{\tilde C} = \mathrm{C} + \mathbf{\hat{\Delta}} \mathbf{S} \mathbf{\hat{\Delta}}^\mathrm{T}$ captures the marginalized chi-squared:
    \begin{equation}
        \chi^2_\mathrm{marg} = (\xi_\mathrm{obs} - \hat\xi_\mathrm{model}(\bm \theta))^\mathrm{T}\  \mathrm{\tilde C}^{-1}\ (\xi_\mathrm{obs} - \hat\xi_\mathrm{model}(\bm \theta)).
    \end{equation}
    
Note that this formulation front-loads the marginalization into a covariance matrix update at the beginning of the analysis, so that the best-fitting $\bm \eta_\mathrm{MAP}$ are not computed per model evaluation. These are instead computed at the end as follows:
    
    \begin{equation}
        \bm \eta_\mathrm{MAP} = (\mathbf{\hat{\Delta}}^\mathrm{T} \mathrm{C}^{-1} \mathbf{\hat{\Delta}} + \mathbf{S}^{-1})^{-1} ( \mathbf{\hat{\Delta}}^\mathrm{T} \mathrm{C}^{-1}r ),
    \end{equation}
\noindent
where $r\equiv \xi_\mathrm{obs} - \hat\xi_\mathrm{model}(\bm \theta_*)$ and $\bm \theta_*$ minimizes $\chi^2_\mathrm{marg}$.

In practice, we add as many templates (rows of $\mathbf{\hat{\Delta}}$) and their associated amplitude parameters ($\bm\eta$) as there are data bins at $r<r_{\rm min}$. Though this introduces many more parameters, we also retain all data at $r<r_{\rm min}$ in the $\chi^2$ fit such that each template amplitude parameter is well constrained by the corresponding data bin. Consequently, this marginalization technique does not add degrees of freedom when compared to a fit without small-scale marginalization and the inclusion of data at $r \ge r_{\rm min}$ only. Rather, it ensures that we correctly model the leakage of small-scale correlations at larger scales caused by the spectral continuum fitting and encoded in the distortion matrix.

\subsection{UV background fluctuations}
\label{subsec:uvb}

Past \lyaf\ analyses were based on the assumption of a homogeneous ionizing UV background (UVB). However, work by \cite{Pontzen2014} and \cite{Gontcho2014} has shown that fluctuations in this ionizing background can give rise to a scale-dependent \lya\ bias, where the relevant scales are given by the mean-free path of UV ionizing radiation, which is expected to be of order hundreds of Mpc at redshift $2.3$. During our validation process, we tested including a model for this effect (as was done in previous analyses, e.g. \cite{2025PhRvD.112h3514A,Cuceu:2025}) and found that it is detected at high statistical significance ($\sim3\sigma$ to $\sim5\sigma$ depending on the exact model setup). Therefore, we decided to include it in our baseline model. We use the simplest version of the model, which only includes the scale-dependent \lya\ bias, as described below.

We model the impact of UVB fluctuations following \cite{Gontcho2014}, which started by introducing fluctuations in the photoinization rate $\Gamma(x)$, defined as $\delta_\Gamma(x)=\Gamma(x)/\bar\Gamma-1$, and showed that they give rise to a scale-dependent \lya\ bias, given by:
\begin{equation}
    b_F(k) = b_\delta + b_\Gamma \frac{(b_s - b_a) W(k/\kappa_0)}{1 + b_a' W(k/\kappa_0)},
\end{equation}
where $W(s)=\arctan(s) / s$, $b_s$ is the bias of the UV sources, $b_a$ is the bias of UV absorbers, $b_a'$ is the response of these absorbers to changes in the ionizing intensity, and $\lambda_0=1/\kappa_0$ is the mean free path of ionizing photons.

We treat $b_\Gamma$ as a free parameter and fit it. We follow \cite{Gontcho2014} and adopt $(b_s-b_a)=1$, as this directly multiplies $b_\Gamma$, and hence we are only sensitive to the product $b_\Gamma (b_s - b_a)$. We also adopt the values $b_a'=-2/3$ and $\lambda_0=300$\hMpc, following \cite{Gontcho2014}. We have tested small changes around these values, and found they do not have a significant impact on our fit.

The prediction of \cite{Gontcho2014} is that $b_\Gamma\simeq0.13$ at $z=2.25$. We measure $b_\Gamma=0.143\pm 0.047$ at $z_\mathrm{eff}=2.33$, which is in excellent agreement with the prediction. We have also tested freeing $\lambda_0$, and found results in agreement with $\lambda_0\sim300$\hMpc, but with very large uncertainties. While the presence of UVB fluctuations in our model significantly improves the fit, they do not have a significant impact on our AP constraints, as discussed in \Cref{subsec:data_validation}, and shown in \validation.

Finally, we note that while only the simplest version of this model is included in our baseline analysis, we have also tested including additional terms and found that the model does not significantly improve with them. These include the shot noise of the UVB sources \cite{Gontcho2014}, and the effect of the HeII re-ionization, which is expected to produce a much weaker scale dependence in the \lya\ bias on scales of $\sim 30$\,\hMpc\ \cite{Gontcho2014}. These extensions will be explored in more detail in future work.

\section{Analysis Validation}
\label{sec:validation}

Our analysis validation process has two major components: validation of the baseline analysis and model using synthetic data sets (mocks), and robustness tests on blinded data. Both of these components are described in detail in \validation. Here, we provide an overview of the validation process and highlight the main results, which demonstrate that our measurements are robust. We present the validation with mocks in \Cref{subsec:mock_validation}, the blinded robustness tests in \Cref{subsec:data_validation}, and discuss our results and conclusions in \Cref{subsec:validation_discussion}.

We primarily focus our validation on what is new in this work relative to DR2 \lya BAO \cite{DR2Lya:2025}, namely the broadband AP constraint, \phis. This is because our analysis approach involves a template with a peak and a smooth component whose coordinates are independently re-scaled (see \Cref{subsec:alpac}), which makes it almost identical to a BAO analysis when it comes to constraining the position of the BAO peak.\footnote{This is also because we do not currently use reconstruction \cite{Padmanabhan:2012} for \lyaf\ BAO analyses.} Our analysis builds upon the DESI DR2 BAO analysis \cite{DR2Lya:2025} by also rescaling the coordinates of the smooth component, although we have checked the BAO results for all of our tests on both mocks and blinded data to ensure there are no significant changes with respect to the results presented in \cite{DR2Lya:2025,Casas2025}. Appendix C of \validation\ presents the BAO results from our blinded data tests, while the BAO tests on mocks are briefly discussed in \Cref{subsec:mock_validation} below, and in more detail in Section IV of \validation.

The DR2 full-shape analysis was blinded in the same manner as the DR1 full-shape analysis described in \cite{Cuceu:2025}. The process consisted of adding a random value to the parameter of interest, namely $\phi_s$, within the likelihood. The random offset was generated from a Gaussian with zero mean and standard deviation of $2\%$, based on the eBOSS measurement of \cite{Cuceu2023a}. The random offset was implemented in our modeling package and automatically kept constant across all of the tests we performed.

The overarching goal of our validation process was to identify a set of analysis configurations that allow us to measure the AP effect primarily at large scales where the linear-theory approach works well. This means that any systematic uncertainties from small-scale non-linearities or contaminants need to be kept at a level well below our statistical uncertainties. If we found this is not possible, we would increase our final uncertainties to account for the systematic error budget. Following previous DESI \lyaf\ analyses \cite{DESI2024.IV.KP6,DR2Lya:2025,Cuceu:2025}, our threshold for validating the measurements was set at $1/3$ of the statistical uncertainty. For \phis, this corresponds to a tolerance of $0.38\%$. For the mock validation, we required the baseline results to be within this threshold and also to display convergence with more conservative choices.\footnote{This is to ensure that we do not cherry-pick an analysis configuration where systematic errors happen to cancel each other out.} For the blinded analysis and modeling variations, we required results to be within this threshold. Any variations exceeding this threshold had to be investigated in detail and addressed before unblinding our measurements. Finally, we also ran a set of data splits to ensure consistency across different subsets of our data, and for these we required results to be within $2\sigma$ of each other.

All decisions on what constitutes our baseline analysis were made using blinded data. Once we converged on this baseline analysis, ensured mock results were within the set tolerance, and investigated and addressed any variations outside the tolerance and data splits in tension, we prepared a detailed technical document describing our tests and validation process.\footnote{Most of the details of this document are presented in the companion paper \validation.} This document was then circulated to the entire DESI collaboration, which had a one-week period to provide comments and feedback before we proceeded to unblind our measurements. We have made no changes to the baseline analysis post-unblinding, and the results presented here are identical to those from when we unblinded.

\subsection{Validation with Mocks}
\label{subsec:mock_validation}

Our validation with mocks was based on two different types of mocks. The first set of \lyaf\ mocks were developed from the \abacus\ simulations by \cite{Hadzhiyska2023} to test the sensitivity of our model for the \lya\ auto-correlation and \lya\ QSO cross-correlation functions to non-linearities in the clustering at small scales. These mocks only simulate the \lyaf\ and quasar clustering and use skewers drawn from a cubic box at redshift $z=2.5$. The second set of mocks are \lpt\ mocks, which were produced with the \colore\ \cite{Ramirez2022} package using Second Order Lagrangian Perturbation Theory (2LPT). These mocks are described in detail in a companion paper \cite{RuizHerrera:2026}, and we only give a brief overview below. Both sets of mocks use the fluctuating Gunn-Peterson approximation \cite[FGPA;][]{Croft:1998} to estimate the \lyaf\ from the matter density field.

The \abacus\ simulations are a suite of large, high-accuracy N-body simulations designed to meet the cosmological simulation requirements of the DESI survey \cite{Maksimova:2021,Garrison:2021}. The \lya\ signal in these mocks was computed by \cite{Hadzhiyska2023} according to four models, which we label M1 to M4 (see Table 1 of \cite{Hadzhiyska2023}). These models correspond to different parameters and methods to estimate the \lya\ optical depth from the smooth matter density field in the \textsc{AbacusSummit} simulations. While the simulations are not as realistic as hydrodynamical simulations, they include the effect of non-linear clustering inherited from the matter field in the original N-body simulation, and the variations in the optical depth modeling give us some sense of the importance of those assumptions for the signal we want to measure. We primarily use these mocks to test our baseline model for the \lya\ auto- and cross-correlation with quasars and to validate our scale cuts. These mocks are also useful for testing biases in the BAO measurements due to non-linear evolution \cite{deBelsunce:2025,Hadzhiyska:2025}.

The \texttt{CoLoRe-2LPT} mocks are light-cone simulations that replace the standard \texttt{CoLoRe} log-normal prescription \cite{Ramirez2022} used in previous \lyaf\ analyses with a 2LPT approximation for both the density and velocity fields. This new prescription captures mild non-linearities and the broadening of the BAO peak, which were absent from previous mocks. The QSO clustering model is also improved, as it has been calibrated with HOD catalogs from the Abacus N-body simulations and includes the Fingers-of-God effect. Line-of-sight 2LPT density and velocity skewers are drawn from these full-sky light-cones and converted into Ly$\alpha$ transmission skewers using a new 2LPT version of \texttt{LyaCoLoRe} \cite{Farr2020_LyaCoLoRe} that has been calibrated to reproduce the Ly$\alpha$ clustering statistics of the latest DESI measurements. We then produce 400 DESI DR2 mocks by post-processing these skewers with the \texttt{quickquasars} script from the \texttt{desisim} package, which applies the DESI survey footprint, generates quasar continua, adds astrophysical contaminants (metals, BALs, and DLAs), and simulates instrumental noise and the spectrograph resolution. Finally, we analyse the mocks with the same pipeline used for the real data. The new mocks are presented in detail in the companion paper \cite{RuizHerrera:2026}.

Our process for validating the analysis using mocks consisted of two steps. First, we used the \abacus\ mocks and the stack of correlations from the 400 \lpt\ mocks to identify an analysis configuration that passes our validation criteria or minimizes the combined statistical and systematic uncertainty. The primary choices during this step were to select a set of scale cuts (discussed below), a configuration for small-scale marginalization (\Cref{subsec:ssm}), and the model and priors for contaminants (\Cref{subsec:model_corr,sec:nuisance}). This first step was conducted in parallel with a wide range of tests on blinded data to ensure that the choices we made also yield robust measurements on real data. In particular, an important goal of this first step was to identify a model that provides good fits to both mocks and real data, while also providing unbiased constraints.

In the second step, we used the baseline model found during the first step to fit all individual \lpt\ mocks and then performed statistical tests on the mock population. The main goal of the second step was to validate our uncertainties. In the remainder of this section, we provide an overview and the main results of the first step in this process, and refer the reader to the companion paper \validation\ for details on the rest of the tests we performed.

As described in \Cref{sec:data}, with the exception of the updates to the DLA catalog, the analysis choices up to and including the measurement of the correlation functions are the same here as in the previous DESI DR2 \lyaf\ analysis \cite{DR2Lya:2025}. This is because we managed to find a set of model configurations that yielded unbiased fits in mocks (up to our threshold), so we did not have to reconsider changes to analysis choices at an earlier stage.

\begin{figure}
    \centering
    \includegraphics[width=1.0\columnwidth,keepaspectratio]{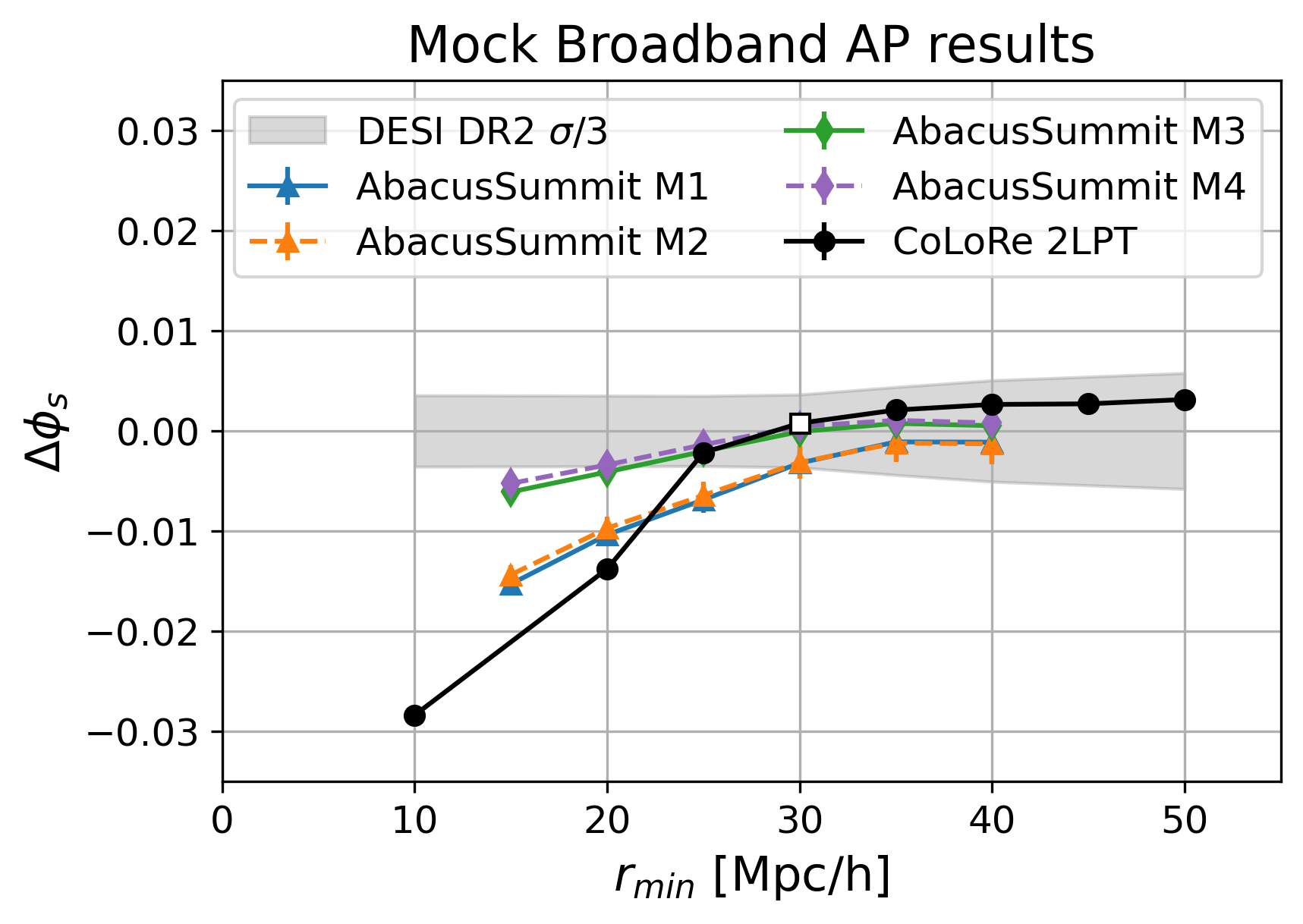}
    \caption{
    Broadband AP constraints in mocks relative to the truth as a function of the minimum separation used in the analysis, $r_\mathrm{min}$. The gray bands indicate the $1/3\; \sigma_\mathrm{DR2}$ threshold obtained from the joint analysis of all four DESI DR2 correlations. We show results from fits to the four different versions of \abacus\ mocks, and from the stacked correlations of 400 \lpt\ mocks. The baseline analysis configuration is shown as an empty square.
    }
    \label{fig:mocks_scale_cuts}
\end{figure}

The most important choice was the minimum scale included in the data we fit, $r_\textrm{min}$.\footnote{One could also consider minimum scale cuts in the line-of-sight or transverse separations, $r_{||}$ and $r_\bot$. We did perform a few tests with scale cuts in $r_{||}, r_\bot$ as part of our validation (see \validation). However, these types of cuts are undesirable in AP analyses, as the goal is to measure the anisotropy in the correlation function, and any significant cuts in $r_{||}, r_\bot$ would remove a large part of the signal.} In \Cref{fig:mocks_scale_cuts}, we show mock results on the broadband AP parameter, \phis, relative to the true value of \phis\ in mocks, for different values of $r_\textrm{min}$. These results are obtained from joint fits of the \lya\ auto-correlation and its cross-correlation with quasars. The gray bands indicate the $1/3$ threshold of the statistical uncertainty in the data. For this test, we also ran the same fits on the real data, so we can compare mock results with the relevant threshold for each $r_\textrm{min}$ value. When including smaller scales around $10$ or $20$\,\hMpc, we observe significant systematic biases in \phis\ of up to $3\%$, and results that rapidly change with different $r_\textrm{min}$. On the other hand, for $r_\textrm{min}\geq30$\,\hMpc, both the \lpt\ mocks and all four types of \abacus\ mocks show converging results within our thresholds. Therefore, we conclude that our model is unbiased within the thresholds we imposed when using scales larger than $30$\,\hMpc.

We show more detailed versions of this test in \validation, where we vary $r_\textrm{min}$ independently for the auto and cross-correlations. We have also tested changing the maximum scale cut $r_\mathrm{max}$ and found that it does not have a significant impact on AP constraints. Based on these results, we chose $r_\textrm{min}=30$\,\hMpc\ for the \lya\ auto-correlation and $r_\textrm{min}=40$\,\hMpc\ for the \lya-QSO cross-correlation. The reason for the more conservative choice in the cross-correlation is due to the fact that we do not have a model to gauge the impact of small-scale non-linearities in the cross-correlation (like we do with the empirical model in \Cref{eq:arinyo} for the auto-correlation). Our choice of minimum scale cut for the cross-correlation is the same as the one used for the DESI DR1 analysis in \cite{Cuceu:2025}, while for the auto-correlation it is slightly more conservative (it was $r_\textrm{min}=25$\,\hMpc\ in \cite{Cuceu:2025}). In terms of the maximum scale cut, we chose $r_\mathrm{max}=200$\,\hMpc, which is the largest scale included in our measurement. In principle, one could also include scales larger than that, but the significant increase in the size of the covariance matrix and diminishing returns when it comes to AP constraints precluded us from attempting it.

We also found that our baseline model provides a good fit to the mock correlations (see Figure 3 in \validation). Based on the quality of the fit and the fact that our baseline model produces unbiased AP constraints given our scale cuts, we conclude that our baseline model and analysis configuration have been validated.

Further tests on AP using mocks are presented in \validation, including statistical tests of the mock population. \validation\ shows that our AP uncertainties are consistent with the distribution of results in individual mocks, which validates the statistical uncertainties of our measurements. Finally, Appendix D of \validation\ shows results from a subset of mocks with a few different fiducial cosmologies, including variations in \lcdm\ and $w_0w_a$CDM parameters, and finds no significant impact on Broadband AP constraints from the choice of fiducial cosmology.

\begin{figure}
    \centering
    \includegraphics[width=0.8\columnwidth,keepaspectratio]{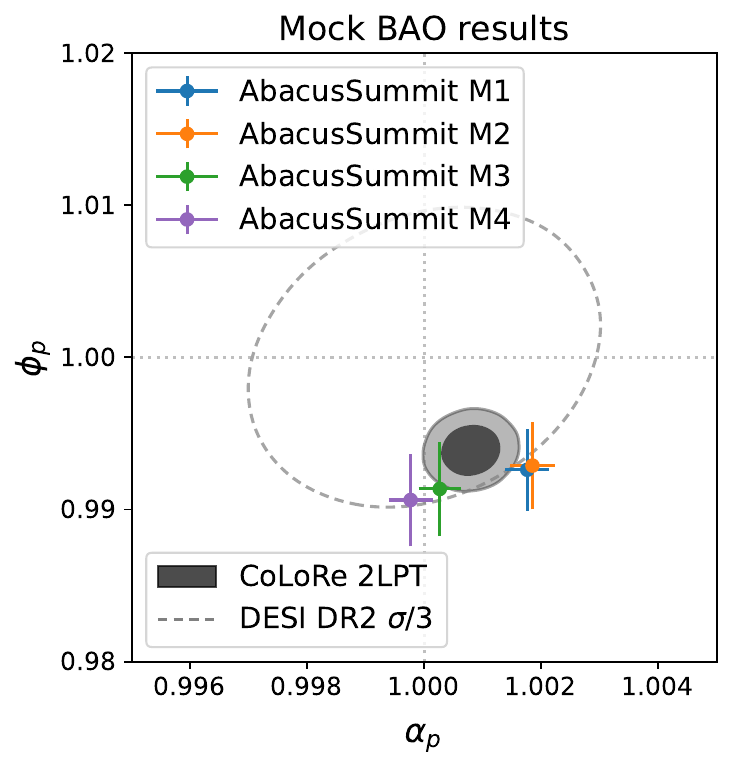}
    \caption{
    BAO constraints in mocks. The gray dashed contour indicates the $1/3\; \sigma_\mathrm{DR2}$ threshold, while the dotted cross indicates the truth in mocks. We show results from fits to the four different versions of \abacus\ mocks as points with error bars, and from the stacked correlations of 400 \lpt\ mocks as the black contours. These results indicate a small but significant BAO bias that is fairly consistent between the different mock versions. We account for this bias in our systematic error budget, and do not adjust the measurement.
    }
    \label{fig:mocks_bao}
\end{figure}

To conclude this section, we briefly discuss BAO results from our mocks. In \Cref{fig:mocks_bao}, we show BAO results from our full-shape fits in mocks.\footnote{BAO results are fairly insensitive to the minimum scale cut used, so we only show results using our baseline model setup here.} We find a small but significant bias present in both \lpt\ and \abacus\ mocks, primarily in the AP direction (\phip). The results from the four types of \abacus\ mocks are not fully consistent with each other in terms of \alphap, but they all indicate a fairly consistent bias in \phip\ of $\sim0.8\%$. The primary cause of the bias in \abacus\ mocks is likely the BAO shift due to non-linear growth as discussed in \cite{Hadzhiyska:2025}. The \lpt\ bias is broadly consistent in direction and magnitude with the biases in \abacus\ mocks. These results indicate that an extra systematic uncertainty is needed to account for this bias in the absence of an analytic model to correct it (for detailed discussions on this, see \cite{deBelsunce:2025,Hadzhiyska:2025}). Indeed, a systematic uncertainty was added to the DESI DR2 BAO measurement to account for the expected bias arising from this BAO shift \cite{DR2Lya:2025}. We revisit this systematic uncertainty in light of the new results here in \Cref{subsec:bao_sys} below.

\subsection{Blinded robustness tests}
\label{subsec:data_validation}

We conducted numerous tests on blinded data. These include several data splits and a large number of analysis variations. The goal of the data splits is to check that different subsets of the data agree with each other, while the goal of the analysis variations is to test the robustness of our measurements against different choices made in our analysis. The full set of blinded data tests is described in detail in \validation. Here, we give an overview of the most important tests.

\begin{figure}
    \centering
    \includegraphics[width=1.0\columnwidth,keepaspectratio]{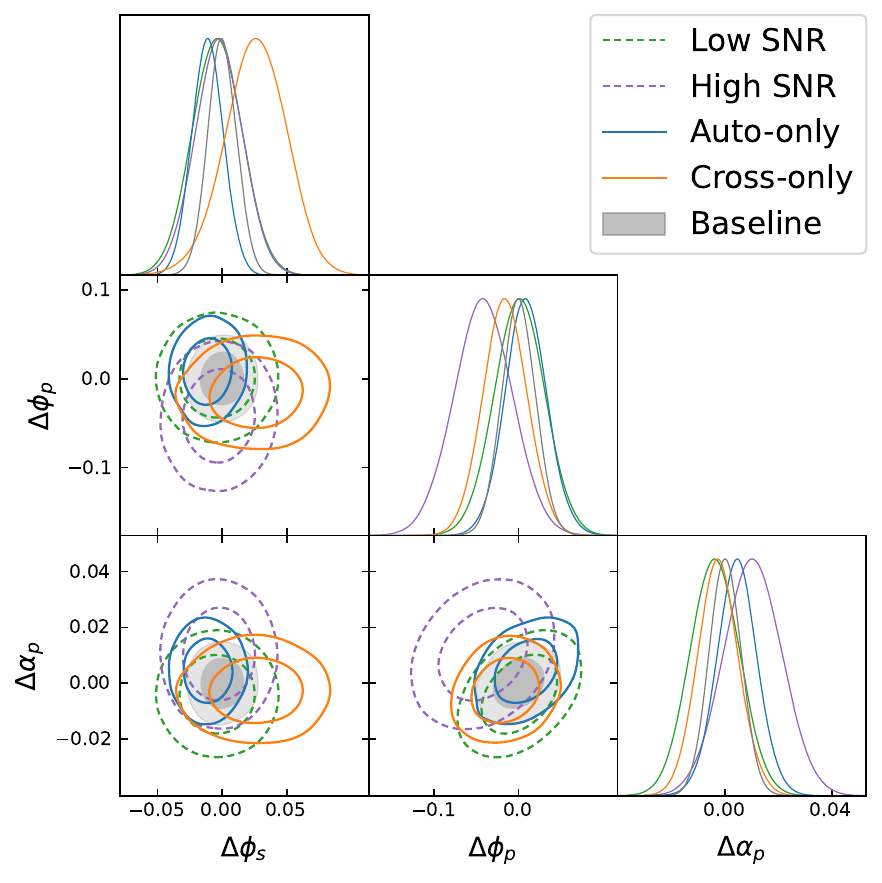}
    \caption{
        Comparison of broadband AP and BAO results from data splits. We show results from the low- (high-) SNR split using green (purple) dashed lines. The constraints from the two Ly$\alpha$ auto-correlations in blue are compared against the constraints from the cross-correlations in orange (orange). Both data splits produce results that are consistent with each other and with the baseline constraints shown in gray.
    }
    \label{fig:data_splits}
\end{figure}

Two of the data splits are shown in \Cref{fig:data_splits}. The first is a cut in the signal-to-noise ratio (SNR), defined as the mean SNR in the rest-frame wavelength range, $1420 < \lambda_{\text{rest}} < 1480$ \AA\ \cite{DESI2024.IV.KP6}. The High SNR sample are quasars with SNR $>4.25$, while the Low SNR sample are those with SNR $<4.25$. The threshold was chosen such that the two samples have roughly similar uncertainties on their measured correlations. Many effects could be different in high versus low SNR spectra, with perhaps the most important ones being our ability to detect and mask contaminants such as DLAs and BALs \cite{brodzeller2025,KP6s9-Martini}, as well as the extent to which noise impacts the quasar continuum fitting. \Cref{fig:data_splits} shows good agreement between the results from the two SNR splits.

The second data split shown in \Cref{fig:data_splits} is between the two \lya\ auto-correlations and the two \lya-QSO cross-correlations. This is a very important test to pass because the auto and cross-correlations are affected differently by most of the important contaminants and there are several systematic effects that are only present in one and not the other. These include the sky contamination for the auto-correlation, and the redshift errors and transverse proximity effect for the cross-correlation. Furthermore, the models for HCD and metal contamination, UV background fluctuations, small-scale non-linearities, and the distortion due to continuum fitting are all different to some extent between the auto- and cross-correlation.\footnote{For most of these it is because they only affect the \lyaf\ and not quasars.} As before, \Cref{fig:data_splits} shows good agreement between the auto-correlations and the cross-correlations.

We conducted additional data splits based on the A and B regions of the \lyaf, into north and south samples based on the imaging surveys used for targeting, and based on the equivalent width of the CIV emission line. These splits are presented in \validation\ and show good agreement with each other. 

We also performed a wide range of analysis variations, testing our choices at each level of the analysis. The first three sets of tests include variations in the data set and catalogs, variations in the continuum fitting process, and variations in the measurement of the correlation functions and covariance matrix. We largely followed previous DESI \lyaf\ analyses in the choices of tests to run in these categories. Notable tests in these categories include changes to the rest-frame wavelength cut, the SNR cut used to mask DLAs, the minimum length of forests, the binning in observed wavelength and in comoving coordinates, the number and size of HEALPix regions used to estimate the covariance matrix, and a number of variations in the configuration of the distortion matrix computation. All of the tests produce results within our $1/3\sigma$ threshold from the baseline. For the full set of tests and results, see Section VB of \validation.

The final set of variations includes those that change parameter priors and the model for the correlation functions. These are fairly distinct from the tests done for BAO analyses because the goal is to test the robustness of the constraint on the broadband AP parameter. First, a number of our nuisance parameters, which are not well constrained by the data, have informative Gaussian priors (see \Cref{sec:nuisance}). We tested all of these by either removing the prior or using a different prior to gauge whether the choice of prior impacts our measurements. In all cases, the results are within our threshold. Secondly, we tested freeing a number of parameters that are fixed in our baseline model. In particular, marginalizing over all parameters in the empirical model for small-scale non-linearities in \Cref{eq:arinyo}) does not significantly impact our measurements and uncertainties. We have also tested removing this correction entirely (i.e., only relying on linear theory) and again found no significant impact on our constraints. This reinforces our conclusion that the analysis configuration primarily targets the larger scale, linear regime.

We tested changing the damping that accounts for redshift errors from a Lorentzian to Gaussian, removing the model for UV background fluctuations, and either fixing the isotropic scale parameter for the smooth component, $\alpha_s=1$, or allowing two different \alphas\ parameters for the auto- and cross-correlation. In all of these cases, the results are within $1/3\,\sigma$ of our baseline measurement. Finally, we have tested that masking out the first $r_{||}$ bin (i.e., imposing a cut $r_{||}>4$\,\hMpc) does not have a significant impact. This is particularly relevant for the sky contamination, which is entirely contained within that first $r_{||}$ bin before the distortion due to the continuum fit (\Cref{subsec:model_corr} and \cite{KP6s5-Guy}).

\subsection{Discussion}
\label{subsec:validation_discussion}

In this section, we discuss our validation results and provide an overview of aspects of our analysis and mocks that are most relevant for these results. Our goal in this analysis was to target large linear scales in the \lyaf\ auto-correlation and its cross-correlation with quasars. For the auto-correlation, we use an empirical model to gauge the impact of small-scale non-linearities (\Cref{subsec:model_corr}). However, when using the small-scale marginalization approach introduced in \Cref{subsec:ssm}, together with the scale cuts discussed above, we find that our AP measurement is insensitive to the presence and parameter values of this model (\Cref{subsec:data_validation}). This confirms that we are not sensitive to small-scale non-linearities, at least in the auto-correlation. On the other hand, for the cross-correlation, we use a simple linear theory model. Because of this, we chose more conservative scale cuts relative to the auto-correlation. We have also shown consistent AP results between the auto- and cross-correlation (\Cref{subsec:data_validation}) and validated our measurements using mocks with realistic quasar clustering (\Cref{subsec:mock_validation}) based on the \abacus\ simulations. This approach works well at the precision of DESI DR2, given the threshold used in our validation. However, its success was not guaranteed, and indeed we expect this approach to encounter a systematic floor as constraints become increasingly precise. Therefore, future analyses using larger data sets, or those attempting to access smaller scales in the \lya\ correlations or power spectra \cite{Belsunce:2024}, are likely to require either an effective field theory approach \cite{Ivanov:2024,Chudaykin:2025,Belsunce:2026} or an emulator approach \cite[e.g.][]{ForestFlow_2025}. 

A major aspect of our validation centers on using DESI DR2 mocks to demonstrate that we can recover the true cosmology with our baseline analysis. Therefore, we also discuss the level of realism in these mocks and whether the approximations used are likely to impact our results. First, there are a number of smaller effects that we know are present in real data but are not present in the mocks. These include CIV metal contamination, the transverse proximity effect, correlated noise due to sky contamination, and UV background fluctuations. We found that ignoring UV background fluctuations and the transverse proximity effect by removing our model for these effects does not significantly affect the AP constraint (see Section VB in \validation). On the other hand, ignoring the sky contamination does have a significant impact (see Appendix C3 in \validation). As this effect mostly contributes to the $r_{||}<4$\,\hMpc\ bins in the auto-correlation \cite{KP6s5-Guy}, we performed a test that masks out these bins, thereby removing most of the sky-contamination effect. This did not have a significant impact on our AP constraint (\validation), which validates our current model for this effect (\Cref{subsec:model_corr}). Finally, for the CIV metal contamination, we currently use an informative prior on the linear bias parameter for CIV absorption. This prior is based on measurements by \cite{KP6s5-Guy} using DESI CIV forest data and is necessary because \lyaf\ correlations cannot constrain this effect. However, the prior we use is already a factor of three wider than the measurement of \cite{KP6s5-Guy}, and the CIV bias parameter does not display a significant correlation with \phis. These effects are discussed in more detail in \validation.

Next, we turn our attention to effects which are modeled in mocks, but are not necessarily as realistic as they could be. This includes the most important \lyaf\ contaminants: HCDs and metals. HCDs are introduced in mocks in a similar fashion to quasars, by sampling the peaks of the density field \cite{RuizHerrera:2026}, but their clustering and especially their column density distribution function are fairly uncertain. However, we find that they have a significant systematic impact on our correlation only at scales smaller than those we fit. When using scales $r>25$\,\hMpc\ we measure an HCD linear bias consistent with 0 at $68\%$ confidence, whereas when we include $r<25$\,\hMpc\ we measure a nonzero value at very high significance ($>5\sigma$). Therefore, we conclude that HCDs are not a significant source of concern given the current precision and scale cuts. In terms of metals, they are simulated in our mocks by rescaling \lyaf\ flux skewers \cite{RuizHerrera:2026,2024arXiv240100303H}, which means their small-scale clustering is the same as that of the \lyaf\ in mocks, but that is likely not the case in real data \cite[e.g.,][]{PerezRafols2023}. To gauge the potential impact of this effect, we allowed greater freedom in the metal-\lya\ cross-correlations\footnote{Which are the primary source of this contamination.} model at small scales, using the empirical model in \Cref{eq:arinyo} with all parameters free, and found no significant impact on AP. Finally, both our sets of mocks rely on the fluctuating Gunn-Peterson approximation (FGPA, \cite{Croft:1998}) to compute \lyaf\ fluctuations from the matter density field. The \abacus\ mocks we use have four versions (labeled M1 to M4), corresponding to different parameters and methods for applying the FGPA, and reproduce \lyaf\ 1D and 3D power spectra from hydrodynamical simulations to various degrees of realism \cite{Hadzhiyska2023}. Therefore, an important part of our validation was to show that all four versions of the \abacus\ mocks produce unbiased constraints within our thresholds, which indicates that our results are not very sensitive to how the FGPA is used. These results are shown in \Cref{fig:mocks_scale_cuts} and discussed in \Cref{subsec:mock_validation}.

Based on the tests presented and discussed above, along with those presented in \validation, we concluded that our \alpac\ measurements are robust and proceeded to unblind our results. Because our validation with mocks produced broadband AP results consistent with the truth within our threshold, combined with the fact that our blinded data tests showed consistent and robust constraints on \phis, we decided not to include an extra systematic uncertainty for this parameter.

\section{Results}
\label{sec:results}

We next turn our attention to the main results of this work. We begin this section by going through the model fits to our correlation function and present the main parameter constraints in \Cref{subsec:ap_res}. After that, we discuss the systematic error budget on the BAO parameters in \Cref{subsec:bao_sys}. Finally, we present our cosmological distance constraints in \Cref{subsec:cosmo_dist}.

\subsection{\alpac\ measurements}
\label{subsec:ap_res}

\begin{figure*}
    \centering
    \includegraphics[width=1.\textwidth,keepaspectratio]{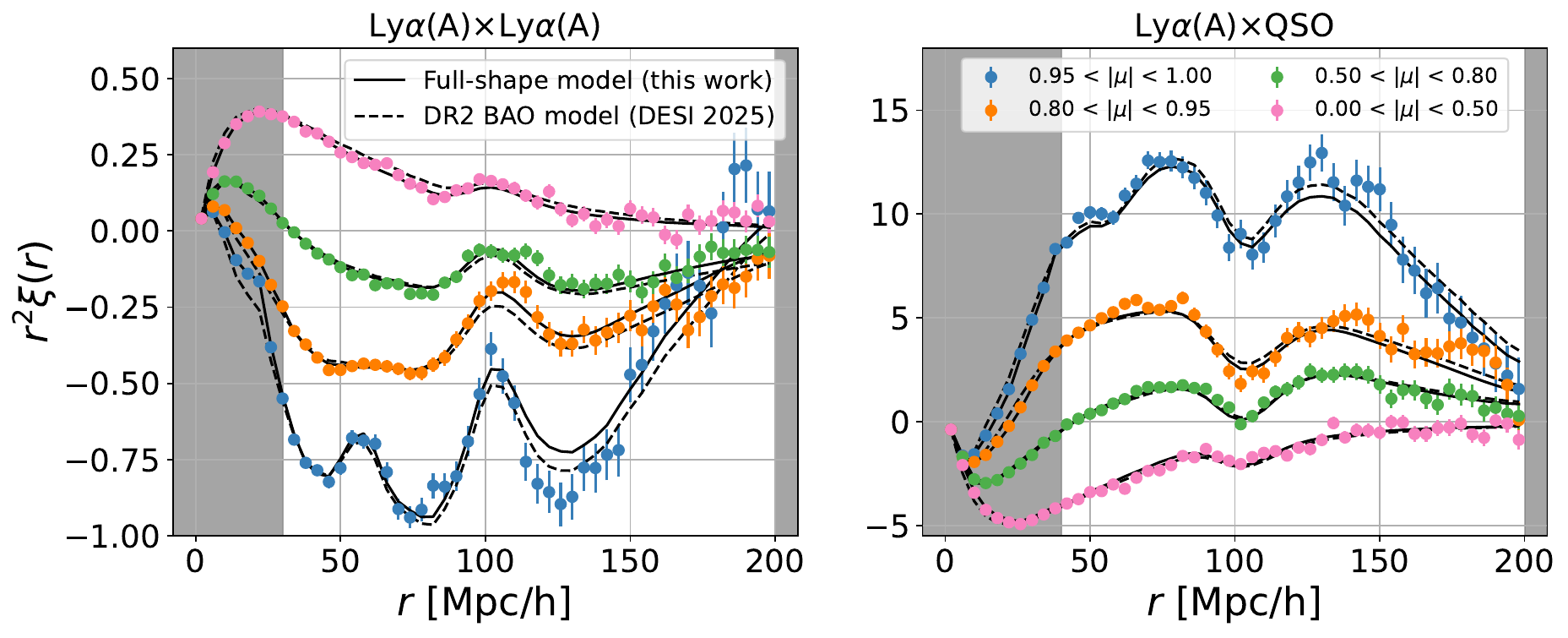}
    \caption{
        Wedge compression of the \lya\ auto-correlation (left) and its cross-correlation with quasars (right). The black lines show the best-fit models, with the continuous line indicating the full-shape model described here and the dashed line showing the best-fit model from the DR2 BAO analysis. Gray bands indicate the scale cuts for the full-shape analysis. While the scale cuts differ slightly between the two fits, this provides a good representation of the evolution of the model.
    }
    \label{fig:data_wedges}
\end{figure*}

\begin{figure*}
    \centering
    \includegraphics[width=0.9\textwidth,keepaspectratio]{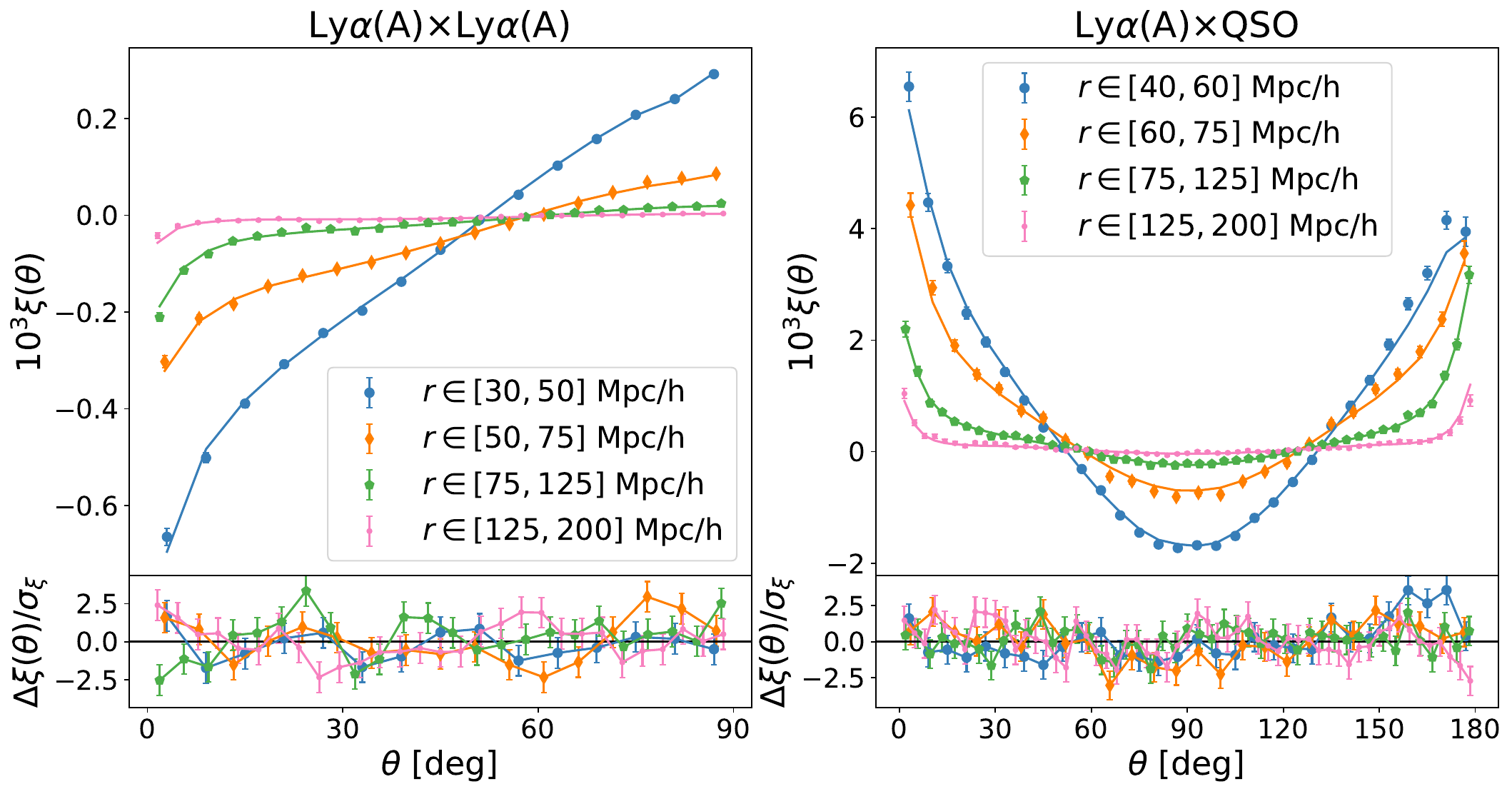}
    \caption{
        Shell compression of the \lya\ auto-correlation (left) and its cross-correlation with quasars (right), along with the best fit model (top) and residuals (bottom). These figures display the anisotropy across four radial shells, with the most relevant ones for our analysis being the smaller-separation shells in blue and orange.
    }
    \label{fig:data_shells}
\end{figure*}

We use the \texttt{Vega} package to compute the model and perform the inference. The likelihood is Gaussian, and we sample the full posterior distribution using the Nested Sampler \texttt{PolyChord} \cite{Handley:2015a,Handley:2015b}. To obtain the best fit model we use the \texttt{iminuit} minimizer \cite{dembinski_2025_17565861,James:1975dr}. The effective number of data points is 11388, computed by subtracting the number of analytically marginalized parameters, 402, from the total number of fitted data points, 11790 (\Cref{sec:data}). The measured correlations and best-fit model are shown using the wedge compression in \Cref{fig:data_wedges} and using the shell compression in \Cref{fig:data_shells}. The model fit is good, with a $\chi^2/(N_\mathrm{data}-N_\mathrm{param})= 11492.0/(11388-21) = 1.011$, and $\mathrm{PTE}=0.20$. Because we effectively implement the scale cuts in the undistorted model via small-scale marginalization and then fit all data points down to zero, the model perfectly fits the data below $r_\mathrm{min}$, but those points do not meaningfully contribute to our parameter constraints.

\Cref{fig:data_wedges} also shows the best-fit model from the DR2 BAO analysis for comparison. While the scale cuts are slightly different ($r_\mathrm{min}=30$\,\hMpc\ for the cross-correlation, versus $40$\,\hMpc\ here), this is a good display of how the changes in the analysis between BAO and full-shape impact the resulting best-fit model. In particular, the model appears to be significantly improved in the transverse wedge ($\mu<0.5$) and around the BAO peak in the two intermediate wedges.

\Cref{fig:data_shells} shows the anisotropy present in our correlations and how well it is fit by the model. This figure also includes the residuals for each shell, which are consistent with our expectations. There are no significant outliers or patterns that could hint at biases in the measured anisotropy. When interpreting the wedge and shell plots, it is important to note that there is significant cross-covariance between neighboring points, which can give rise to apparent features due to noise. Outliers in these measured correlations were studied in section IVA of the DR2 BAO paper \cite{DR2Lya:2025} using the BAO model. They concluded that there is no evidence of significant disagreement between the data and their best-fit model. This conclusion carries over to our analysis as well, given that our model improves upon theirs.

Our main measurements are summarized in \Cref{tab:results}, where we focus on the broadband AP measurement and the BAO constraints. The full set of results, including nuisance parameters, can be found in \Cref{sec:nuisance}. The primary new result of this work is the broadband AP constraint, $\phi_\mathrm{s} = 1.007 \pm 0.011$, obtained from the smooth component of the \lyaf\ correlations. This represents a $1.1\%$ constraint on the distance ratio $D_M/D_H$ at $\zeff=2.33$, which corresponds to a factor of $\sim1.8$ improvement with respect to the constraint from the DR2 BAO analysis ($2\%$ constraint on AP; \cite{DR2Lya:2025}) and a $\sim45\%$ improvement with respect to the DR1 full-shape analysis ($1.6\%$ constraint on AP; \cite{Cuceu:2025}).

In \Cref{tab:results} we also report the BAO measurements obtained from our full-shape fit, which include the AP parameter measured from the BAO feature, $\phi_\mathrm{p} = 1.001 \pm 0.020$, and the isotropic BAO scale, $\alpha_\mathrm{p} = 1.001 \pm 0.006$. The AP measurement from BAO is consistent with our new broadband AP result, which represents another useful consistency check. This is because the two are constrained by different components of the correlation function and are fairly independent, with a correlation coefficient $\rho < 1\%$. Our final results will represent the combination of the broadband AP and BAO measurements, and are described in \Cref{subsec:cosmo_dist} below. As discussed in \Cref{subsec:validation_discussion}, we did not include an extra systematic uncertainty on \phis. However, following DESI DR2 BAO, we do include a systematic uncertainty in the BAO measurements, which are described in the next section.

\begin{table}
    \centering
    \begin{tabular}{c|c|c}
    Description & Distance ratio & Measurement \\
    \hline \hline
    Broadband AP & $D_M/D_H$ & $\phi_\mathrm{s} = 1.007 \pm 0.011$ \\
    AP from BAO & $D_M/D_H$ & $\phi_\mathrm{p} = 1.001 \pm 0.020$ \\
    Isotropic BAO & $(D_M D_H)^{1/2} / r_d$ & $\alpha_\mathrm{p} = 1.001 \pm 0.006$ \\
    \end{tabular}
    \caption{Main DR2 \lyaf\ full-shape results. We report the broadband AP constraint (the primary new result of this work), together with the AP and isotropic BAO constraints from the same dataset. For each parameter, the middle column indicates the corresponding distance ratio that is constrained by that parameter. Parameter values in the final column correspond to the ratio of these distances with respect to their fiducial values.}
    \label{tab:results}
\end{table}

\subsection{BAO systematic uncertainties}
\label{subsec:bao_sys}

At the statistical precision of DESI DR2, BAO constraints from the \lyaf\ require an assessment of possible shifts in the BAO peak position caused by non-linear evolution. Such shifts can arise when the fitting model does not fully capture the relevant non-linear effects \cite{Sinigaglia:2024,deBelsunce:2025,Hadzhiyska:2025}. In the DR2 BAO analysis, this contribution was treated as a theoretical systematic uncertainty on the BAO distance parameters, rather than as a direct correction to the measured values. This choice is motivated by the fact that no reconstruction is applied to the \lyaf\ field, so residual BAO shifts may remain non-negligible at sub-percent precision \cite{DR2Lya:2025}.

Section VIA of the DR2 BAO paper reviews recent \lya\ BAO-shift studies based on the \abacus\ mocks and on effective field theory (EFT) modeling calibrated to hydrodynamical simulations. These studies broadly indicate shifts of order a few $\times 10^{-3}$, but they differ in both magnitude and sign, and each currently has only moderate statistical significance. Given this lack of convergence, the DR2 BAO analysis adopted a conservative covariance-level systematic term, with $\Delta\alpha_\parallel=0.3\%$ and $\Delta\alpha_\perp=0.3\%$, added in quadrature to the statistical covariance, and reported both statistical-only and statistical+systematic constraints separately.

Given the results we presented in \Cref{subsec:mock_validation}, we decided to take a slightly different approach for our BAO systematic error budget. As we saw, both \abacus\ and \lpt\ mocks indicate a systematic uncertainty that is significantly larger in the AP direction (\phip) than that used in the DR2 BAO paper. Therefore, we decided to update the BAO systematic error budget here to account for this fact.

We follow \cite{DR2Lya:2025} and add the systematic uncertainties to the $2\times2$ covariance matrix of the BAO parameters, but use the \alphap\ and \phip\ directions (as opposed to $\alpha_{||}$ and $\alpha_\bot$):
\begin{equation}
    C(\alpha_p,\phi_p) = C(\alpha_p,\phi_p)_\mathrm{stat} + C(\alpha_p,\phi_p)_\mathrm{sys},
\end{equation}
with
\begin{equation}
C(\alpha_p,\phi_p)_\mathrm{sys} = 
\begin{bmatrix}
\Delta \alpha_p^2 & 0 \\
0 & \Delta \phi_p ^2
\end{bmatrix},
\end{equation}
where we choose the systematic uncertainties in the isotropic BAO and AP directions to be:
\begin{eqnarray}
    \Delta \alpha_p &= 0.15\%, \\
    \Delta \phi_p &= 0.8\%.
\end{eqnarray}
This corresponds to greater uncertainty in the AP direction compared to the DR2 BAO analysis, and smaller uncertainty in the isotropic BAO direction.

We note that the exact magnitude and direction of the BAO shift are still fairly uncertain, as we do not know how close to reality the shift observed in our two sets of mocks is. The fact that \lpt\ and \abacus\ mocks give consistent results represents an important step forward in our understanding of this effect, but we note that predictions based on EFT fits in \abacus\ mocks and hydrodynamical simulations still disagree on the exact direction of the shift \cite{deBelsunce:2025,Hadzhiyska:2025}. Due to this disagreement, we chose not to adjust our results by the shift measured in \Cref{subsec:mock_validation}. However, both our results and those of \cite{deBelsunce:2025,Hadzhiyska:2025} indicate that the shift is at most at the level of about $1/3$ of our current statistical uncertainty in BAO. This means that the systematic error and the corresponding systematic uncertainty used here have a small impact on the resulting BAO constraints. Finally, the broadband AP constraint presented here is significantly more precise than the BAO measurement, further diluting the importance of this choice.

\subsection{Cosmological distances}
\label{subsec:cosmo_dist}

The primary result presented here is based on the broadband AP signal (\phis) measured using the smooth component of the \lyaf\ correlation functions. This provides a tight constraint on the distance ratio $D_M/D_H$ at $\zeff=2.33$, given by:
\begin{equation}
    \textbf{broadband AP}:\; D_M / D_H (z_\mathrm{eff}) = 4.578 \pm 0.052.
\end{equation}
When combining this with the BAO AP constraint (\phip), including the systematic uncertainty discussed above, we measure the AP signal from the full shape of \lya\ correlations:
\begin{equation}
    \textbf{full shape AP}:\; D_M / D_H (z_\mathrm{eff}) = 4.572\pm 0.046,
\end{equation}
which corresponds to a $1.0\%$ constraint of the \alpac\ effect.

Our final results are given by the combination of the full-shape AP and isotropic BAO constraints (including the systematic uncertainty). This means we are using all the information from our three main parameters: \phis, \phip, \alphap. The corresponding distance constraints are: 
\begin{equation}
\begin{array}{ll}
\textbf{DESI \lya}& \\
\textbf{full-shape}&
\end{array}
\left\{
\begin{array}{ll}
    D_M (z_\mathrm{eff})/r_d &= 39.32 \pm 0.33, \\
    D_H (z_\mathrm{eff})/r_d &= 8.600 \pm 0.066, \\
    \rho(D_M/r_d, D_H/r_d) &= 0.225,
\end{array}
\right. \label{eq:dmdh_dr2}
\end{equation}
where $\rho$ is the correlation coefficient between the two parameters. These results correspond to a $0.84\%$ constraint on $D_M/r_d$, and a $0.77\%$ constraint on $D_H/r_d$.

\begin{figure}
    \centering
    \includegraphics[width=1.0\columnwidth,keepaspectratio]{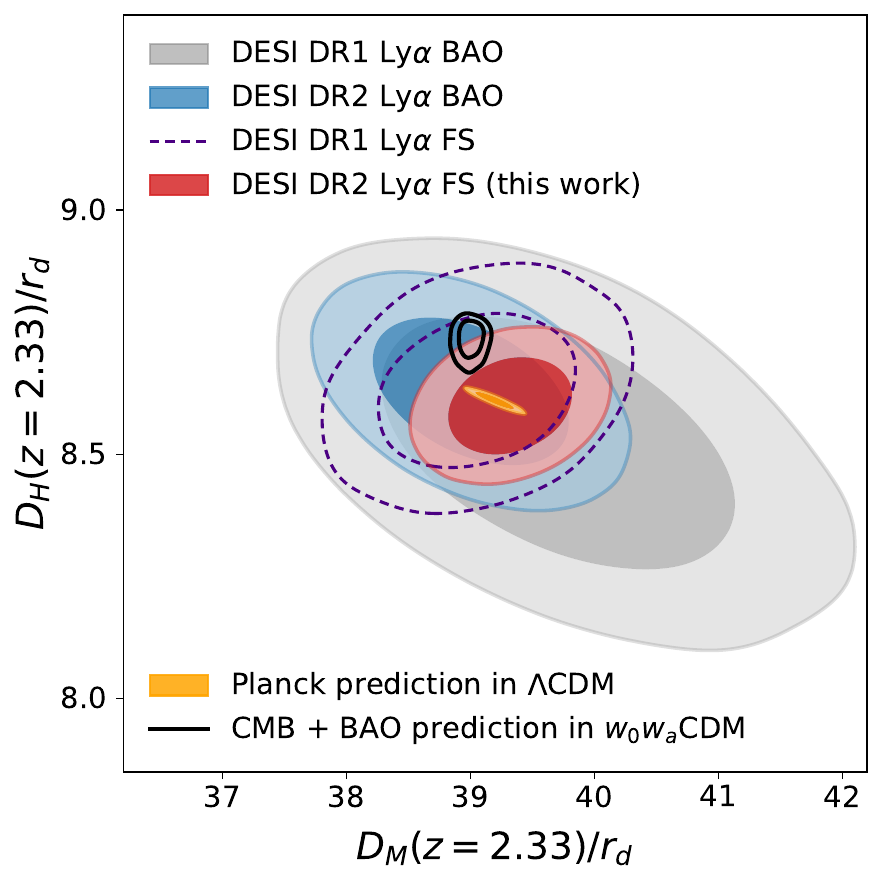}
    \caption{
    Constraints on the distance pair $(D_M/r_d,\,D_H/r_d)$ at $\zeff=2.33$ from \lyaf\ measurements, shown as 68\% and 95\% confidence contours. The red filled contours give the DESI DR2 full-shape result from this work, compared with DESI DR2 BAO (blue), DESI DR1 full-shape (purple dashed), and DESI DR1 BAO (gray). The orange contour shows the Planck CMB prediction in $\Lambda$CDM, while the black contour shows the CMB+BAO prediction in $w_0w_a$CDM. The \lya\ full-shape results provide the tightest distance constraints from large-scale structure at $z > 1$, and the uncertainty is now small enough to begin to separate between \lcdm\ and $w_0w_a$CDM predictions from other data sets in this space.}
    \label{fig:dmdh}
\end{figure}

The DESI instrument was designed to achieve combined precisions of about $0.9\%$ in both $D_M/r_d$ and $D_H/r_d$ in the redshift range of the \lyaf\ ($1.9 < z < 4.0$) from the original 5-year survey \cite{DESI2016a.Science,DESI2023a.KP1.SV}. Those forecasts were based on BAO-only, as no \lyaf\ 3D full-shape analyses had been performed at the time. Therefore, with the new constraints presented here, we have already exceeded the original DESI requirements using only the first 3 years of DESI data.

We compare our results with previous DESI \lyaf\ measurements in \Cref{fig:dmdh}. The first DESI results from DR1 BAO are shown in gray. When they were released two years ago, they were the state-of-the-art distance constraints in the redshift interval $1 < z < 4$, narrowly improving over the previous best constraints from SDSS \cite{dMdB2020}. Since then, DESI has also released DR1 \lyaf\ full-shape measurements (dashed purple) and DR2 BAO measurements (blue), both of which significantly improved these distance constraints and were in excellent agreement with each other. The measurement we have presented here (red) provides yet another significant improvement and corresponds to a $6.7$-fold reduction in the area of the contours compared with the original DESI DR1 BAO constraint.

The precision gain from the new measurement allows us to begin to separate between the \lcdm\ and $w_0w_a$CDM predictions of $D_M(z_\mathrm{eff})/r_d$ and $D_H(z_\mathrm{eff})/r_d$ from other cosmological tracers at more than $1\sigma$. This is illustrated in \Cref{fig:dmdh}, where we also show the \lcdm\ prediction from Planck (orange) and the $w_0w_a$CDM prediction from the joint CMB\footnote{See \Cref{subsec:external_data} for the combination of CMB results used here.} and DESI BAO analysis (black).\footnote{The prediction from the joint $w_0w_a$CDM analysis that also includes Supernovae is not meaningfully different, as SNe largely contribute to improvements at lower redshifts ($z < 1.5$).} The DR2 \lyaf\ constraints are in excellent agreement with the \lcdm\ prediction, but show a roughly $2\sigma$ tension with the $w_0w_a$CDM prediction. We will investigate this tension and its implications in the following Section.

To summarize, we have provided our final measurements in \Cref{eq:dmdh_dr2}, which represent significant improvements in the precision of cosmological distances at $z=2.33$. These measurements allow us for the first time to meaningfully distinguish among different state-of-the-art dark energy constraints with the \lyaf\ alone.

\section{Cosmological Constraints}
\label{sec:cosmo}

In this section, we discuss the cosmological implications from the Ly$\alpha$ forest full-shape measurements and define the datasets that are used to perform joint constraints on cosmological parameters.

\subsection{Other data sets}
\label{subsec:external_data}
We complement the high-redshift information accessible to the Ly$\alpha$ forest  by including BAO measurements from galaxy and quasar clustering. We use the DESI DR2 BAO data presented in \cite{DESIDR2:2025}. This dataset includes measurements of $D_\text{V}/r_\text{d}$ at $0.1<z<0.4$ for the Bright Galaxy Sample (BGS), joint measurements of $D_\text{M}/r_\text{d}$ and $D_\text{H}/r_\text{d}$ for the Luminous Red Galaxies (LRG) at $0.4<z<0.6$ and $0.6<z<0.8$, the combined LRG and Emission Line Galaxies (ELG) tracer at $0.8<z<1.1$, the ELG at $1.1<z<1.6$, and the Quasar (QSO) clustering at $0.8<z<2.1$. For comparison, we also include the Ly$\alpha$ BAO measurements from \cite{DR2Lya:2025}, when necessary. Throughout, we refer to this combination as DESI BAO, and to the combination with the new AP Ly$\alpha$ measurement as either DESI BAO+Ly$\alpha$ FS or simply as DESI.

We also include external data to DESI to constrain parameters beyond the late-time expansion history. In particular, anisotropies in the CMB provide a wealth of information that can help constrain the complete set of base cosmological parameters. Following the previous main DESI analyses, we use data from the \Planck\ satellite \cite{Planck-2018-overview,Planck-2018-cosmology} based on the PR4 release that uses the NPIPE maps \cite{Carron:2022}. Additionally, we now include new data from ground-based CMB experiments. In particular, we include the Atacama Cosmology Telescope (ACT) data release 6 (DR6) \cite{AtacamaCosmologyTelescope:2025blo,AtacamaCosmologyTelescope:2025vnj,AtacamaCosmologyTelescope:2025nti} and the South Pole Telescope (SPT)-3G D1 \cite{SPT-3G:2025bzu}. We combine the three CMB surveys to measure temperature (TT), polarization (EE), and temperature-polarization cross (TE) spectra,  as follows. For the low-$\ell$ range ($\ell<30$) we use the \texttt{Commander} likelihood for TT \cite{Planck-2018-likelihoods}, and the the \texttt{Sroll2} likelihood for EE \cite{2020_Sroll2_Pagano}, both based on \emph{Planck}. For the high-$\ell$ range ($\ell\geq 30$), the CMB surveys have overlapping measurements and we choose to truncate the high-$\ell$ CMB likelihoods before combining them to avoid the need to model cross-correlations between CMB surveys. We also use foreground-marginalized versions of the CMB likelihoods to avoid leaving foreground nuisance parameters unconstrained by this truncation. We use the lite-version of the \texttt{CamSpec} TTTEEE likelihood \cite{Planck-2020-NPIPE} as presented in \cite{Jense:2025wyg} over the range [($\ell<1500, TT$), ($\ell<1000, TE$), ($\ell<600, EE$)] and complement it with the ACT-lite likelihood \cite{AtacamaCosmologyTelescope:2025blo} evaluated from that multipole range up to $\ell<6500$ (see also \cite{DESI+ACT:2025gwf} for an alternative truncation scheme). Given the small footprint of SPT-3G D1, cross-correlations with the other CMB surveys are less relevant, such that the data can be combined without truncation. We use the SPT-lite likelihood as presented in \cite{SPT-3G:2025bzu}. Additionally, we include information from the 4-point lensing reconstruction and use the unified CMB lensing likelihood based on \Planck, ACT, and SPT presented in \cite{ACT:2025qjh_FrankQu_unified_CMBlensing}.

Finally, we include Type Ia supernovae data from the recently reanalyzed DES-Dovekie sample \cite{DES:2025sig_DESDovekie_cosmo}, which now comprises a sample of 1820 SNe (197 SNe at $z<0.1$ and 1623 SNe at $0.1<z<1.13$). This sample has 1718 SNe in common with the previous DES sample originally presented in \cite{DES:2024tys} and around 100 SNe that were not previously included. The differences between the two samples are a result of new light-curve fits after updates to the calibration, SALT model, re-run simulations, and other changes as described in \cite{Popovic:2025glk_DESDovekie_reassesment,DES:2025sig_DESDovekie_cosmo}. Overall, this new SNe sample led to a $\sim 1\sigma$ shift towards a smaller $\Omega_\text{m}$ value in $\Lambda$CDM compared to the original DESY5-SNe sample. The level of agreement with other SNe datasets, such as Pantheon+ \cite{Brout:2022} and Union3.1 (originally Union3 \cite{Rubin:2023jdq}), has improved significantly in the recent updated analysis \cite{Hoyt:2026fve_SNeReanalysis}. Given this agreement, we only present results for the DES-Dovekie sample, as it provides the tightest constraints, and defer combinations with other SNe samples to future work.

Finally, to calibrate the sound horizon, $r_\mathrm{d}$, we also use a measurement of the baryon density, $\Omega_\mathrm{b} h^2=0.02218 \pm 0.00055$, from \cite{BBN:2024}. This is based on measurements of the primordial deuterium abundance \cite{Cooke:2018} and Big Bang Nucleosynthesis (BBN). When combining with this measurement, we will refer to it as BBN.

We perform cosmological likelihood calculations using the \texttt{cobaya} package \cite{Torrado21} for Markov Chain Monte Carlo parameter sampling. The priors on the cosmological parameters are identical to the ones used in \cite{DESIDR2:2025}. We use a Gelman-Rubin convergence criterion of $R-1<0.01$ \cite{Gelman92}. Distances and CMB spectra are computed using the Einstein-Boltzmann solver \texttt{camb} \cite{Lewis00}. We evaluate the quality of the fits in terms of the $\chi^2_\mathrm{MAP}$ at the maximum a posteriori (MAP) points of each chain, which are found using the \texttt{bobyqa} algorithm \cite{BOBYQAPowell, BOBYQACartis2018, BOBYQACartis2021}.

\subsection{Matter era distance}
\label{subsec:matter_distance}

The \lya\ forest enables accurate distance measurements well into the matter-dominated era. This provides an excellent opportunity to test $\Lambda$CDM at an epoch, $z_\mathrm{eff}=2.33$, when the contribution of dark energy to the cosmic energy density is greatly suppressed. The distances of \Cref{eq:dmdh_dr2} can be directly compared with the $\Lambda$CDM predictions derived from DESI BAO measurements of galaxy and quasar clustering at lower redshifts. The \lya\ measurements are consistent with these predictions, though with a $1.4\sigma$ offset from the prediction for the transverse comoving distance of $D_\mathrm{M}(z_\mathrm{eff})/r_\mathrm{d}=38.82\pm 0.16$. Combining DESI BAO with \lya\ FS improves the $\Lambda$CDM constraint by $30\%$ to $D_\mathrm{M}(z_\mathrm{eff})/r_\mathrm{d}=38.85\pm 0.11$.
However, unlike \Cref{eq:dmdh_dr2}, the inferred distance is not independent of assumptions about dark energy at lower redshifts. Allowing for dynamical dark energy and adopting the CPL parameterization for an evolving dark energy equation of state \cite{Chevallier:2001,Linder2003}, $w(a)=w_0+w_a(1-a)$, we find a constraint of $D_\mathrm{M}(z_\mathrm{eff})/r_\mathrm{d}=38.91\pm 0.12$, which corresponds to a $0.5\sigma$ increase compared to $\Lambda$CDM.

Once combined with acoustic scale information from the CMB, these constraints on $D_\mathrm{M}(z_\mathrm{eff})/r_\mathrm{d}$ provide a useful diagnostic for physics at higher redshifts \cite{Eisenstein04,Weiner26}. Similar to the BAO feature in the \lya\ forest, the acoustic peaks in the power spectra of CMB temperature and polarization anisotropies yield a precise determination of the transverse comoving distance $D_\mathrm{M}(z_\mathrm{d})/r_\mathrm{d}\equiv1/\theta^\mathrm{CMB}_\mathrm{d}$, where $z_\mathrm{d}\approx1060$ is the redshift at the drag epoch. Taking the difference with the constraint on $D_\mathrm{M}(z_\mathrm{eff})/r_\mathrm{d}$ from DESI defines a long baseline during the matter era, $D_\mathrm{M}(z_\mathrm{d},z_\mathrm{eff})/r_\mathrm{d}\equiv D_\mathrm{M}(z_\mathrm{d})/r_\mathrm{d} - D_\mathrm{M}(z_\mathrm{eff})/r_\mathrm{d}$, known as the matter era distance interval \cite{Weiner26}.

\begin{figure}
    \includegraphics[width=1.0\columnwidth,keepaspectratio]{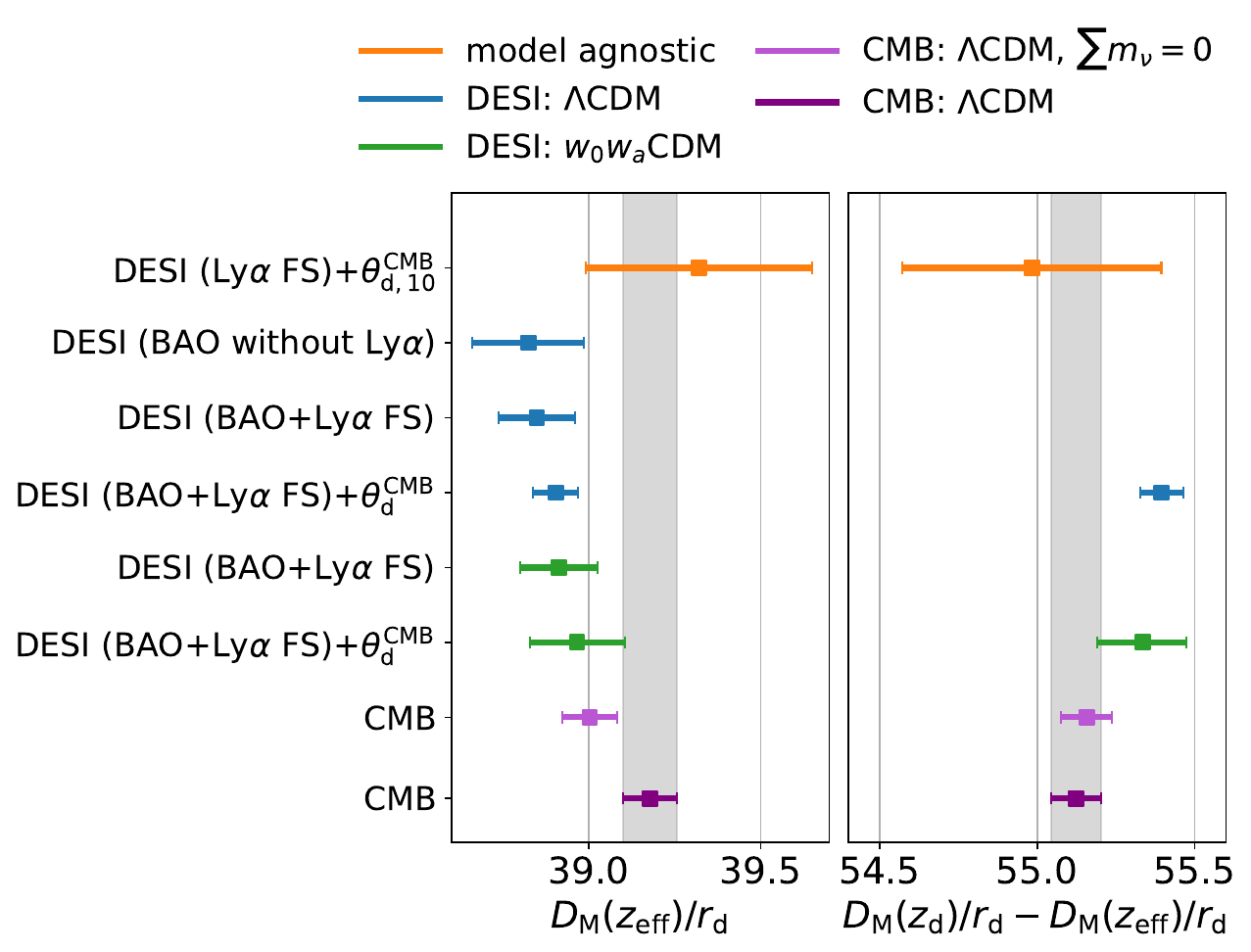}
    \caption{
    Constraints on the transverse comoving distance, $D_\mathrm{M}(z_\mathrm{eff})/r_\mathrm{d}$, and the matter era distance interval, $D_\mathrm{M}(z_\mathrm{d},z_\mathrm{eff})/r_\mathrm{d}\equiv D_\mathrm{M}(z_\mathrm{d})/r_\mathrm{d} - D_\mathrm{M}(z_\mathrm{eff})/r_\mathrm{d}$, for various data combinations and models, showing the posterior means and 68\% credible intervals. The top row shows model-agnostic constraints from \lya\ FS combined with a broad Gaussian prior on $\theta^\mathrm{CMB}_\mathrm{d}$, with $10$ times the uncertainty of that obtained assuming $\Lambda$CDM (orange). The next three rows show the geometric constraints for $\Lambda$CDM from DESI BAO without \lya, combined with \lya\ FS, and combined with an informative prior on $\theta^\mathrm{CMB}_\mathrm{d}$ (blue). This last constraint is in $2.6\sigma$ tension with the $\Lambda$CDM constraint derived from the full CMB likelihood (purple). Adopting a $w_0w_a$CDM background reduces the tension (green), as does reducing the sum of neutrino masses (light purple).
    }
    \label{fig:medi}
\end{figure}

\Cref{fig:medi} shows constraints on $D_\mathrm{M}(z_\mathrm{eff})/r_\mathrm{d}$ and $D_\mathrm{M}(z_\mathrm{d},z_\mathrm{eff})/r_\mathrm{d}$ for various data combinations and models. The top row shows constraints that are directly derived from \Cref{eq:dmdh_dr2} and a broad Gaussian prior on $\theta^\mathrm{CMB}_\mathrm{d}$.\footnote{Constraints on $\theta^\text{CMB}_{\mathrm{d}}$ are model dependent. For instance, uncertainties increase significantly when varying the sum of neutrino masses. In order to account for this model dependence, we adopt a prior $\mathcal{N}(1.06042,0.0028^2)$, with a width 10 times greater than the uncertainty of the marginalized constraint from the baseline CMB likelihood, $\theta^\mathrm{CMB}_\mathrm{d}=1.06042\pm0.00028$, assuming $\Lambda$CDM. We refer to this prior as $\theta^\text{CMB}_{\mathrm{d},10}$.} This approach is insensitive to assumptions about dark energy at lower redshifts, but the uncertainty is currently too large to place meaningful constraints on extensions such as massive neutrinos. The constraints can be tightened considerably by adding information from DESI BAO at lower redshifts, but at the expense of introducing greater model dependence. The fourth row in \Cref{fig:medi} shows the constraints obtained assuming $\Lambda$CDM from \lya\ FS combined with DESI BAO at lower redshifts and an informative prior on $\theta^\mathrm{CMB}_\mathrm{d}$. This constraint relies entirely on geometric information and should be confronted with the $\Lambda$CDM prediction from the full CMB likelihood, including lensing reconstruction, shown at the bottom of \Cref{fig:medi}. There is a $2.6\sigma$ tension between the two constraints, which can be expressed either in $D_\mathrm{M}(z_\mathrm{d},z_\mathrm{eff})/r_\mathrm{d}$ or in $D_\mathrm{M}(z_\mathrm{eff})/r_\mathrm{d}$, since the latter dominates the uncertainty.

These distance measurements offer an alternative lens to analyze joint constraints from DESI and CMB on extended cosmological models. For masses allowed by current data, neutrinos become non-relativistic after recombination, increasing the matter density after some redshift $z_\mathrm{nr}<z_\mathrm{d}$. This increases the CMB prediction for $D_\mathrm{M}(z_\mathrm{eff})/r_\mathrm{d}$ more than for $D_\mathrm{M}(z_\mathrm{d})/r_\mathrm{d}$, thus reducing the interval $D_\mathrm{M}(z_\mathrm{d},z_\mathrm{eff})/r_\mathrm{d}$. The late-time geometric constraints from DESI are entirely unaffected, but once combined with $\theta^\mathrm{CMB}_\mathrm{d}$, the prediction for the interval $D_\mathrm{M}(z_\mathrm{d},z_\mathrm{eff})$ increases for larger neutrino masses. Larger neutrino masses thus increase the tension in both directions. As a result, combinations of DESI and CMB data yield strong upper limits on the neutrino mass, even pushing towards negative effective neutrino masses if allowed \cite{Craig24,Green24,Elbers_24,DESI_Elbers25}, in order to compensate for the difference in the matter density \cite{Loverde24,Weiner26}.

While physical neutrino masses increase the tension with the CMB, dynamical dark energy can partially resolve it. This is best understood in terms of the low-redshift distance ratio $D_\mathrm{M}(z_\mathrm{eff})/r_\mathrm{d}$, since $D_\mathrm{M}(z_\mathrm{d})/r_\mathrm{d}$ is much less sensitive to low-redshift assumptions about dark energy. The geometric constraints from DESI, combined with $\theta^\mathrm{CMB}_\mathrm{d}$, define a sharp degeneracy direction in the $w_0$-$w_a$ plane (see \Cref{subsec:extended_models}). Moving away from $\Lambda$CDM along this degeneracy increases the DESI constraint on $D_\mathrm{M}(z_\mathrm{eff})/r_\mathrm{d}$ by $0.5\sigma$. At the same time, moving in that direction reduces the CMB prediction for $D_\mathrm{M}(z_\mathrm{eff})/r_\mathrm{d}$, alleviating the tension and the need for smaller neutrino masses.

\subsection{\lcdm\ results}
\label{subsec:lcdm}

\begin{figure}
    \includegraphics[width=\columnwidth]{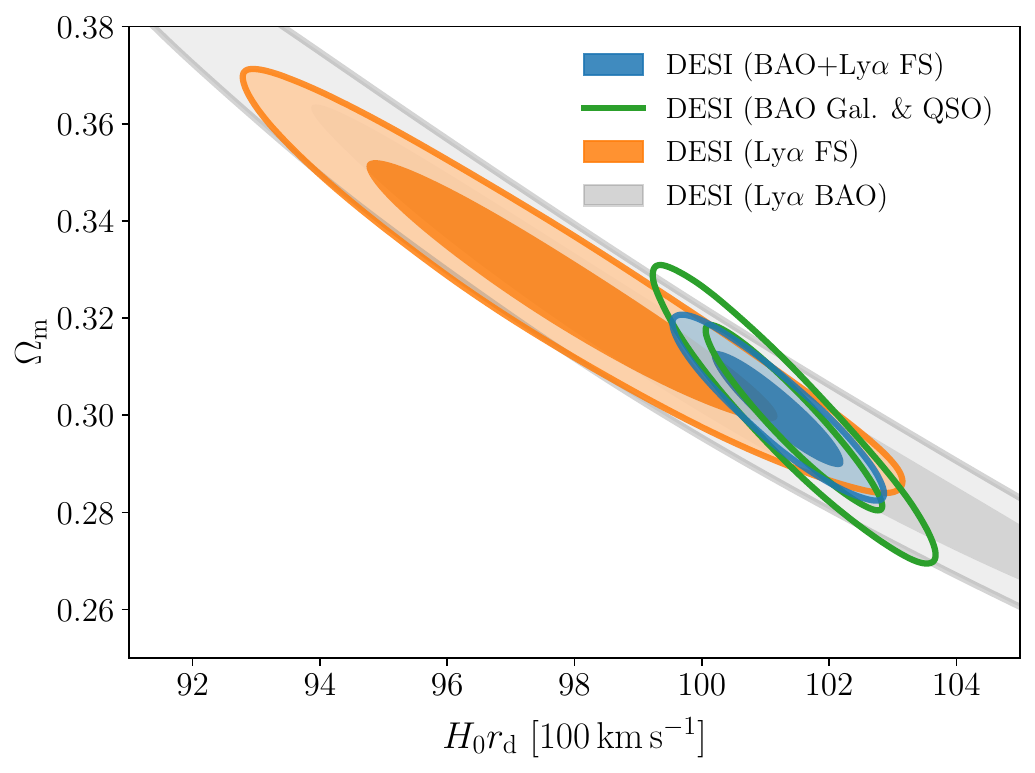}
    \vspace{0.5em}
    \includegraphics[width=\columnwidth]{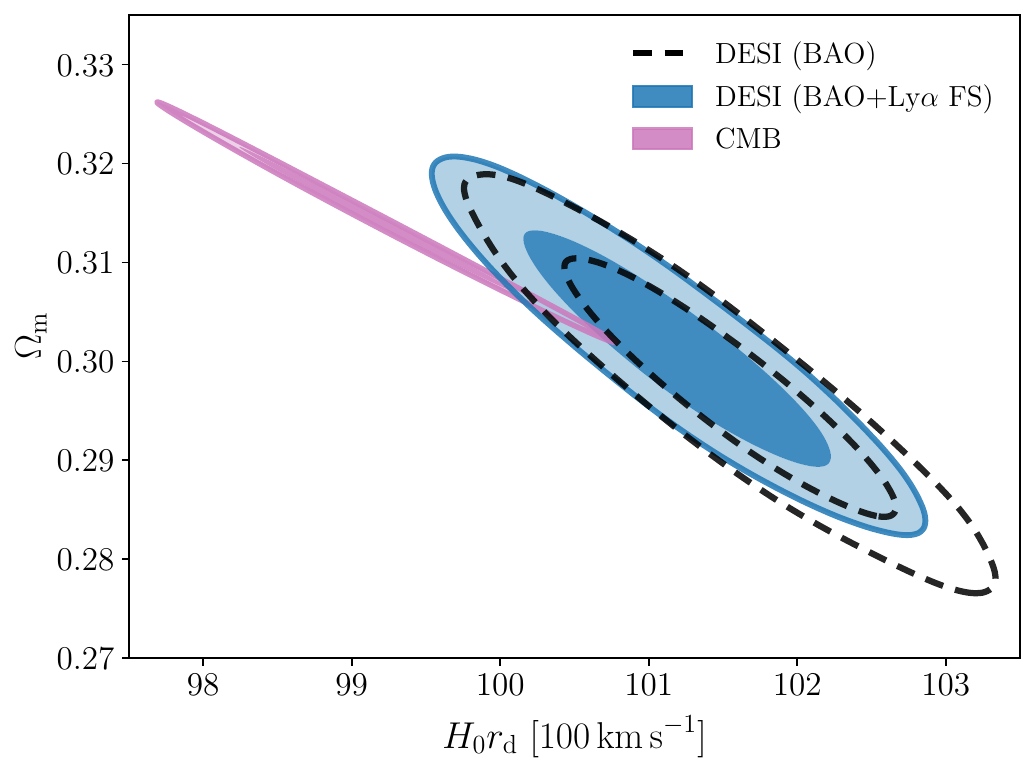}
    \caption{
        \textbf{Top:} Marginalized 68\% and 95\% confidence regions representing the posterior constraints on $\Omega_\mathrm{m}$ and $H_0r_d$. The light gray and green contours represent the individual constraints from Ly$\alpha$ BAO and BAO from galaxy and quasar clustering, respectively. The orange contour shows the updated constraint from the Ly$\alpha$ FS analysis and the blue contour the represents the updated DESI BAO+Ly$\alpha$ FS results. \textbf{Bottom:} Constraints comparing the updated DESI results with respect to the CMB prediction in the \lcdm\ model. The DESI BAO constraints (same as the DR2 BAO results presented in \cite{DESIDR2:2025}) are shown as the black dashed contour, while the CMB constraints are shown using the pink contour. Adding the new \lyaf\ measurement to DESI BAO improves the combined DESI result by $8\%$ (blue) and leads to a slight shift towards the CMB, reducing the tension between the two.
    }
    \label{fig:lcdm}
\end{figure}

We next turn our attention to results in the flat \lcdm\ model. When using only DESI data, the parameters we can constrain are the matter fraction, $\Omega_\mathrm{m}$, and the product of the Hubble constant and the size of the acoustic scale, $H_0r_\mathrm{d}$ \cite{DESIDR2:2025}. When inferring these quantities from the \lyaf\ results alone, we find:
\begin{equation}
\left.
 \begin{aligned}
\Omega_\mathrm{m} &= 0.325\pm 0.018, \\
hr_\mathrm{d} &= (97.9\pm 2.1) \text{ Mpc},
 \end{aligned}
\ \right\} \ \ \mbox{\text{\leftparbox{2cm}{\lya\ FS}}}\label{eq:lya_lcdm}
\end{equation}
with a correlation coefficient of $\rho=-0.96$ and $h=H_0/(100 \text{ km} \text{ s}^{-1} \text{ Mpc}^{-1})$. These results are consistent with the DESI BAO measurements at lower redshifts, as illustrated in the top panel of \Cref{fig:lcdm}. When compared to the combined DESI BAO constraints excluding \lya, we find them statistically consistent within 1$\sigma$ (orange versus green in \Cref{fig:lcdm}). Also, the new Ly$\alpha$ FS constraints represent about a factor of 2 precision improvement in $\Omega_\text{m}$ with respect to its BAO counterpart. This means that the new \lyaf\ result can be safely combined with DESI BAO measurements at $z<2$ within \lcdm. The joint DESI results are:
\begin{equation}
\left.
 \begin{aligned}
\Omega_{\mathrm{m}} &= 0.3012\pm 0.0079, \\
hr_\mathrm{d} &= (101.19\pm 0.68) \text{ Mpc},
 \end{aligned}
\ \right\} \ \ \mbox{\text{\leftparbox{2.5cm}{DESI BAO+\lya\ FS}}}\label{eq:desi_lcdm}
\end{equation}
with a correlation coefficient of $\rho=-0.91$. These constraints correspond to a $9\%$ improvement in $\Omega_\mathrm{m}$ and $7\%$ in $hr_\mathrm{d}$ when compared to DESI BAO results alone.

\Cref{fig:lcdm} shows the updated DESI results (blue), which consist of the combination of \lyaf\ full-shape constraints and BAO from galaxy clustering. In the lower panel, we compare the new results with the DESI BAO-only constraints (dashed black) and with the CMB constraints (pink). Adding the new \lyaf\ measurement leads to slightly tighter constraints as described above, and also to a small shift that brings the DESI result closer to the CMB one. The discrepancy between the CMB combination used here and DESI BAO is $2.4\sigma$\footnote{Computed using the Gaussian approximation for the two sets of 2-dimensional results. See Eq. 18 of \cite{DESIDR2:2025}.} (black versus pink), and this is reduced to $2.2\sigma$ if DESI BAO+\lya\ FS (blue versus pink) is instead considered. The combination DESI BAO+\lya\ FS and CMB leads to $\Omega_\text{m} = 0.3042 \pm 0.0033$. In addition, the corresponding spectral index constraint is $n_\mathrm{s} = 0.9739\pm 0.0029$, which is roughly a $0.4\sigma$ shift lower compared to the value found by \cite{Ferreira:2025lrd} in the context of implications for inflation, using the DESI BAO+ACT combination. Moving forward, we also refer to the combination of DESI \lya\ full-shape and galaxy BAO constraints simply as DESI, unless explicitly comparing to DESI BAO.

The discrepancy between CMB and DESI BAO results leads to models with a more flexible background expansion history, such as $w_0w_a$CDM, being preferred over \lcdm\ \cite{DESIDR2:2025,DR2extendedDE}. Therefore, at first glance, we expect a reduction in the tension between the CMB and DESI to directly lead to a reduction in the preference for these models compared to \lcdm. That is indeed what happens when comparing $w_0w_a$CDM and \lcdm, as we will see in the next section. However, the situation is more nuanced, because the new \lya\ result is at high redshift ($z=2.33$), whereas the DESI BAO measurements driving the tension are at low redshift ($z<1$). This means that \lcdm\ might still not be the model that best fits the joint DESI and CMB data. We explore the implications of this in \Cref{subsec:extended_models} below.

\begin{table*}

\centering
\begin{tabular*}{\textwidth}{@{\extracolsep{\fill}} lllllll}
\toprule
Model/Dataset & $\Omega_\mathrm{m}$ & $H_0$ [km s$^{-1}$ Mpc$^{-1}$] & $10^3\Omega_\mathrm{K}$ & $\sum m_\nu$ [eV] & $w$ or $w_0$ & $w_a$ \\
\midrule
$\mathbf{\Lambda}$\textbf{CDM} &  &  &  &  &  &  \\
DESI (Ly$\alpha$ FS) & $0.325\pm 0.018$ & --- & --- & --- & --- & --- \\
DESI & $0.3012\pm 0.0079$ & --- & --- & --- & --- & --- \\
DESI (Ly$\alpha$ FS)+BBN & $0.326\pm 0.018$ & $66.5\pm 1.3$ & --- & --- & --- & --- \\
DESI+BBN & $0.3020\pm 0.0075$ & $68.47\pm 0.57$ & --- & --- & --- & --- \\
DESI+CMB & $0.3042\pm 0.0033$ & $68.11\pm 0.24$ & --- & --- & --- & --- \\
\hline
$\mathbf{w_0w_a}$\textbf{CDM} &  &  &  &  &  &  \\
DESI & $0.363^{+0.033}_{-0.017}$ & --- & --- & --- & $-0.43^{+0.31}_{-0.14}$ & $<-1.68$ \\
DESI+CMB & $0.343^{+0.019}_{-0.022}$ & $64.5\pm 1.9$ & --- & --- & $-0.54^{+0.19}_{-0.21}$ & $-1.39^{+0.61}_{-0.50}$ \\
DESI+CMB+DES-Dovekie & $0.3134\pm 0.0052$ & $67.37\pm 0.54$ & --- & --- & $-0.821\pm 0.054$ & $-0.65\pm 0.20$ \\
\hline
$\mathbf{\Lambda}$\textbf{CDM+}$\mathbf{\Omega_\textbf{K}}$ &  &  &  &  &  &  \\
DESI & $0.289\pm 0.011$ & --- & $51.00\pm32.0$ & --- & --- & --- \\
DESI+CMB & $0.3038\pm 0.0034$ & $68.51\pm 0.33$ & $2.10\pm1.1$ & --- & --- & --- \\
DESI+CMB+DES-Dovekie & $0.3050\pm 0.0032$ & $68.43\pm 0.30$ & $2.20\pm1.0$ & --- & --- & --- \\
\hline
$\mathbf{\Lambda}$\textbf{CDM+}$\mathbf{\sum m_\nu}$ &  &  &  &  &  &  \\
DESI+CMB & $0.3017\pm 0.0033$ & $68.36\pm 0.26$ & --- & $< 0.0592$ & --- & --- \\
DESI+CMB+DES-Dovekie & $0.3030\pm 0.0033$ & $68.26\pm 0.25$ & --- & $< 0.0654$ & --- & --- \\
\hline
$\mathbf{w_0w_a}$\textbf{CDM+}$\mathbf{\sum m_\nu}$ &  &  &  &  &  &  \\
DESI+CMB & $0.343^{+0.021}_{-0.024}$ & $64.5\pm 2.0$ & --- & $< 0.166$ & $-0.53^{+0.21}_{-0.25}$ & $-1.42^{+0.78}_{-0.58}$ \\
DESI+CMB+DES-Dovekie & $0.3129\pm 0.0054$ & $67.40\pm 0.55$ & --- & $< 0.128$ & $-0.827\pm 0.056$ & $-0.62^{+0.25}_{-0.20}$ \\
\bottomrule
\end{tabular*}
\caption{
Summary of cosmological parameter constraints from DESI \lya\ FS and DESI BAO + \lya\ FS (labeled as `DESI' here) alone and in combination with external datasets. We quote the marginalized posterior means and the $68\%$ credible intervals for all parameters, except for the sum of neutrino masses, $\sum m_\nu$, for which we quote the $95\%$ upper bound. The unreported $\sum m_\nu$ values here corresponds to a fixed 0.06~eV value.
\label{tab:cosmo_res}}
\end{table*}

We next turn to comparisons with results from SNe analyses. The DES Dovekie SNe sample constrains $\Omega_\mathrm{m}=0.330 \pm 0.015$ \cite{Popovic:2025glk_DESDovekie_reassesment}, while the latest constraints using the Union sample are: $\Omega_\mathrm{m}=0.334^{+0.025}_{-0.024}$ \cite{Rubin:2026,Hoyt:2026fve_SNeReanalysis}. Both of these are in good agreement with the \lyaf\ measurement in \Cref{eq:lya_lcdm}. Notably, the DESI \lya\ result alone is $~30\%$ tighter than the Union one, and only $20\%$ weaker than the tightest SNe constraint from the DES Dovekie sample. Adding the new \lyaf\ constraint to the rest of the DESI BAO measurements results in slightly larger values of $\Omega_\mathrm{m}$, reducing the tension with respect to DES Dovekie from $1.9\sigma$\footnote{The tension was originally $2.9\sigma$ in \cite{DESIDR2:2025}, when comparing results from DESI BAO and the original DES Y5 SNe sample, and was reduced to $1.9\sigma$ with the DES Dovekie sample \cite{Popovic:2026}.} to $1.7\sigma$, and with respect to Union from $1.4\sigma$ to $1.3\sigma$.

To measure the Hubble constant, we need to calibrate the BAO ruler by constraining $r_\mathrm{d}$. Assuming standard pre-recombination physics, and a fixed neutrino sector with $\sum m_\nu=0.06$ eV, the only missing quantity is a constraint on the baryon density $\Omega_\mathrm{b}h^2$. Using the $\Omega_\mathrm{b}h^2$ prior from BBN \cite{BBN:2024}, we can obtain a constraint on the Hubble constant that is independent of CMB anisotropies. For the \lyaf\ full-shape measurement, this corresponds to:
\begin{equation}
\left.
H_0 = (66.5\pm 1.3)\;\kmsMpc \quad\mbox{\text{\leftparbox{3cm}{(\lya\ FS+BBN).}}}
  \right.
\end{equation}
This represents a $\sim2\%$ constraint on the Hubble constant, derived entirely from high-redshift data\footnote{The BBN constraints are based on primordial deuterium abundance measurements from DLAs in spectra of high-redshift quasars \cite[e.g.,][]{Cooke:2018,pdg:2022}.} and independent of CMB anisotropies. The measurement is consistent with both CMB and DESI BAO + BBN results, and is $1.7\sigma$ away from the tightest cosmic distance ladder result from the Chicago-Carnegie Hubble Program \cite{Freedman:2025}, and in $4.3\sigma$ tension with the latest results from SH0ES \cite{Riess:2025}. This tension rises to $4.6\sigma$ with respect to the baseline result from the $H_0$ Distance Network Collaboration \cite{H0DN26}.

Our full set of parameter constraints is presented in \Cref{tab:cosmo_res}, and also includes the combination of DESI+BBN and DESI+CMB results. Both of these are consistent with the results in \cite{DESIDR2:2025}, but with slightly smaller uncertainties.

\subsection{Extended models}
\label{subsec:extended_models}

In the following, we perform analyses of extended models beyond $\Lambda$CDM, covering evolving dark energy, spatial curvature, and a free neutrino mass.

\subsubsection{Dark energy}

The unknown nature of dark energy, as well as the current trends in cosmology, make the search for constraining the equation of state of dark energy and its redshift dependence of particular interest. We model the equation of state relating the pressure and the energy density of dark energy, $P=w\rho c^2$, with the Chevallier-Polarski-Linder (CPL) parameterization, which uses $w(a)=w_0 + (1-a)w_a$ where $a=1/(1+z)$. It is worth noting that \cite{DR2extendedDE} showed that an alternative redshift-dependence of the dark energy equation of state or a non-parametric approach like binning or Gaussian processes converges to the same conclusions as the CPL parameterization. We refer to this model in the following as $w_0w_a$CDM. In particular, a sequence of DESI papers \cite{DESI2024.VI.KP7A,DESI2024.VII.KP7B,DESIDR2:2025} have reported a preference for evolving dark energy from joint analyses with external data. Currently and as shown in \Cref{sec:cmb}, the DESI BAO results show a $\sim 3\sigma$ preference in favor of evolving dark energy once combined with CMB data. On the other hand, further reanalyses on the side of the SNe datasets \cite{DES:2025sig_DESDovekie_cosmo,Hoyt:2026fve_SNeReanalysis} mitigated discrepancies between different SNe samples, leaving results essentially unchanged regardless of which SNe dataset is chosen, with a preference for evolving dark energy ranging from 3$\sigma$ to 3.5$\sigma$ for the DESI+CMB+SNe combination. 

\begin{figure}
    \includegraphics[width=1.0\columnwidth,keepaspectratio]{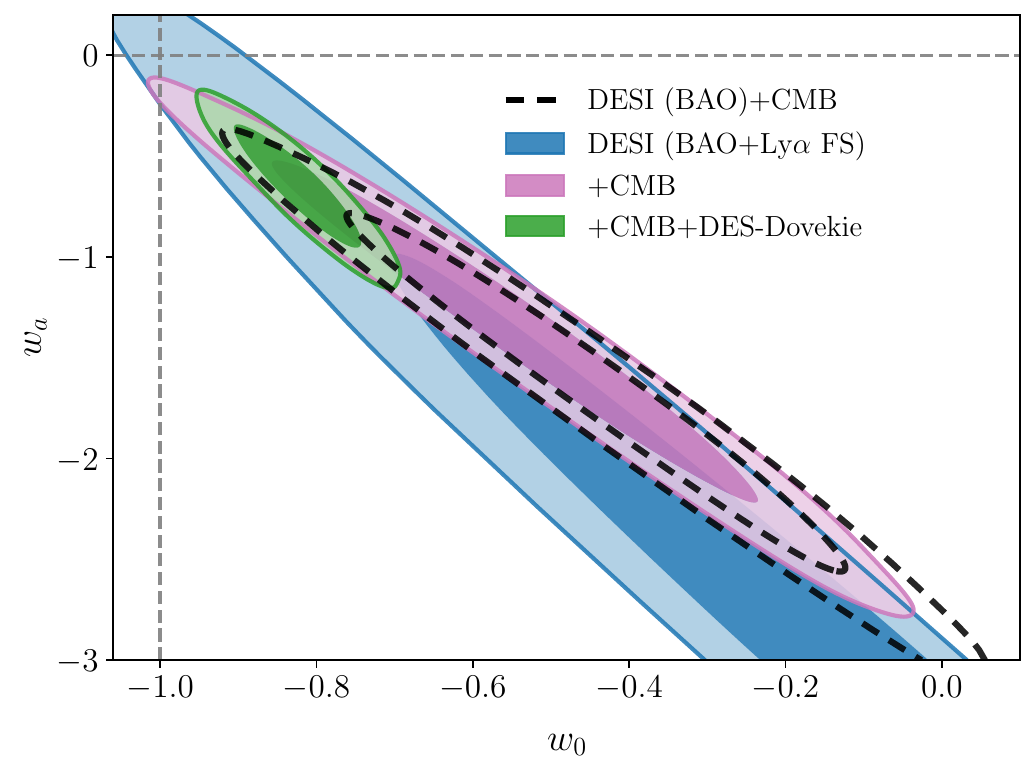}
    \caption{
        68\% and 95\% confidence contours on the dark energy parameters $w_0$ and $w_a$ from DESI alone (blue) and in combination with other data sets. The intersection of the dashed gray lines represents the $\Lambda$CDM model. We show the previous DESI DR2 BAO constraints with the updated CMB combination used here in dashed black contours and its equivalent with the new DESI Ly$\alpha$ measurement in the purple contour. The updated DESI and CMB data show a preference of $2.7\sigma$ for $w_0w_a$CDM compared to \lcdm. This preference increases to $3.2\sigma$ when also including SNe from the DES-Dovekie sample (green).
    }
    \label{fig:w0wa}
\end{figure}

For this analysis, the high-redshift measurements of cosmological distances accessible to the Ly$\alpha$ forest allow us to set constraints on the expansion history deep into the matter-domination era. In particular, constraining the AP parameter with sufficient precision at high-redshift can potentially help distinguish between a $\Lambda$CDM and a $w_0w_a$CDM model (illustrated in \Cref{fig:summary_models_vs_data} below). Using the new \lya\ measurements, the DESI constraints on $w_0$ and $w_a$ are now
\begin{equation}
\left.
 \begin{aligned}
w_0 &= -0.43^{+0.31}_{-0.14}, \\
w_a &< -1.68.
 \end{aligned}
\ \right\} \ \ \mbox{\text{\leftparbox{4cm}{DESI BAO+\lya\ FS}}}\label{eq:desi_w0wacdm}
\end{equation}
This result is illustrated in \Cref{fig:w0wa} (blue contours). 

To compare the fit with respect to \lcdm, we estimate the $\Delta \chi^2_\text{MAP} \equiv -2\Delta \ln \mathcal{L}$ evaluated by maximizing the posteriors of each model and comparing the fit for the $w_0w_a$CDM model to a $w_0w_a$CDM model with $w_0=-1$ and $w_a=0$. We find that $\Delta \chi^2_\text{MAP}=-5.6$ for DESI (BAO+Ly$\alpha$ FS), corresponding to a $1.9\sigma$ preference for $w_0$ and $w_a$ distinct from a cosmological constant. This means that the inclusion of the Ly$\alpha$ FS data has slightly increased the $\Delta \chi^2_\text{MAP}$ difference between the two models with respect to DESI BAO constraints. On the other hand, the individual $\chi^2$ values have actually increased, indicating that the fit to the data has deteriorated mildly compared to the fit using DESI BAO alone. 

The flexibility of the $w_0w_a$CDM model allows it to provide a good fit to both DESI measurements at late-times and the new \lya\ forest measurements at $z>2$. However, the situation changes when also considering CMB measurements of the acoustic scale at recombination. The acoustic scale is defined as $\theta_*=r_*/D_\text{M}$ and its measurement sets an anchor at high redshift, roughly at $z_*\simeq 1090$. This high-redshift anchor removes some of the freedom in the fit for the dark energy parameters, forcing the fit to not deviate too far from the \emph{Planck} value at recombination (although we notice that the acoustic scale inferred by DESI is consistent with the CMB measurement). The DESI+CMB joint fit measures
\begin{equation}
\left.
 \begin{aligned}
w_0 &= -0.54^{+0.19}_{-0.21}, \\
w_a &= -1.39^{+0.61}_{-0.50},
 \end{aligned}
\ \right\} \ \ \mbox{\text{\leftparbox{4cm}{DESI BAO+\lya\ FS +CMB}}}\label{eq:desi+cmb_w0wacdm}
\end{equation}
with a $\Delta \chi^2_\text{MAP}=-9.8$ relative to \lcdm. This corresponds to a preference for evolving dark energy of $2.7\sigma$. Therefore, adding the new \lya\ constraints leads to a roughly $\sim0.3\sigma$ shift towards $\Lambda$CDM when compared to the DESI (BAO)+CMB constraint, as illustrated in \Cref{fig:w0wa} (dashed black versus pink contours). This is consistent with the updated AP measurement being in better agreement with \Planck\ CMB, as well as the DESI posteriors on $\Lambda$CDM measuring a slightly higher $\Omega_\text{m}$ value. A high value of $\Omega_\mathrm{c} h^2$ inferred by the CMB can also be correlated with the evidence for evolving dark energy from the DESI+CMB combination. 

On the other hand, it has been pointed out that the value of the optical depth to reionization, $\tau$, from the CMB is also correlated with the preference for evolving dark energy \cite{Sailer:2025lxj,Jhaveri:2025neg}. A much higher value of the optical depth would effectively lead to a lower value of $\Omega_\text{m}$, making DESI and the CMB consistent. We perform a test of this scenario within the $\Lambda$CDM model by dropping the low-$\ell$ EE CMB likelihood. This leads to $\Omega_\text{m}=0.2998\pm 0.0035$ and $\tau=0.0905\pm 0.0096$ for the combination of DESI and remaining CMB data. However, this is incompatible with the $\Omega_\text{m}$ value inferred by the SNe samples and of course inconsistent with the CMB low-$\ell$ E mode measurements by \emph{Planck}. It would also raise the amplitude of clustering, which will be tested with forthcoming galaxy clustering and lensing analyses. Independently of this, high optical depth scenarios can be tested with astrophysical probes of reionization \cite[e.g][]{Sims25,Elbers25,Aggarwal26,GarciaGallego25,Kageura26,Cain26}. We compare the best-fit model from this test with the $w_0w_a$CDM fit and various other models in \Cref{subsec:cosmo_summary} below.

The inferred dark energy equation of state $w(z)$ still favors a phantom-crossing-like behaviour, indicating a $w(z)<-1$ during matter-domination and a $w(z)>-1$ at low redshift, where we only have one $D_\text{V}/r_\text{d}$ measurement from the BGS tracer. Thus, to extend our analysis we use the DES-Dovekie SNe sample to provide constraints from the distance modulus and constrain the low-redshift information not fully accessible by BAO. In combination with DESI and CMB, we obtain
\begin{equation}
\left.
 \begin{aligned}
w_0 &= -0.821\pm 0.054, \\
w_a &= -0.65\pm 0.20.
 \end{aligned}
\ \right\} \ \ \mbox{\text{\leftparbox{4cm}{DESI BAO+\lya\ FS +CMB+DES-Dovekie}}}\label{eq:desi+cmb+sne_w0wacdm}
\end{equation}
The associated preference for evolving dark energy under this combination is 3.2$\sigma$ with $\Delta \chi^2_\text{MAP}=-12.9$. The use of alternative SNe datasets shows a significance of 3.0$\sigma$ ($\Delta \chi^2_\text{MAP}=-11.5$) and 3.5$\sigma$ ($\Delta \chi^2_\text{MAP}=-15.4$) for the updated Pantheon+ and Union3.1 samples, respectively. The complete set of constraints corresponding to these combinations and beyond is shown in \Cref{tab:cosmo_res}. 

We can also compute the pivot redshift $z_\text{p}$ at which the dark energy equation of state, $w_\text{p}=w(z_\text{p})$, is best constrained. While the pivot redshift depends on the dataset combination and we expect $w_\text{p}$ to be close to $-1$ since the pivot redshift is close to the crossing redshift, this quantity can still give us an insight into the precision at which we are constraining the dark energy equation of state. In particular, we find that $w_\text{p} = -1.030 \pm 0.040$ with $z_\text{p}=0.55$ for DESI+CMB, and $w_\text{p} = -0.981 \pm 0.022$ with $z_\text{p}=0.32$ for DESI+CMB+DES-Dovekie. This represents roughly an 8\% precision increase with respect to the values quoted in \cite{DESIDR2:2025}. Finally, we notice that the figure-of-merit (FoM) based on $w_0$-$w_a$ from DESI+CMB+DES-Dovekie reaches 227.4, which is 4\% greater with respect to the FoM if we only use DESI BAO.

\subsubsection{Neutrino masses}
\label{subsec:neutrinos}

\begin{figure}
    \includegraphics[width=1.0\columnwidth,keepaspectratio]{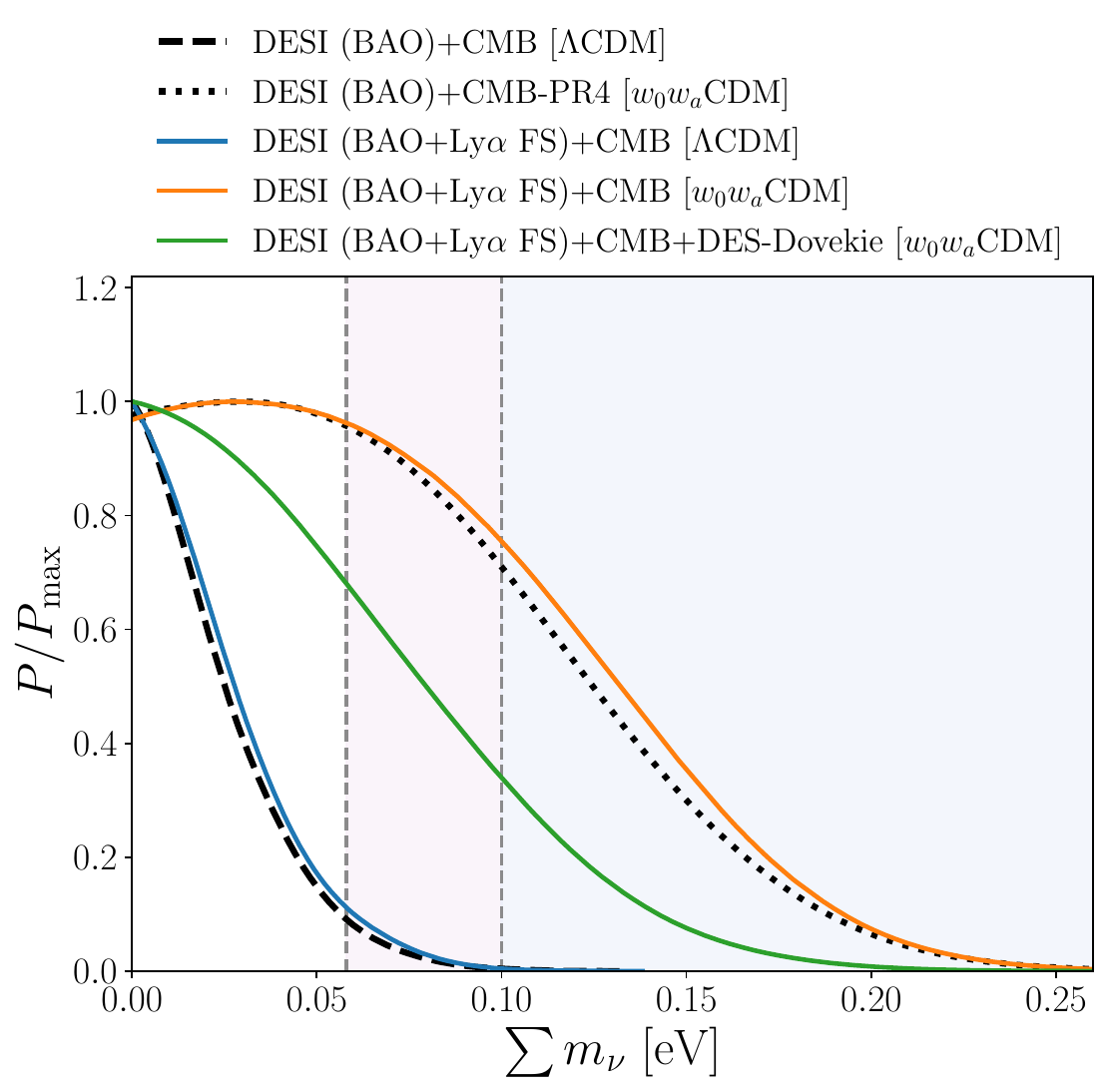}
    \caption{
        Marginalized posterior distributions for the sum of neutrino masses, $\sum m_\nu$, from combinations of DESI, CMB, and SN data for $\Lambda$CDM and $w_0w_a$CDM. The constraints from DESI DR2 BAO without \lya\ FS combined with CMB data are shown in black for both $\Lambda$CDM (dashed) and $w_0w_a$CDM (dotted). The addition of \lya\ FS shifts the posteriors slightly towards larger masses (blue and orange). Including DES-Dovekie SN data tightens the $w_0w_a$CDM posterior (green), which remains compatible with the lower limits from neutrino oscillations \cite{Esteban:2024eli}.
    }
    \label{fig:neutrinos}
\end{figure}

We also present updated constraints on the sum of neutrino masses. With the analysis of DR2 BAO, DESI already placed tight constraints on the sum of neutrino masses, assuming $\Lambda$CDM with three degenerate neutrino species and the standard physical prior that $\sum m_\nu>0\,\mathrm{eV}$. In combination with CMB data from \emph{Planck} PR4 and CMB lensing reconstruction from \emph{Planck} and ACT DR6, DESI found $\sum m_\nu<0.0642\,\mathrm{eV}$ (95\%) \cite{DESIDR2:2025,DESI_Elbers25}. Compared to this analysis, we now additionally include new data from ground-based CMB experiments (see \cref{subsec:external_data}). Using DESI DR2 BAO without \lya\ FS, this yields a stronger upper limit of $\sum m_\nu<0.0565\,\mathrm{eV}$ (95\%). However, with the addition of \lya\ FS, this relaxes slightly to
\begin{align}
\sum m_\nu < 0.0592\,\mathrm{eV}. \qquad \mbox{\text{\leftparbox{3.5cm}{($\Lambda$CDM; 95\%):\\ DESI~BAO+\lya~FS +CMB}}}\label{eq:desi+cmb_mnu}
\end{align}
\noindent
This $2\sigma$ upper limit is nearly identical to the lower limit from neutrino oscillations under the normal mass ordering, $\sum m_\nu>0.059\,\mathrm{eV}$ \cite{Esteban:2024eli}. Moreover, the inverted ordering is excluded at more than $3\sigma$. However, this should not be interpreted as strong evidence for the normal ordering, since most of the posterior volume is inconsistent with both mass orderings. Comparing the two cases, we find a modest $\Delta\chi^2_\mathrm{MAP}=-5.2$ in favor of the normal ordering. We show the one-dimensional marginalized posterior distributions for various data combinations in \Cref{fig:neutrinos}.

As explained in \cref{subsec:matter_distance}, the upper limit is relaxed when we adopt a $w_0w_a$CDM background. Combining DESI BAO, \lya\ FS, and CMB data, we find an upper limit of $\sum m_\nu<0.166\,\mathrm{eV}$ (95\%). \Cref{fig:neutrinos} demonstrates that the posterior distribution now peaks in the positive mass range and reveals the slight shift towards larger neutrino masses after the addition of the \lya\ FS measurement. Additionally including SN data tightens this limit again to
\begin{align}
\sum m_\nu < 0.128\,\mathrm{eV}, \qquad \mbox{\text{\leftparbox{3.5cm}{($w_0w_a$CDM; 95\%):\\ DESI~BAO+\lya~FS +CMB+DES-Dovekie}}}\label{eq:desi+cmb_w0wa}
\end{align}
which is very close to the result presented in \cite{DESIDR2:2025}. The forthcoming full-shape analysis of galaxy and quasar clustering from DESI DR2 will yield even greater constraining power. Crucially, this will also bolster the robustness of the constraints by providing information from neutrino free streaming \cite{DESI_Elbers25,DESI_ForeroSanchez26}.

\subsubsection{Curvature}

Cosmological distance measurements also allow us to test deviations from flatness within $\Lambda$CDM. The spatial curvature density parameter, $\Omega_\text{K}$, can affect the measurements of $\Omega_\text{m}$ across the redshift tracers, and even impact measurements of neutrino masses, relaxing the neutrino mass bounds (see \cite{Chebat:2025kes,Pulido-Hernandez:2026hcs}, for instance). While cosmological data have been historically found to be consistent with a flat $\Lambda$CDM model, \cite{Chen:2025mlf} have pointed out that a negative spatial curvature value (or equivalently, a positive $\Omega_\mathrm{K}$ value) can alleviate some of the current cosmological tensions, with the high-redshift window probed by Ly$\alpha$ being key to set constraints that can distinguish between deviations from flatness, evolving dark energy, or the  $\Lambda$CDM model. Therefore, here we evaluate the implications of the new Ly$\alpha$ FS measurement regarding spatial curvature. 

\begin{figure}
    \includegraphics[width=1.0\columnwidth,keepaspectratio]{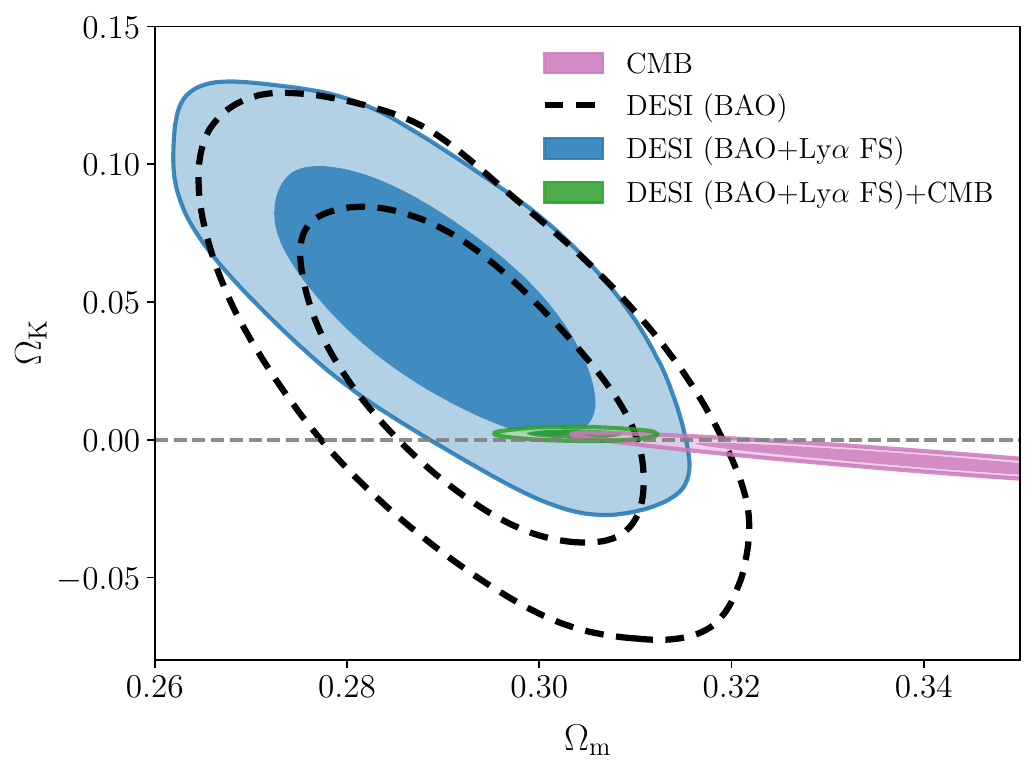}
    \caption{
    Marginalized constraints in the $\Omega_\mathrm{m}$--$\Omega_\mathrm{K}$ plane for a model with free curvature (i.e, $\Lambda{\rm CDM}+\Omega_\mathrm{K}$). We show DESI results with and without the new \lyaf\ constraints (blue versus dashed black). Both results are consistent with a flat Universe, and adding \lya\ FS leads to a $22\%$ improvement in the $\Omega_\mathrm{K}$ precision. Combining DESI and CMB data breaks the degeneracy present in CMB constraints (pink), leading to a tight measurement of $\Omega_\mathrm{K}$ that remains close to zero within $\sim 2\sigma$ (green). 
    }
    \label{fig:omk}
\end{figure}

In this scenario, CMB measurements are more susceptible to exhibiting geometrical degeneracies between $\Omega_\mathrm{m}$ and $\Omega_\mathrm{K}$. Fortunately, this is partially mitigated by CMB lensing and the inclusion of DESI distance measurements which can help break the degeneracy. The updated DESI constraints on spatial curvature are 
\begin{equation}
\left.
10^3\Omega_\text{K} = 51\pm 32,\; \quad\mbox{\text{\leftparbox{3.5cm}{DESI BAO+\lya\ FS}}}
  \right.
\end{equation}
where we find that the addition of Ly$\alpha$ FS leads to a 22\% improvement in the $\Omega_\text{K}$ precision with respect to DESI BAO. When CMB data are included, the DESI+CMB combination yields 
\begin{equation}
\left.
10^3\Omega_\text{K} = 2.1\pm 1.1,\; \quad\mbox{\text{\leftparbox{3.5cm}{DESI BAO+\lya\ FS + CMB}}}
  \right.
\end{equation}
in agreement with previous BAO+CMB analyses \cite{AtacamaCosmologyTelescope:2025blo, AtacamaCosmologyTelescope:2025nti, SPT-3G:2025bzu}. The corresponding confidence contours for $\Omega_\text{K}$ and $\Omega_\text{m}$ are shown in \Cref{fig:omk}. We find a mild statistical preference for an open Universe, with the flat case lying approximately at the boundary of the $2\sigma$ confidence region. More precisely, a $\Delta \chi^2_\text{MAP}=-4.3$, corresponding to a 2.0$\sigma$ preference over flatness. 

Finally, the inclusion of the DES-Dovekie SNe data has a minor impact on the curvature constraints, leading to $10^3\Omega_\mathrm{K} = 2.2\pm 1.0$. As expected, the inclusion of SNe data does not lead to a noticeable improvement, since the geometric $\Omega_\text{K}$-$\Omega_\text{m}$ degeneracy has already been largely broken by the BAO+CMB combination, and DESI provides higher precision on curvature constraints since it is able to access higher redshifts. 

Overall, despite the mild $2\sigma$ fluctuation toward positive curvature values, the constraints are consistent with flatness, and upcoming DESI data releases will help to further constrain spatial curvature.

\subsection{Summary and discussion}
\label{subsec:cosmo_summary}
\begin{figure*}
    \centering
   \includegraphics[width=1.0\textwidth,keepaspectratio]{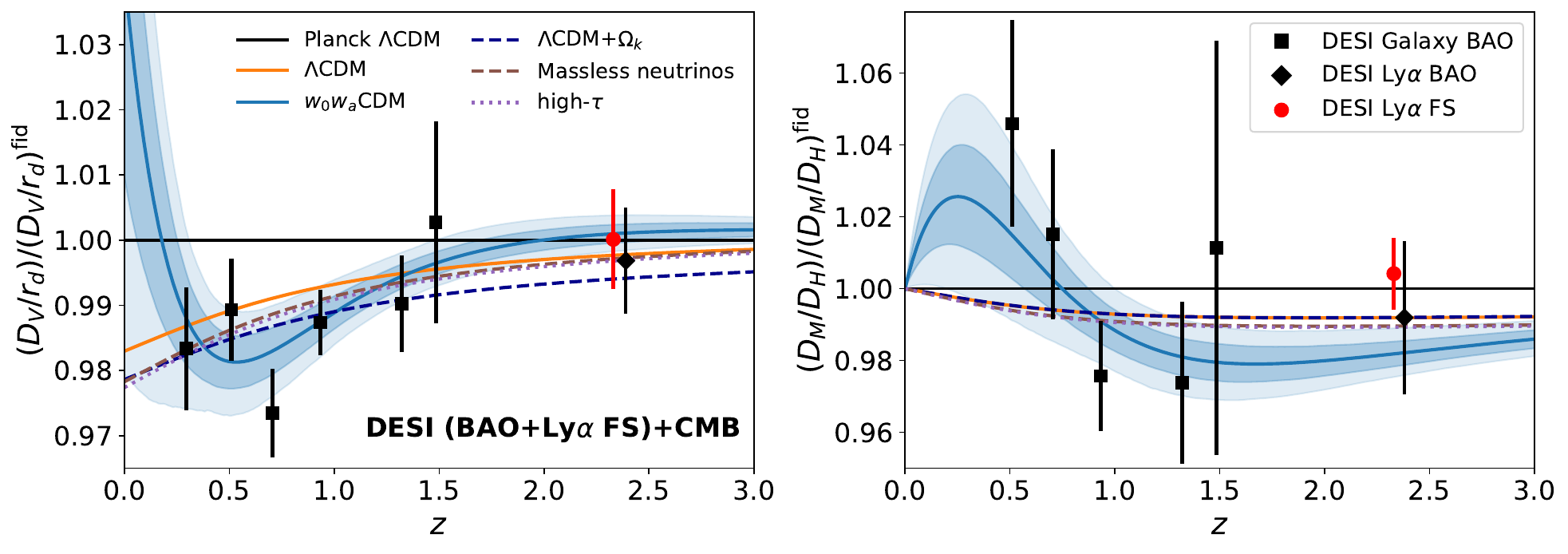}
   \caption{
       Measurements of cosmological distances, $D_\text{V}$ in $r_\text{d}$ units and the ratio $D_\text{M}/D_\text{H}$, relative to the fiducial \emph{Planck} $\Lambda$CDM model (horizontal black line). We show the DESI DR2 BAO measurements and the new DESI DR2 Ly$\alpha$ FS measurement. The DESI $+$ CMB best-fit $w_0w_a$CDM model that provides the highest improvement with respect to $\Lambda$CDM is shown in blue, with the color bands representing the 68\% and 95\% confidence regions. We also show other DESI $+$ CMB best-fit models tested in this work as different colored lines. These models include \lcdm\ (orange), free curvature (dashed blue), massless neutrinos (dashed brown), and the high-$\tau$ scenario (dotted), which is a $\Lambda$CDM fit where we drop the low-$\ell$ EE CMB likelihood.
   }
   \label{fig:summary_models_vs_data}
\end{figure*}

We now summarize the results found across the different models. The best-fit to the DESI data is presented for each model in \Cref{fig:summary_models_vs_data}, where we use the \emph{Planck} $\Lambda$CDM model as reference and fit all other models to DESI+CMB. Overall, the new $D_\text{M}/D_\text{H}$ Ly$\alpha$ FS measurement at $z_\text{eff}=2.33$ is consistent with the \emph{Planck} $\Lambda$CDM model within 0.5$\sigma$. For extensions, this $z_\text{eff}=2.33$ measurement is consistent with models such as DESI+CMB $\Lambda$CDM, with or without spatial curvature, to within 1.1$\sigma$. This measurement, however, does not favor evolving dark energy, with the best-fit for $w_0w_a$CDM being around 2$\sigma$ apart from the Ly$\alpha$ FS data point. We summarize the improvements of the fit with respect to flat $\Lambda$CDM for various data combinations and extensions considered in this work in \Cref{tab:significances}.

Together with the new \lya\ FS measurement, the inclusion of CMB data anchors the high-redshift fits to be close to \emph{Planck} $\Lambda$CDM. However, as we approach lower redshifts, the values of the spherically-averaged distance, $D_\text{V}/r_\text{d}$, are systematically low with respect to \emph{Planck} $\Lambda$CDM. Such a deficit in the distances measured by BAO can be explained by a variety of models, with $w_0w_a$CDM mildly outstanding since it has the flexibility to better fit the LRG2 BAO measurement. Models such as $\Lambda$CDM$+\Omega_\text{K}$, $\Lambda$CDM with three massless neutrinos instead of the baseline neutrino with a mass of $0.06\,\mathrm{eV}$, and allowing for a high value of the optical depth $\tau$ (by excluding low-$\ell$ CMB polarization data) provide quite similar fits in $D_\mathrm{V}/r_\mathrm{d}$ and particularly $D_\text{M}/D_\text{H}$.
However, with the exception of $w_0w_a$CDM, none of these models are able to capture the data points in the AP measurements below $z<1$. The upcoming DESI full-shape analysis from galaxy and quasar clustering at these redshifts may shed further light on this feature, while the DR3 BAO analysis could provide further insights into the AP measurement at $z_\text{eff}=2.33$.

\begin{table}
   \centering
   \begin{tabular}{lcc}
   \toprule
   \textbf{Datasets} & $\mathbf{\Delta\chi}^2_\textbf{MAP}$ & \textbf{Significance} \\
   \midrule
   \multicolumn{3}{l}{$\mathbf{w_0w_a}$\textbf{CDM}} \\
   \midrule
   DESI & $-5.6$ & $1.9\sigma$ \\
   DESI+CMB & $-9.8$ & $2.7\sigma$ \\
   DESI+DES-Dovekie & $-6.7$ & $2.1\sigma$ \\
   DESI+CMB+DES-Dovekie & $-12.9$ & $3.2\sigma$ \\
   \addlinespace[5pt]
   \midrule
   \multicolumn{3}{l}{$\mathbf{\Lambda}$\textbf{CDM +} $\mathbf{\Omega_\textbf{K}}$} \\
   \midrule
   DESI & $-2.4$ & $1.5\sigma$ \\
   DESI+CMB & $-4.3$ & $2.0\sigma$ \\
   DESI+DES-Dovekie & $-4.3$ & $2.0\sigma$ \\
   DESI+CMB+DES-Dovekie & $-4.8$ & $2.2\sigma$ \\
   \bottomrule
   \end{tabular}
   \caption{Summary of the effective $\chi^2_\text{MAP}$ differences compared to $\Lambda$CDM for different models and datasets. The differences are listed for the best-fit $w_0w_a$CDM model relative to a $\Lambda$CDM model with $w_0=-1$ and $w_a=0$, and similarly for the best-fit model with spatial curvature compared to flat $\Lambda$CDM. The corresponding frequentist significance for the preference over flat $\Lambda$CDM is listed in the third column.}
   \label{tab:significances}
\end{table}

\section{Conclusions}
\label{sec:conclusions}

We have presented a measurement of the \alpac\ (AP) effect using the full shape of \lyaf\ correlation functions measured from the second DESI data release (DR2). Our data set includes over 820,000 \lyaf\ spectra and the positions of over 1.2 million quasars, which are used to compute the \lya-QSO cross-correlation. Our combined AP measurement corresponds to a $1\%$ constraint on the ratio $D_M/D_H$ at an effective redshift $z_\mathrm{eff}=2.33$, where $D_M$ is the comoving angular diameter distance and $D_H$ is the ratio of the speed of light and the Hubble parameter at $z_\mathrm{eff}$. This is the tightest constraint on $D_M/D_H$ at redshifts $z>1$ so far.

Our analysis is based on a template fit of the \lya\ auto-correlation and its cross-correlation with quasars. Following \cite{Cuceu2021}, our template is split into a peak component and a smooth component, similar to the approach used in BAO analyses. This allows us to separate the AP information from the broadband and the BAO peak, which means our analysis builds on the DESI DR2 BAO analysis by also extracting AP information from the broadband of \lyaf\ correlations (\Cref{subsec:alpac}). Relative to previous \lya\ full-shape and BAO analyses, we have also added two major improvements to our model for \lyaf\ correlations. These include a framework for marginalizing over small-scale information that is spread to large scales due to the quasar continuum fitting procedure (\Cref{subsec:ssm}), and a model for a scale-dependent \lyaf\ bias arising from larger-scale fluctuations in the UV background (\Cref{subsec:uvb}).

We have validated our analysis with two types of mocks and with a wide range of tests on blinded data, which are presented in detail in the companion paper \validation. For our validation, we use a threshold of one-third of the statistical uncertainty to determine whether a test passes or fails, in line with previous DESI \lya analyses \cite{DESI2024.IV.KP6,DR2Lya:2025}. The two types of mocks we use include \lyaf\ mocks from the \abacus\ simulations \cite{Hadzhiyska2023} and a new set of mocks based on second-order Lagrangian Perturbation Theory (2LPT), which are presented in a companion paper \cite{RuizHerrera:2026}. Our model configuration and scale cuts are based on tests with these two types of mocks as described in \Cref{subsec:mock_validation}. We target large scales in the \lyaf\ correlation functions, where a linear-theory approach provides a good fit and unbiased constraints. The tests with mocks show that our baseline analysis recovers unbiased AP constraints within our threshold (\Cref{subsec:mock_validation}). Finally, we also present BAO results from the new mocks, which led us to update the systematic error budget on BAO with respect to the one used in the DESI DR2 BAO analysis \cite{DR2Lya:2025} to include a slightly larger uncertainty in the AP direction (\Cref{subsec:bao_sys}).

In \Cref{subsec:data_validation} and the companion work \validation, we present the validation with blinded data. This includes several data splits and a large set of analysis variations designed to test the robustness of our measurement to different analysis and modeling choices. All data splits are consistent with each other, and we do not find significant outliers in our set of analysis variations. In particular, modeling variations show consistent results for various choices of how contaminants and small-scale non-linearities are treated, for more conservative scale cuts, and when using different priors. Based on these results, and the validation with mocks, we conclude that our AP result is robust (\Cref{subsec:validation_discussion}). The validation process was performed on blinded data, and the baseline analysis choices were decided before unblinding the measurement, with no further changes after unblinding.

Our \lyaf\ results are presented in \Cref{subsec:ap_res}, where we show the model fits to the \lya\ correlation functions and the broadband AP and BAO results. The full-shape model provides a good fit to the correlation functions, and we find that AP constraints from the broadband are consistent with those from the BAO peak. Our new AP constraint is a factor of $2$ tighter compared to the BAO measurement from the same data set \cite{DR2Lya:2025}. We use these constraints, along with the extra systematic uncertainty on BAO (\Cref{subsec:bao_sys}), to infer the corresponding cosmological distances in \Cref{subsec:cosmo_dist}. We present the $D_M/D_H$ constraint from our broadband AP measurement, along with the combined broadband AP and BAO constraints on the ratios $D_M(z_\mathrm{eff})/r_d$ and $D_H(z_\mathrm{eff})/r_d$, where $r_d$ is the size of the acoustic scale at the drag epoch and $z_\mathrm{eff}=2.33$. Our measurements correspond to $0.84\%$ and $0.77\%$ constraints on $D_M/r_d$ and $D_H/r_d$, respectively. The precision of these results exceeds the requirements of the full DESI survey in the redshift range $1.9<z<4.0$ \cite{DESI2016a.Science,DESI2023a.KP1.SV}, with just three years of data. 

We present the cosmological interpretation of our measurement alone and in combination with DESI BAO measurements at lower redshifts and other external data sets in \Cref{sec:cosmo}. The \lya\ full-shape (FS) measurements are consistent with both DESI BAO and CMB predictions within \lcdm. Furthermore, the inferred matter fraction, $\Omega_\mathrm{m}$, constraint from the \lyaf\ is consistent with SNe results. Combining our measurement with DESI BAO constraints from galaxy clustering leads to an $\sim8\%$ improvement with respect to DESI DR2 BAO (which includes \lya\ BAO), and a small shift to larger values of $\Omega_\mathrm{m}$. This leads to a decrease in the tension between DESI and the CMB from $2.4\sigma$ to $2.2\sigma$. We also combine our \lyaf\ measurement with a BBN prior on the baryon density in order to calibrate the acoustic scale. This results in a $2\%$ constraint on the Hubble constant, $H_0=66.5\pm1.3$, entirely from high-redshift data (\lya+BBN) and independent of CMB anisotropies.

\Cref{subsec:extended_models} presents our constraints on extended models beyond \lcdm\ from the DESI BAO+\lya\ FS combination. When considering models with an evolving dark energy equation of state parametrized by $w_0$ and $w_a$, the new \lyaf\ AP measurement is in $\sim2\sigma$ tension with the $w_0w_a$CDM prediction of AP at $z=2.33$ for both DESI+CMB and DESI+CMB+SNe. Therefore, adding \lya\ FS leads to a small decrease in the preference for $w_0w_a$CDM relative to \lcdm\ when combined with other data sets. For DESI+CMB, the preference decreases from $3.2\sigma$ to $2.7\sigma$, while for DESI+CMB+SNe, it has a milder impact, decreasing the tension from $3.4\sigma$ to $3.1\sigma$. We also present updated results on the sum of the neutrino masses, $\sum m_\nu$, where we find that the \lya\ result leads to a very mild relaxation of the upper bound on $\sum m_\nu$, due to its preference for a higher $\Omega_\mathrm{m}$ value. For models with free curvature, $\Omega_\mathrm{K}$, we find that DESI BAO+\lya\ FS is consistent with a flat Universe within $1.5\sigma$, while the DESI+CMB constraint is $2\sigma$ away from a flat Universe, in line with \cite{DESIDR2:2025}. Our cosmological results are compared and discussed in \Cref{subsec:cosmo_summary}.

\lyaf\ measurements provide the opportunity to test cosmological models in a unique regime at $z>2$, which has so far been difficult to access with large galaxy surveys. Our measurements come from a time when the Universe was still deep in the matter domination era, providing an anchor to low-redshift results for constraining the expansion history and the evolution of dark energy, and to the CMB for constraining physics during matter domination (\Cref{subsec:matter_distance}). The DESI DR2 results allow the \lyaf\ to begin to separate at the $\sim2\sigma$ level between state-of-the-art dark energy constraints for the first time. These measurements will continue to improve with future DESI data releases, and future methodology improvements may also allow us to access information on smaller scales leading to even tighter constraints.


\section{Data Availability}
The data used in this analysis will be made public along with Data Release 2 (details in \url{https://data.desi.lbl.gov/doc/releases/}). The data points corresponding to the figures from this paper will be available in a Zenodo repository.

\begin{acknowledgments}

This material is based upon work supported by the U.S. Department of Energy (DOE), Office of Science, Office of High-Energy Physics, under Contract No. DE–AC02–05CH11231, and by the National Energy Research Scientific Computing Center, a DOE Office of Science User Facility under the same contract. Additional support for DESI was provided by the U.S. National Science Foundation (NSF), Division of Astronomical Sciences under Contract No. AST-0950945 to the NSF’s National Optical-Infrared Astronomy Research Laboratory; the Science and Technology Facilities Council of the United Kingdom; the Gordon and Betty Moore Foundation; the Heising-Simons Foundation; the French Alternative Energies and Atomic Energy Commission (CEA); the Secretariat of Science, Humanities, Technology and Innovation (SECIHTI) of Mexico; the Ministry of Science, Innovation and Universities of Spain (MICIU/AEI/10.13039/501100011033), and by the DESI Member Institutions: \url{https://www.desi.lbl.gov/collaborating-institutions}. Any opinions, findings, and conclusions or recommendations expressed in this material are those of the author(s) and do not necessarily reflect the views of the U. S. National Science Foundation, the U. S. Department of Energy, or any of the listed funding agencies.

The authors are honored to be permitted to conduct scientific research on I'oligam Du'ag (Kitt Peak), a mountain with particular significance to the Tohono O’odham Nation.

\end{acknowledgments}

\bibliographystyle{apsrev4-2-5authors}
\bibliography{main}

\appendix

\section{Further details on SSM}
\label{sec:ssm_app}

\begin{figure}
\includegraphics[width=1.0\columnwidth,keepaspectratio]{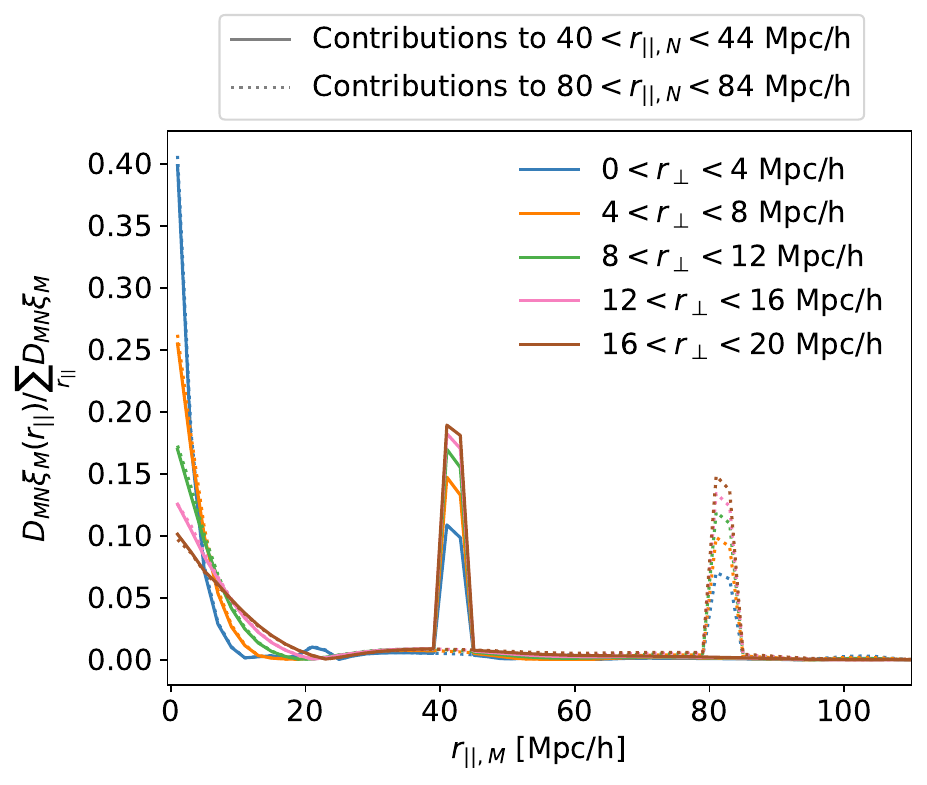}
    \caption{
    Fractional contributions to one bin of the distorted model $\xi_N$ from one transverse bin in the undistorted model $\xi_M$ as a function of the line-of-sight separation in the undistorted space, $r_{||,M}$. Different colors display different choices of the transverse bin, which is the same for both $\xi_M$ and $\xi_N$. We show contributions to two different distorted $r_{||,N}$ bins using continuous and dotted lines, respectively.
    }
    \label{fig:dmat}
\end{figure}

In this work, we introduced a new method for small-scale marginalization (SSM) of distorted \lyaf\ correlation functions, as described in \Cref{subsec:ssm}. Here, we wish to provide more details on the motivation and consequences of this new method.

We begin with the motivation for introducing the new SSM method. As shown in \Cref{eq:dm_xi,eq:dmat1,eq:dmat2}, we use a distortion matrix that multiplies our undistorted model $\xi_M$, such that the resulting distorted model $\xi_N$ can then be fitted to the data. The distortion matrix is built from the projection matrix of each forest in \Cref{eq:etas}, and results in a mixing of the information along the line-of-sight. We showcase this mixing in \Cref{fig:dmat}, which displays the fractional contributions from each undistorted line-of-sight bin $r_{||,M}$, to a given distorted line-of-sight bin $r_{||,N}$, for a fixed transverse bin $r_\bot = r_{\bot,M}=r_{\bot,N}$. Contributions to the $40<r_{||,N}<44$\hMpc\ bin are shown using continuous lines, while those to the $80<r_{||,N}<84$\hMpc\ bin are shown using dotted lines. Different colors indicate different choices of $r_{\bot}$.

\begin{figure}
\includegraphics[width=1.0\columnwidth,keepaspectratio]{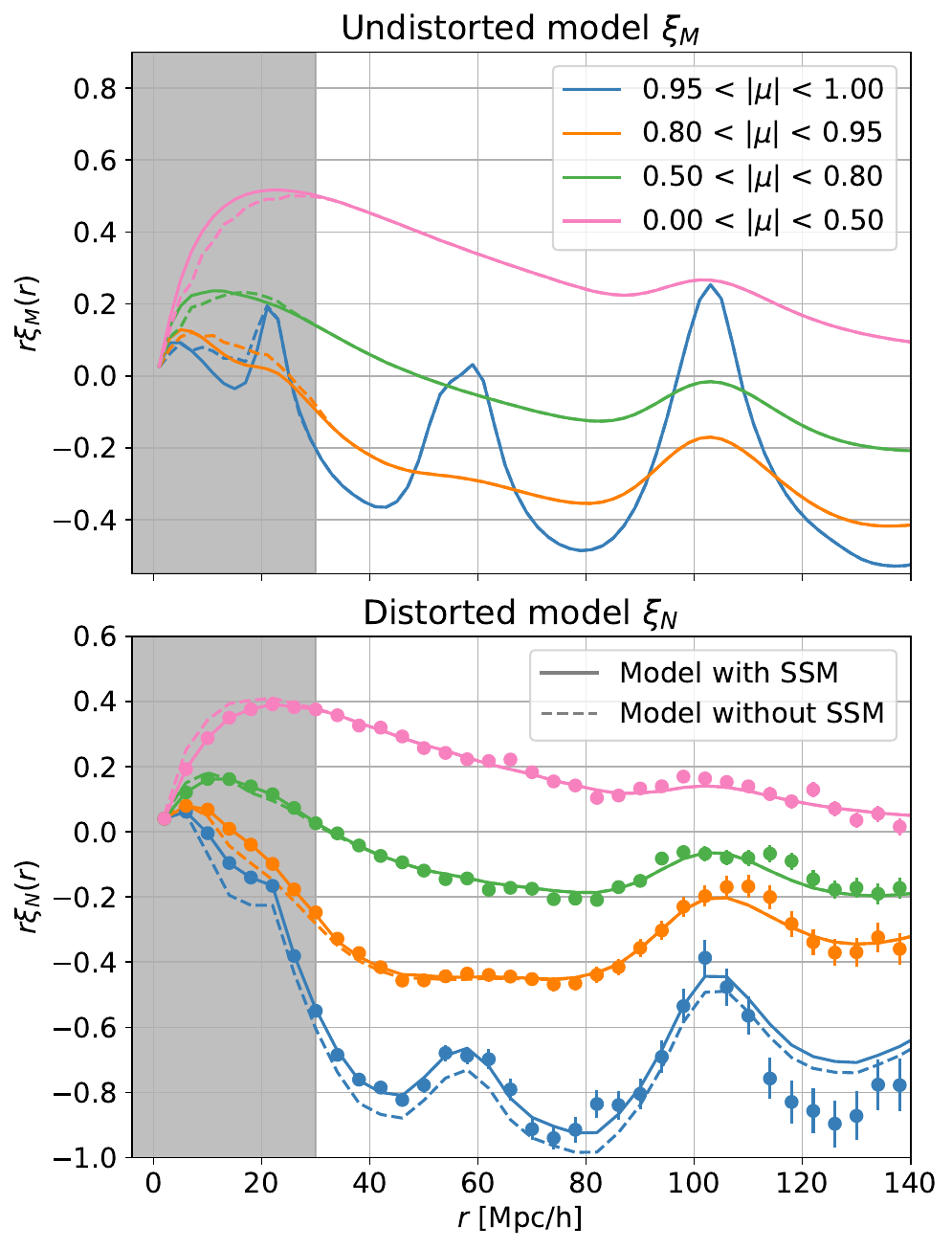}
    \caption{
    Best-fit model compressed into wedges as a function of $\mu$ and plotted in the undistorted space (top) and in the distorted space (bottom). We show the same model with and without the small-scale marginalization (SSM) terms as continuous and dashed lines, respectively. The data is also shown in the bottom panel.
    }
    \label{fig:ssm_wedges}
\end{figure}

\Cref{fig:dmat} shows that for each distorted bin $r_{||,N}$, a significant amount of information comes from small scales in the undistorted correlation $r_{||,M}<20$\hMpc. For the first $r_\bot$ bin shown in blue ($0<r_\bot<4$\hMpc), about $70\%$ of the information comes from the small scales, and only $\sim20\%$ comes from $r_{||,M}=r_{||,N}$. By the fifth $r_\bot$ bin shown in brown ($0<r_\bot<4$\hMpc), the contribution from small scales drops to $\sim45\%$ while that from $r_{||,M}=r_{||,N}$ increases to $\sim40\%$. On the other hand, comparing the dotted and continuous lines of the same color shows that the contribution from small scales is almost identical for very different choices of $r_{||,N}$. While not displayed in \Cref{fig:dmat}, when fitting small scales in $\xi_N$, the vast majority of the information, between $95\%$ and $99\%$ for $r_\mathrm{min}=30$\hMpc, comes from those same scales in $\xi_M$ (i.e., from $r<r_\mathrm{min}$).

This motivates our approach of introducing one free parameter to marginalize over each small-scale bin in $\xi_M$, while fitting the corresponding $\xi_N$ bins to the real data. Each small-scale ($r_{||,M}$, $r_{\bot,M}$) bin is marginalized over, and has a corresponding fitted ($r_{||,N}$, $r_{\bot,N}$) data bin. This ensures that only the large scales in our model are used to fit the data, and that we do not lose information due to the large number of extra degrees of freedom. Our method can be thought of as using the distortion matrix to deconvolve the data, and then applying scale cuts to the deconvolved data.

We show the impact of SSM on our best-fit model in \Cref{fig:ssm_wedges}. The continuous and dashed lines represent the same model, with and without the SSM terms.\footnote{This means that the model without SSM is not re-fitted to the data, which is why it is not a good fit to the line-of-sight wedge in the bottom panel.} The top panel shows the undistorted model, where the marginalized terms only affect scales $r<r_\mathrm{min}$ (displayed using the gray band). The bottom panel shows the distorted model and the data. In this case, the marginalized small-scale terms have an impact on all scales, especially in the line-of-sight wedge (blue). Furthermore, due to how our method is set up, the $r_N<r_\mathrm{min}$ data bins are perfectly fitted by our model with SSM.

We have validated the new SSM method using mocks, as discussed in \Cref{subsec:mock_validation} and \validation. In the case of contaminated \lpt\ mocks, we found that using SSM does not lead to a significant difference in AP constraints for $r_\mathrm{min}\geq30$\hMpc. On the other hand, for \abacus\ we found percent-level biases in AP, which disappear when using SSM, as seen in \Cref{fig:mocks_scale_cuts}.

Finally, we note that the SSM method relies on a choice of prior for the marginalization (the diagonal elements of the $\mathbf{S}$ matrix in \Cref{eq:ssm}). For numerical stability, we use a finite value $\mathbf{S}_{ii}=100$, which is significantly larger than the amplitude of the model bins that are marginalized (by about 4 orders of magnitude). We have tested that changing this value by a factor of 2 in either direction does not impact our results.

\section{Nuisance parameters}
\label{sec:nuisance}

\begin{table}

\centering
\begin{tabular}{lll}
\toprule
Parameter & Priors & Results \\
\midrule
$\alpha_p$ & $\mathcal{U}[0.50, 1.50]$ & $1.0007\pm 0.0061$ \\
$\phi_p$ & $\mathcal{U}[0.50, 1.50]$ & $1.001\pm 0.020$ \\
$\phi_s$ & $\mathcal{U}[0.50, 1.50]$ & $1.007\pm 0.011$ \\
\midrule
$\alpha_s$ & $\mathcal{U}[0.50, 1.50]$ & $1.044\pm 0.036$ \\
$f$ & $\mathcal{U}[0.00, 2.00]$ & $1.43^{+0.17}_{-0.20}$ \\
$b_{F}$ & $\mathcal{U}[-1.00, 0.00]$ & $-0.167^{+0.010}_{-0.015}$ \\
$\beta_{F}$ & $\mathcal{U}[0.00, 3.00]$ & $1.193^{+0.083}_{-0.12}$ \\
$b_{Q}$ & $\mathcal{U}[0.00, 6.00]$ & $3.49\pm 0.22$ \\
$\sigma_z\;[{\rm Mpc/h}]$ & $\mathcal{U}[0.00, 15.00]$ & $4.8\pm 1.6$ \\
$\Delta r_{||}\;[{\rm Mpc/h}]$ & $\mathcal{N}(0.0, 1.0)$ & $0.34\pm 0.26$ \\
$\xi_0^{\rm TP}$ & $\mathcal{U}[0.00, 2.00]$ & $<0.24$ \\
$b_{\rm HCD}$ & $\mathcal{U}[-0.50, 0.00]$ & $>-0.027$ \\
$\beta_{\rm HCD}$ & $\mathcal{N}(0.500, 0.090)$ & $0.498\pm 0.089$ \\
$L_{\rm HCD}\;[{\rm Mpc/h}]$ & $\mathcal{N}(5.0, 2.0)$ & $4.9\pm 1.8$ \\
$10^3 b_{{\rm SiII(1190)}}$ & $\mathcal{U}[-500.00, 0.00]$ & $-3.62^{+0.44}_{-0.38}$ \\
$10^3 b_{{\rm SiII(1193)}}$ & $\mathcal{U}[-500.00, 0.00]$ & $-3.12^{+0.42}_{-0.36}$ \\
$10^3 b_{{\rm SiIII(1207)}}$ & $\mathcal{U}[-500.00, 0.00]$ & $-9.50\pm2.1$ \\
$10^3 b_{{\rm SiII(1260)}}$ & $\mathcal{U}[-500.00, 0.00]$ & $-3.05^{+0.43}_{-0.38}$ \\
$10^3 b_{{\rm CIV(eff)}}$ & $\mathcal{N}(-19.0, 5.0)$ & $-19.80\pm5.1$ \\
$10^4 a_{\rm noise}$ & $\mathcal{U}[0.00, 100.00]$ & $2.24\pm0.2$ \\
$b_\gamma$ & $\mathcal{U}[-1.00, 1.00]$ & $0.143\pm 0.046$ \\
\bottomrule
\end{tabular}
\caption{Priors and parameter constraints (posterior mean and $68\%$ credible intervals) for the full set of parameters, including the main parameters used for cosmology (first three) and the nuisance parameters.}
\label{tab:nuisance}
\end{table}

Our model has a total of 21 free parameters, excluding those fitting small-scale bins which are analytically marginalized (\Cref{subsec:ssm}). We present the full set of priors and parameter constraints in \Cref{tab:nuisance}. We denote uniform priors as $\mathcal{U}[\mathrm{min}, \mathrm{max}]$, and Gaussian priors with mean $\mu$ and standard deviation $\sigma$ as $\mathcal{N}[\mu,\sigma]$. The choice of priors closely follows that of the DR2 BAO analysis \cite{DR2Lya:2025} and the DR1 \lya\ full-shape analysis \cite{Cuceu:2025}. The impact of the choice of Gaussian priors has been tested in the companion work \validation, and is discussed in \Cref{subsec:validation_discussion}. 

Several parameters have fairly different values compared to those obtained in the DR2 BAO analysis. This is because of a few different factors, including the change in scale cuts, the improved model, and the fact that we vary the scale parameters of the smooth component (which are kept fixed in BAO analyses). Finally, we note that nuisance parameter results are presented for completeness purposes, and advise against interpretation of these values in other contexts. This is because we have not validated their constraints, and throughout the analysis we simply treated them as parameters to be marginalized over.

\section{Updates on the CMB likelihood}
\label{sec:cmb}

\begin{figure}
\includegraphics[width=1.0\columnwidth,keepaspectratio]{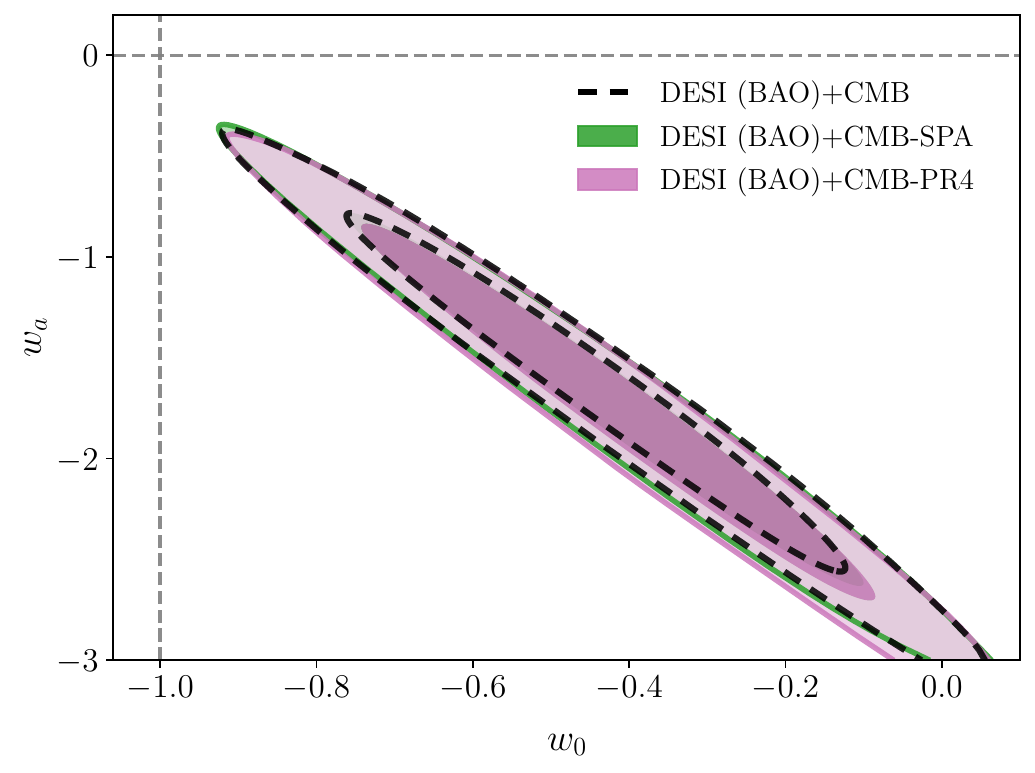}
    \caption{
     Constraints on the dark energy equation of state parameters, $w_0$ and $w_a$, from DESI BAO (without \lya\ FS) in combination with different CMB datasets. The contours represent the 68\% and 95\% credible intervals. The CMB-PR4 combination represents the likelihood choice from \cite{DESIDR2:2025}, the CMB-SPA combination is used in \cite{SPT-3G:2025bzu}, and the CMB combination is the baseline choice described in \Cref{subsec:external_data}.
    }
    \label{fig:cmb_likelihoods}
\end{figure}

Here, we provide a comparison of our baseline CMB combination with respect to other relevant CMB choices adopted in recent cosmological analyses. \Cref{fig:cmb_likelihoods} shows the comparison of constraints on $w_0w_a$CDM when DESI BAO is combined with CMB. We opt not to include the Ly$\alpha$ FS data here for a direct comparison to \cite{DESIDR2:2025}. With DESI BAO and CMB data, the preference for evolving dark energy is around 3$\sigma$ with minor 0.1-0.2$\sigma$ variations depending on the choice of CMB dataset.

\end{document}